\documentclass{aa}  

\usepackage{graphicx}
\usepackage[varg]{txfonts}
\usepackage{longtable} % for 'longtable' environment
\usepackage{pdflscape} % for 'landscape' environment
\usepackage{rotating}
\usepackage{lscape}
\usepackage{gensymb}
\usepackage[utf8]{inputenc}
\DeclareUnicodeCharacter{2212}{-}
\usepackage{float}
\usepackage{color}
\usepackage{afterpage}
\usepackage{orcidlink}
\usepackage{enumitem}
\usepackage{ulem}

\begin{document}

   \title{Identifying tidal disruption events among radio transient galaxies}

   \author{M. Kunert-Bajraszewska
          \inst{1}
          \orcidlink{0000-0002-6741-9856}
          \and
          D. Kozieł-Wierzbowska
          \inst{2}
          \orcidlink{0000-0003-4323-0984}
          \and
          D. Stern
          \inst{3}
          \orcidlink{0000-0003-2686-9241}
          \and
          A. Krauze
          \inst{1}
          \orcidlink{0009-0002-5915-4592}
          \and
           N. Zafar
          \inst{1}
          \orcidlink{0009-0005-2076-9217}
          \and
          T. Connor
          \inst{3,4}
          \orcidlink{0000-0002-7898-7664}
          \and
          M. J. Graham
          \inst{5}
          \orcidlink{0000-0002-3168-0139}
          }

   \institute{Institute of Astronomy, Faculty of Physics, Astronomy and Informatics, NCU, Grudzi\k{a}dzka 5/7, 87-100, Toru\'n, Poland\\
              \email{magda@astro.uni.torun.pl}
         \and
             Astronomical Observatory of Jagiellonian University, ul. Orla 171, 30-244 Krak\'ow, Poland
         \and    
             Jet Propulsion Laboratory, California Institute of Technology, 4800 Oak Grove Drive, Pasadena, CA 91109, USA
         \and
             Center for Astrophysics, Harvard\ \&\ Smithsonian, 60 Garden St., Cambridge, MA 02138, USA
         \and
             California Institute of Technology, 1200 E. California Blvd, Pasadena, CA 91125, USA\\
        }

  \date{Received ...; accepted ....}

 \abstract
{We present the optical and infrared properties of a sample of 24 radio transient sources discovered in the Very Large Array Sky Survey (VLASS). Previous studies of their radio emission showed that these sources resemble young gigahertz-peaked spectrum (GPS) radio sources, but are less powerful and characterized by low-power jets. The bursts of radio activity in most cases are likely due to intrinsic changes in the accretion processes. However, for a few sources in this sample, we cannot rule out the possibility that their radio variability results from a tidal disruption event (TDE).
In this work, we extend our analysis to the optical and infrared regimes, confirming that our sample of radio transients is not homogeneous in terms of their optical and infrared properties either. The host galaxies of most of these sources are massive ellipticals with emission dominated by active galactic nuclei (AGN). They host supermassive black holes (SMBHs) with masses typical of radio-loud AGNs ($\rm >10^7\,M_{\odot}$), but exhibit very low accretion activity.
In contrast, the sources for which a TDE origin is suspected are either pure star-forming galaxies or show significant ongoing star formation, similar to radio-selected optically-detected TDEs. Additionally, two of them exhibit infrared flares characteristic of TDEs, while the remaining sources do not display significant variability outside the radio regime. Moreover, the evolution of their radio brightness in the W3/radio diagnostic diagram—which we employ in our analysis—also sets our TDE candidates apart from the rest of the sample and resembles the radio variability seen in optically discovered TDEs with radio emission.
Finally, based on our findings, we hypothesize that the mid-IR/radio relation can serve as a tool to distinguish between radio transients caused by TDEs and those originating from intrinsic AGN variability.
}


   \maketitle
%
%-------------------------------------------------------------------

\section{Introduction}

Active Galactic Nuclei (AGN) are characterized by energetic emission across the electromagnetic spectrum. Some exhibit strong radio emission, classifying them as radio-loud, while others are categorized as radio-quiet. The transition of an AGN from a radio-quiet to a radio-loud state is a subject of growing interest as it provides insight into the mechanisms governing AGN activity and the interplay between the central supermassive black hole (SMBH) and its surrounding galactic environment.

Modern optical-UV time-domain surveys have led to the discovery of two types of variable accretion onto SMBHs, both of which can trigger bursts of radio activity and potentially lead to the formation of radio jets. One proposed mechanism for this variability involves changes in the accretion rate, which can result in episodic radio activity \citep{Reynolds, Czerny, MKB10, An}. According to this model, AGNs may experience transient states, oscillating between radio-quiet and radio-loud phases. These transitions can occur over relatively short timescales, $\rm <10^5\,years$, reflecting changes in accretion dynamics or jet geometry.
A second mechanism is a tidal disruption event (TDE), in which a star passing near a black hole is torn apart by tidal forces. The stellar debris then accretes onto the black hole, which often has been previously dormant, triggering a short-lived accretion episode lasting up to several years and producing a luminous flare across multiple wavelengths. It is currently estimated that up to $\sim40\%$ of TDEs exhibit flares in the radio band \citep{Horesh2021, Sfaradi2022,Goodwin2022, Cendes2024}. However, in most cases, these flares are associated with thermal, non-relativistic low luminosity outflows whose underlying mechanism remains poorly understood \citep{vanVelzen2011,Alexander2020, Gezari2021}. In rare cases, a TDE can generate a powerful relativistic radio jet \citep{Eftekhari2018, Mattila2018}. Additionally, an increasing number of objects have been observed to exhibit signatures of both AGN activity and TDEs. Most recently, a TDE has even been proposed as a possible cause of episodic radio activity lasting thousands of years, potentially leading to the development of small, jetted radio structures \citep{Readhead2024, Sullivan2024}.

The ability to study transient radio emission has recently improved due to new multi-epoch radio surveys such as the Caltech-NRAO Stripe 82 Survey \citep[CNSS;][]{Mooley, Mooley2019}, the Very Large Array Sky Survey \citep[VLASS;][]{Lacy2020PASP}, and the Australian Square Kilometre Array Pathfinder Variables and Slow Transients Survey \citep[ASKAP VAST;][]{Murphy2013}. These surveys enable both the detection of radio emission in optically identified TDEs \citep{Cendes2024, Somal2024} and the characterization of pure radio transients. Recent studies suggest that the latter are most likely caused by intrinsic changes in accretion properties, similar to what is observed in changing-state AGNs \citep{MKB2020, Nyland, Wolowska}, but are characterized by low-power jets and short activity timescales. However, for a subset of these objects, the possibility that they are associated with radio-emitting TDEs cannot be ruled out \citep{MKB2025}. Ultimately, distinguishing between TDEs and AGN flares remains challenging without a better understanding of the mechanisms responsible for radio emission in TDEs \citep{Merloni2015,Yang2019,Chan2019,Somal2024}.

In this paper, we report on the host galaxies of a sample of 24 slow radio transients discovered using the VLASS survey. All of these sources exhibit a significant increase in flux density (>200\%) over several decades. \citet{MKB2025} reported on their radio evolution following the onset of radio activity. Here, we present our original spectroscopic observations along with an analysis of optical and infrared data for these sources.

Throughout this paper, we assume a concordance cosmology with $\rm H_0= 70\,km\,s^{−1}~Mpc^{−1}$, $\rm \Omega_M= 0.3$, and $\rm \Omega_{\Lambda}= 0.7$.

\section{The sample and its observation}
\label{sec:sample}
\begin{figure*}[t]
\centering
\includegraphics[scale=0.4]{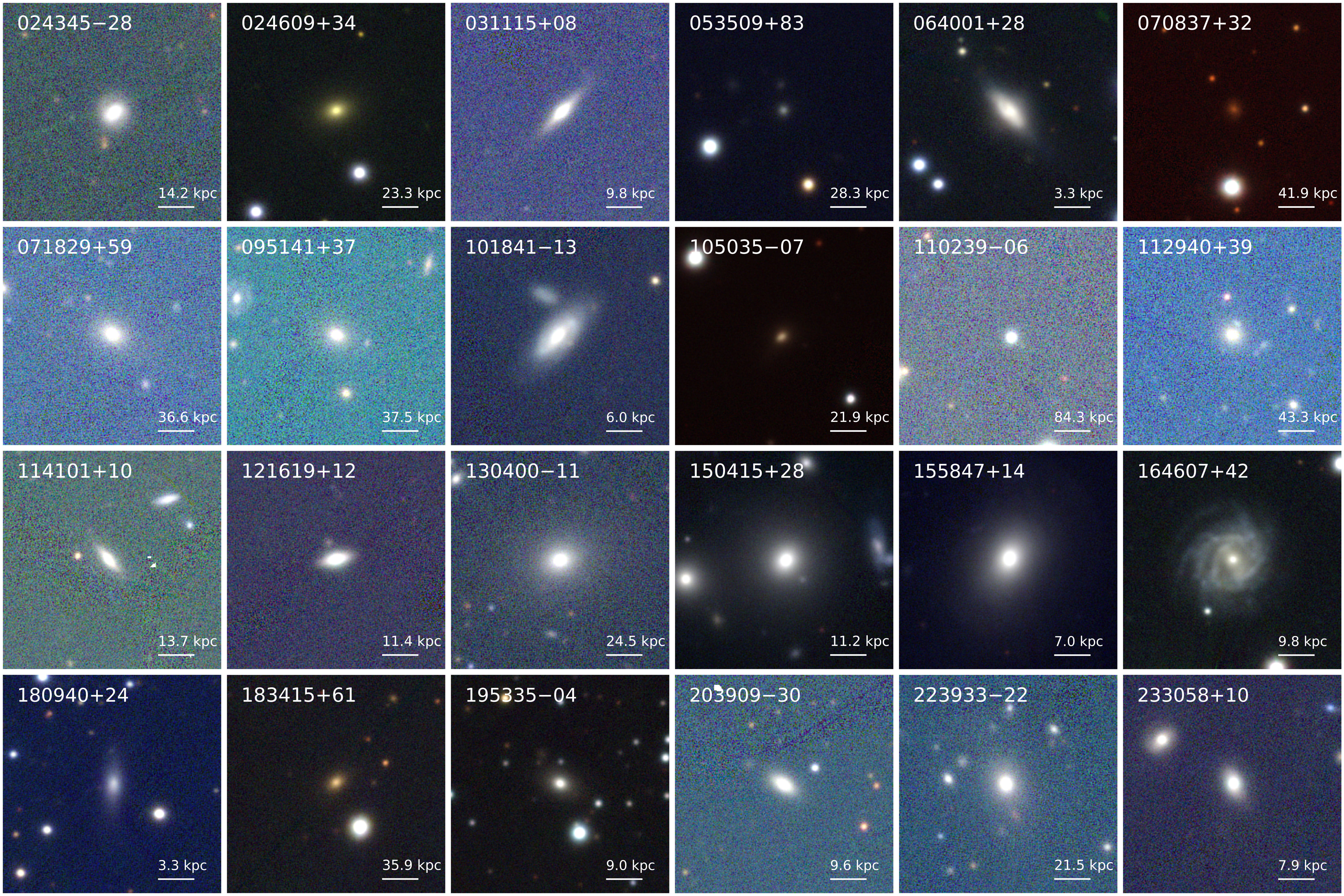}
\caption{Cutout stack  {\it g/r/i} images $(1'\times1')$ from Pan-STARRS. The target source is located at the center of each image. The scale bar is size 10$''$.
}
\label{figure:cut-outs}
\end{figure*}

The criteria for selecting the sample of 24 radio transients are described in detail in \citet{MKB2025} together with a discussion of their radio properties. Here we present 
analysis of their optical and infrared properties based on our spectroscopic observations and those available from other surveys. Basic data about the sources are provided in Table \ref{table:basic} and Figure \ref{figure:cut-outs} shows their Panoramic Survey Telescope \& Rapid Response System (Pan-STARRS) cutout images. The sample contains 23 galaxies and one quasar.

Between 2021 and 2023, we obtained recent optical spectra of all the sources in this sample using a combination of the Double Spectrograph (DBSP) mounted on the Palomar 200-inch Hale Telescope and the Low-Resolution Imaging Spectrograph \citep[LRIS;][]{Oke1995} mounted on the Keck~I telescope. The sources 130400$−$11, 155847$+$14, 164607$+$42 and 203909$−$30 were observed using LRIS, while the rest were observed using DBSP. The dates of the observations are given in Table~\ref{table:host_galaxies}. Both instruments are double spectrographs, where a dichroic splits the light between optimized blue and red arms.  The DBSP observations used the 1.5\arcsec\, wide longslit, the 5500~\AA\, dichroic, the 600/4000 blue grating (resolving power $R \equiv \lambda/\Delta \lambda = 1200$), and the 316/7500 red grating ($R = 750$). The LRIS observations used the 1.5\arcsec\, wide longslit, the 5600~\AA\, dichroic, the 600/3400 blue grating (resolving power $R \approx 1000$), and the 400/8500 red grating ($R \approx 1000$). 
We processed the data using the IRAF package \citep{Tody86,Tody93} and our own custom pipeline with standard procedures, including bias correction, flat-fielding, cosmic ray removal, and sky subtraction. The spectra were flux-calibrated using standard stars from \citet{Massey1990}, observed on the same nights.
The reduced spectra are presented in Appendix \ref{app:opt_spec}, Figure~\ref{figure:optical_spectra}.

Between 2001 and 2008, SDSS obtained spectroscopic observations for six of the 24 objects in our sample. These objects—095141$+$37, 112940$+$39, 114101$+$10, 121619$+$12, 155847$+$14, and 164607$+$42—are shown in Figure \ref{figure:optical_sdss_spectra}. All spectra have been corrected for Galactic extinction using the reddening map of \citet{Schlafly} and shifted to the rest-frame wavelength based on their SDSS redshifts.

\begin{table}
\small
\caption{The sample of 24 transient radio sources.}
\begin{tabular}{c c c c l c}
\hline\hline
\noalign{\smallskip}
ID &Name & RA    & DEC                        & \multicolumn{1}{c}{\it z} & {\it r}  \\
   &  & h~m~s & $\degr$~$\arcmin$~$\arcsec$&   & (mag)   \\
 (1) & (2)   & (3)                        &\multicolumn{1}{c}{(4)}& \multicolumn{1}{c}{(5)}&(6) \\  
\noalign{\smallskip}
\hline
1& 024345$-$28&02 43 45.7&$-$28 40 39.4	&0.075	& 17.99\\
2& 024609$+$34&02 46 09.3&$+$34 08 20.3	&0.131	& 18.22 \\
3& 031115$+$08&03 11 15.5&$+$08 51 32.6	&0.050	& 18.17\\
4& 053509$+$83&05 35 09.9&$+$83 45 41.3	&0.165	& 18.47 \\
5& 064001$+$28&06 40 01.8&$+$28 03 07.1	&0.016	& 17.66\\
6& 070837$+$32&07 08 37.8&$+$32 04 20.6	&0.275	& 19.87\\
7& 071829$+$59&07 18 29.3&$+$59 42 52.0	&0.229	& 18.79\\
8& 095141$+$37&09 51 41.6&$+$37 03 34.6	&0.236	& 18.85\\
9& 101841$-$13&10 18 41.4&$-$13 00 06.5	&0.030	& 17.58\\
10& 105035$-$07&10 50 35.7&$-$07 45 02.1	&0.122	& 17.86\\
11& 110239$-$06&11 02 39.2&$-$06 11 50.0	&1.393$\ast$& 18.37\\
12& 112940$+$39&11 29 40.0&$+$39 00 46.5	&0.288	& 19.62\\
13& 114101$+$10&11 41 01.7&$+$10 28 21.4	&0.072	& 18.96\\
14& 121619$+$12&12 16 19.6&$+$12 40 27.7	&0.059	& 17.24\\
15& 130400$-$11&13 04 00.7&$-$11 58 57.5	&0.139	& 17.63\\
16& 150415$+$28&15 04 15.9&$+$28 29 47.9	&0.058	& 16.54\\
17& 155847$+$14&15 58 47.7&$+$14 12 13.4	&0.035	& 16.53\\
18& 164607$+$42&16 46 07.0&$+$42 27 37.5	&0.050	& 18.03\\
19& 180940$+$24&18 09 40.5&$+$24 36 40.9	&0.016	& 19.00\\
20& 183415$+$61&18 34 15.4&$+$61 24 00.6	&0.223	& 19.01\\
21& 195335$-$04&19 53 35.2&$-$04 52 03.5	&0.046	& 17.46\\
22& 203909$-$30&20 39 09.1&$-$30 45 20.4	&0.049	& 18.18\\
23& 223933$-$22&22 39 33.2&$-$22 31 26.8	&0.119	& 17.87\\
24& 233058$+$10&23 30 58.9&$+$10 00 30.2	&0.040	& 17.47\\
\noalign{\smallskip}
\hline
\end{tabular}
\vspace{0.1 in}
\tablefoot{
The columns are marked as follows: (1) the identifying number used in figures; (2) source name; (3,4) VLA coordinates in J2000.0; (5) spectroscopic redshift, $\ast$ indicates the only quasar on this list; (6) magnitude in {\it r} ﬁlter from Pan-STARRS.}
\label{table:basic}
\end{table}

\subsection{Emission line measurements}
\label{emission_lines}
The spectra of 23 galaxies in our sample are contaminated by host galaxy starlight. To account for this, we processed them using the STARLIGHT code\footnote{\url{www.starlight.ufsc.br}}, which fits their continua and measures the stellar velocity dispersion, $\sigma_{*}$. STARLIGHT is an inverse stellar population synthesis code that reconstructs a galaxy's stellar populations by fitting a pixel-by-pixel model to the observed spectrum. To model the optical spectra of our radio sources, we employed two sets of synthetic simple stellar populations (SSSPs) from \citet{Bruzual.Charlot.2003a}, containing 45 and 150 components, respectively, using the extinction law of \citet{Cardelli1989} with $\rm R = 3.1$. The final results are based on the fits using 45 components, while the 150 component SSSPs fits were used for verification.
Additionally, to estimate the stellar velocity dispersion, we performed separate fits for the blue and red portions of the spectra. This procedure ensured that instrumental dispersion, which varies between instruments and spectral ranges, was properly accounted for. We assumed no contribution from AGN emission to the continuum, which may lead to an overestimation of the young stellar population fraction in Seyfert 1 galaxies but does not affect velocity dispersion or black hole mass estimates. A summary of the stellar population modelling for our sources is provided in Appendix \ref{app:notes} while the comparison between observed spectra and STARLIGHT models is given in Appendix \ref{app:opt_spec}.

For the only quasar in our sample, 110239$−$06, whose continuum is dominated by AGN emission, we used the PyQSOFit tool to model its continuum and Fe spectrum. To remove the Fe emission contribution, we applied UV iron templates from \citet{Vestergaard2001}, convolving them with a Gaussian profile to account for kinematic broadening of Fe lines.

Emission line fluxes were measured via Gaussian fitting after subtracting the modelled stellar spectrum or quasar continuum and Fe emission. These measurements are listed in Table \ref{table:emission_line_measurements}. A line was considered detected if its signal-to-noise ratio (S/N) was $\geq 3$. 
The uncertainty in line flux measurements was estimated as the maximum uncertainty, calculated as half the difference between the highest and lowest values in the series of measurements.

\subsection{Optical and infrared light curves}
\label{optical_infrared_lightcurves}

To investigate potential brightness variations in the optical and infrared domains for the studied sources, we constructed light curves using publicly available data from the BHTOM\footnote{https://bh-tom2.astrolabs.pl/} database \citep{Wyrzykowski2024}. This platform aggregates measurements from the following surveys: the Catalina Real-Time Transient Survey \citep[CRTS;][]{Drake}, Zwicky Transient Facility \citep[ZTF;][]{Graham2019}, ASAS-SN Sky Patrol \citep{Shappee2014, Hart2023}, and the Wide-field Infrared Survey Explorer \citep[WISE;][]{Wright}, including its continuation under the NEOWISE mission \citep{Mainzer2014}. Specifically, the ZTF data correspond to forced PSF photometry, ASAS-SN and CRTS provide standard aperture photometry, and WISE/NEOWISE use PSF photometry.
We do not apply explicit corrections for Galactic extinction or host galaxy subtraction. The relatively low extinction values for most of our sources ($ \lesssim 0.3$ mag) have a negligible impact on the temporal variability trends. Host subtraction is not feasible using the available photometry alone and would require dedicated image differencing, which is beyond the scope of this study.
We also do not apply any additional SNR or photometric quality cuts beyond the default filters implemented in the respective survey catalogues. Nevertheless, the variability analysis procedure described below—based on monthly binning and outlier removal—effectively mitigates the impact of low-quality measurements.

WISE operated from 2010 to 2011, observing in four infrared bands: W1 (3.4 $\mu$m), W2 (4.6 $\mu$m), W3 (12 $\mu$m), and W4 (22 $\mu$m). Since 2013, the mission has continued under the name NEOWISE, with observations limited to the W1 and W2 bands. 
In Table~\ref{table:wise_radio_measurements}, we list infrared brightness measurements for our sources based on the AllWISE catalogue, which combines WISE observations from 2010–2011 and provides higher-quality, averaged photometry in all four bands. These values are used in Figures~\ref{figure:wise_color_color} and~\ref{figure:flux_wise_radio}, while our additional analysis of the infrared properties is based on light curves in the W1 and W2 bands from the ongoing NEOWISE mission.

We first visually inspected these light curves to identify any significant ($\rm >0.2~mag$) brightness variations or flares. No variability was detected in most sources. The exceptions were the quasar 110239$−$06, which exhibits typical AGN variability \citep{Caplar2017}, and three galaxies (064001$+$28, 101841$−$13, and 180940$+$24) that show significant infrared brightness variations, with possible optical variability also observed in 064001$+$28.

To quantitatively confirm these findings, we partially followed the method of \citet{Zhang2022}. For each source and each filter, we first computed monthly-averaged brightness values, considering only months with at least five measurements. We then removed outliers exceeding 3$\sigma$ within each month and recalculated the monthly averages. Finally, we determined the apparent amplitude of variability ($\rm V_{mag}$) for each source and compared it to the typical photometric error. The $\rm V_{mag}$ was defined as the standard deviation of the magnitudes from the monthly-binned photometric points, while the typical error level was estimated using the root mean square (RMS) of the photometric uncertainties. This approach confirmed significant infrared brightness variations in the three previously identified sources. However, it remains insensitive to small optical brightness changes in 064001$+$28, which we discuss further in Section \ref{sec:variability}.

\begin{table*}[ht]
\caption{Physical parameters of the host galaxies of 24 radio transient sources.}
\centering
\small
\begin{tabular}{l r c r c c r c r c c}
\hline\hline
\noalign{\smallskip}
\multicolumn{1}{c}{Name} & \multicolumn{1}{c}{Epoch}& \multicolumn{1}{c}{Morph} & \multicolumn{1}{c}{$\rm log M_{*}$} & 
\multicolumn{1}{c}{$\rm A_v$} & \multicolumn{1}{c}{$\rm \sigma_{*}$} & \multicolumn{1}{c}{$\rm log M_{BH}$} & \multicolumn{1}{c}{$\rm log L_{bol}$}& \multicolumn{1}{c}{$\rm \lambda_{Edd}$} & BPT& WHAN\\
& & Class& \multicolumn{1}{c}{(M$_{\odot}$)} & \multicolumn{1}{c}{(mag)} & \multicolumn{1}{c}{($\rm km\,s^{-1}$)} & \multicolumn{1}{c}{(M$_{\odot}$)} & \multicolumn{1}{c}{$\rm (erg\,s^{-1})$}&\multicolumn{1}{c}{$\rm (10^{-2})$}& Class& Class\\
\multicolumn{1}{c}{(1)} & \multicolumn{1}{c}{(2)}   & \multicolumn{1}{c}{(3)} &\multicolumn{1}{c}{(4)}& \multicolumn{1}{c}{(5)} & \multicolumn{1}{c}{(6)} &\multicolumn{1}{c}{(7)} & (8) & \multicolumn{1}{c}{(9)}& (10)& (11)\\
\noalign{\smallskip}
\hline
\noalign{\smallskip}
024345$-$28 & Sep 2021& S & 9.98$\pm$0.41 & 0.83 & 111$\pm$17 & 7.36$\pm$0.30 & $43.07^{+0.02}_{-0.03}$ & 0.39 &LINER & sAGN\\
\noalign{\smallskip}
024609$+$34 & Sep 2021& E & 11.37$\pm$0.29 & 0.96 & 200$\pm$31 & 8.49$\pm$0.30 & $-$& \multicolumn{1}{c}{$-$}& $-$ & retired\\
\noalign{\smallskip}
031115$+$08 & Sep 2021& L/S & 10.13$\pm$0.30 & 0.26 & 54$\pm$10 & 5.98$\pm$0.34 & $-$&\multicolumn{1}{c}{$-$}& $-$&$-$\\
\noalign{\smallskip}
053509$+$83 & Sep 2021& E & 10.65$\pm$0.38 & 0 & 193$\pm$80 & 8.42$\pm$0.79 & $-$&\multicolumn{1}{c}{$-$}& $-$&$-$ \\
\noalign{\smallskip}
064001$+$28 & Sep 2021& L/S & 9.93$\pm$0.11 & 0.68 & 59$\pm$11 & 6.17$\pm$0.36 & $<40.95$
& $<0.05$ & SF & SF\,/\,sAGN \\
\noalign{\smallskip}
070837$+$32 & Jan 2023& E & 11.05$\pm$0.30 & 0.54 & 128$\pm$6 & 7.65$\pm$0.09 &$-$& \multicolumn{1}{c}{$-$}& $-$&sAGN\\
\noalign{\smallskip}
071829$+$59 & Jan 2023& E & 11.55$\pm$0.13 & 0.43 & 320$\pm$73 & 9.38$\pm$0.44 & $>43.60$& $>0.02$&LINER&sAGN \\
\noalign{\smallskip}
095141$+$37 & Dec 2003& E & 11.21$\pm$0.52 & 0 & 179$\pm$3 & 8.28$\pm$0.27 & $<40.10$
&$<0.05$& \multicolumn{1}{c}{$-$}&SF \\
\noalign{\smallskip}
            & Feb 2022& E & 11.62$\pm$0.19 & 0.33 & 198$\pm$23 & 8.47$\pm$0.22 & $<42.96$
            &$<0.03$& $-$&sAGN \\
\noalign{\smallskip}
101841$-$13 & Feb 2022& S & 10.23$\pm$0.12 & 0.83 & 108$\pm$6 & 7.31$\pm$0.11& $42.77^{+0.03}_{-0.05}$&0.22& AGN&sAGN\\
\noalign{\smallskip}
105035$-$07 & May 2023& D & 11.00$\pm$0.49 & 0.08 & 224$\pm$27 & 8.71$\pm$0.23 & $-$&\multicolumn{1}{c}{$-$}& $-$&$-$\\
\noalign{\smallskip}
110239$-$06	& Jan 2023&$-$& $-$   &$-$&$-$  & 8.97$\pm$0.08& $46.01^{+0.08}_{-0.10}$& 8.43&\multicolumn{1}{c}{$-$}&$-$\\
\noalign{\smallskip}
112940$+$39 & Mar 2005& E & 11.52$\pm$0.22 & 0 & 299$\pm$11 & 9.25$\pm$0.07 &$-$& \multicolumn{1}{c}{$-$}& $-$&$-$\\
\noalign{\smallskip}
            & Jan 2023& E & 11.48$\pm$0.31 & 0.38 & 225$\pm$59 & 8.72$\pm$0.50 &$-$&\multicolumn{1}{c}{$-$} & $-$&$-$\\
\noalign{\smallskip}
114101$+$10 & May 2003& L & 9.75$\pm$0.44 & 0.12 & 77$\pm$9 & 6.68$\pm$0.21 & $41.63^{+0.05}_{-0.11}$&0.07& LINER& wAGN\,/\,sAGN\\
\noalign{\smallskip}
            & Jan 2023& L & 10.30$\pm$0.47 & 0.92 & 60$\pm$23 & 6.21$\pm$0.71 &$-$& \multicolumn{1}{c}{$-$}& $-$&wAGN\\
\noalign{\smallskip}
121619$+$12 & May 2004& L & 9.53$\pm$0.42 & 0 & 60$\pm$8 & 6.21$\pm$0.25 & $>41.51$&$>0.15$ &$-$&$-$\\
\noalign{\smallskip}
            & Jan 2023& L & 10.01$\pm$0.41 & 0 & 81$\pm$21 & 6.77$\pm$0.50 & $-$&\multicolumn{1}{c}{$-$}& $-$&$-$\\
\noalign{\smallskip}
130400$-$11 & Jun 2023& E & 10.98$\pm$0.39 & 0 & 269$\pm$16 & 9.06$\pm$0.11 & $-$&\multicolumn{1}{c}{$-$}& $-$&$-$\\
\noalign{\smallskip}
150415$+$28 & Sep 2021& E & 11.38$\pm$0.57 & 0 & 258$\pm$18 & 8.97$\pm$0.13 & $-$&\multicolumn{1}{c}{$-$}& $-$&$-$\\
\noalign{\smallskip}
155847$+$14 & Apr 2008& E & 10.16$\pm$0.39 & 0 &  199$\pm$14  & 8.48$\pm$0.14 & $>41.48$&$>0.01$& LINER&retired\\
\noalign{\smallskip}
            & Sep 2021& E & 10.80$\pm$0.53 & 0 & 153$\pm$26 & 7.99$\pm$0.25 & $ 41.71$&$ 0.01$& LINER & wAGN\\
\noalign{\smallskip}
164607$+$42 & Jun 2001& S & 10.41$\pm$0.22 & 0 & 41.1$\pm$18 & 5.48$\pm$0.83 & $<41.34$
&$<0.55$& SF&sAGN\\
\noalign{\smallskip}
            & Sep 2021& S & 10.94$\pm$0.51 & 0.25 & 63$\pm$14 & 6.29$\pm$0.43 & $<42.01$
            &$<0.40$& Composite&sAGN\\
\noalign{\smallskip}
180940$+$24 & Sep 2021& S? & 9.17$\pm$0.46 & 0.26 & 81$\pm$8 & 6.76$\pm$0.18& $-$&\multicolumn{1}{c}{$-$}& SF&SF\\
\noalign{\smallskip}
183415$+$61 & Jul 2021& E & 11.36$\pm$0.43 & 0 & 241$\pm$42 & 8.85$\pm$0.33 & $>42.19$&$>0.01$& $-$&wAGN\\
\noalign{\smallskip}
195335$-$04 & Sep 2021& E & 9.99$\pm$0.50 & 0.39 & 58$\pm$13 & 6.14$\pm$0.42 & $>41.94$&$>0.48$& $-$&$-$\\
\noalign{\smallskip}
203909$-$30 & Sep 2021& E? & 9.55$\pm$0.59 & 0 & 34$\pm$10 & 5.11$\pm$0.54 & $-$&\multicolumn{1}{c}{$-$}& $-$&$-$\\
\noalign{\smallskip}
223933$-$22 & Sep 2021& E & 11.23$\pm$0.51 & 0.79 & 237$\pm$11 & 8.81$\pm$0.08 &$-$& \multicolumn{1}{c}{$-$}& $-$&$-$\\
\noalign{\smallskip}
233058$+$10 & Sep 2021& E? & 10.01$\pm$0.55 & 0.18 & 297$\pm$48 & 9.24$\pm$0.31 &$-$& \multicolumn{1}{c}{$-$}& SF&SF\\
\noalign{\smallskip}
\hline

\end{tabular}
\vspace{0.1 in}
\tablefoot{
The columns are marked as follows: (1) source name; (2) observation epoch; (3) morphological classification of the host galaxy: S - spiral, E - elliptical, L - lenticular, D - distorted ; (4) stellar mass; 
(5) stellar extinction, '0' indicates a very small extinction value that the STARLIGHT program was unable to reliably estimate; (6) velocity dispersion; (7) log of the black hole mass; (8) log of the bolometric luminosity; (9) Eddington ratio; (10)/(11) BPT/WHAN classification.}
\label{table:host_galaxies}
\end{table*}

\section{Calculations of physical parameters}
\label{sec: physical_parameters}

The black hole mass of quasar was estimated from the MgII$\lambda$2799 line and luminosity at 3000$\rm \AA$, using the following relation \citep{Trakhtenbrot}:
\begin{equation}
{\rm \frac{\mathrm{M_{BH}}}{\mathrm{M_{\odot}}}=5.6\times10^6 \left( \frac{\lambda L_{3000}}{10^{44} ~\mathrm{erg~s}^{-1}} \right)^{0.62}
\left[ \frac{\mathrm{FWHM(MgII)}}{10^3~\mathrm{km~s}^{-1}} \right]^2} .\\
\end{equation}

Then the bolometric AGN luminosity was calculated (Table \ref{table:host_galaxies}) using $\rm \lambda L_{3000}$ and a conversion factor of 5.3 to convert from monochromatic to bolometric luminosity \citep{Runnoe}.

In the case of sources with a significant host contribution in the spectra, we used the established scaling relation $\rm M_{BH}-\sigma_{*}$ \citep{Kormendy} to get their black hole masses:

\begin{equation}
\rm
log \left(\frac{M_{BH}}{M_{\odot}} \right)=8.49+4.38\times log \left(\frac{\sigma_{*}}{200\,kms^{-1}} \right).
\end{equation}

To calculate the bolometric luminosities of the studied sources, we used $\rm [O,III]$ line measurements corrected for nebular reddening. The corrections assumed an intrinsic Balmer decrement of $\rm H\alpha/H\beta = 2.85$ and applied the extinction law of \citet{Fitzpatrick1999} together with the formula from \citet{Stasinska2001}.
The bolometric luminosities were then estimated using the relation from \citet{Trump2015}:

\begin{equation}
\rm
\frac{L_{bol}}{10^{40}\,erg\,s^{-1}}=112\times \left(\frac{L_{[O,III]}}{10^{40}\,erg\,s^{-1}}\right)^{1.2} .
\end{equation}

For galaxies where reddening corrections were not feasible, we used the observed $\rm [O,III]$ fluxes, treating the resulting $\rm L_{bol}$ values as lower limits. Conversely, for sources where the BPT/WHAN classification indicates a mix of star formation and AGN activity, the derived $\rm L_{bol}$ values may be overestimated. For this reason, for the three sources 064001$+$28, 095141$+$37, and 164607$+$42, we report $\rm L_{bol}$ as upper limits.
We did not compute bolometric luminosities for sources unambiguously classified as star-forming galaxies by both the BPT and WHAN diagrams, namely 180940$+$24 and 233058$+$10.

The resulting $\rm L_{bol}$ estimates are presented in Table \ref{table:host_galaxies}. As with the line fluxes, we report maximum uncertainties for $\rm \sigma_{*}$ and $\rm M_{BH}$.

To calculate the flux density in the W3 band we used the following relation:

\begin{equation}
\rm
S_{\nu}[Jy]=S_{\nu 0}10^{-0.4m_{Vega}},
\end{equation}

\noindent
where $\rm S_{\nu0}$ is the zero magnitude flux density and $\rm m_{Vega}$ is the
WISE magnitude of the object. We assumed that the MIR
spectral energy distributions do not significantly deviate from a
power law ($S_{\nu} \propto \nu^0$) and thus we use the zero magnitude flux
density $S_{\nu0} = 31.674$ for W3 \citep{Jarrett2011}.

The total stellar mass was estimated using the mass-to-light ratios of the stellar populations identified in the galaxy. To correct for aperture effects, we compared the total galaxy flux in the Pan-STARRS {\it i}-band with the flux obtained by convolving the galaxy spectrum with the Pan-STARRS {\it i}-band transmission curve and integrating it. However, it is important to note that this correction implicitly assumes that the stellar populations within and beyond the spectroscopic aperture are comparable. The reported uncertainty for $\rm M_{*}$ is the maximum error.

\section{Results and discussion}
\label{sec:discussion}
In this paper, we examine the optical and infrared properties of 24 recently discovered transient radio sources, all of which have shown a significant increase in radio flux density over the past two decades. Their radio characteristics are discussed in detail in \citet{MKB2025}. The sample comprises 23 galaxies and one quasar (Table \ref{table:basic}), all relatively bright, with an $r$-band magnitude of $r < 20$ mag. With the exception of the quasar, these objects are located in the local Universe ($ z < 0.3$).

\begin{figure*}[th!]
   \centering
   \includegraphics[scale=0.5]{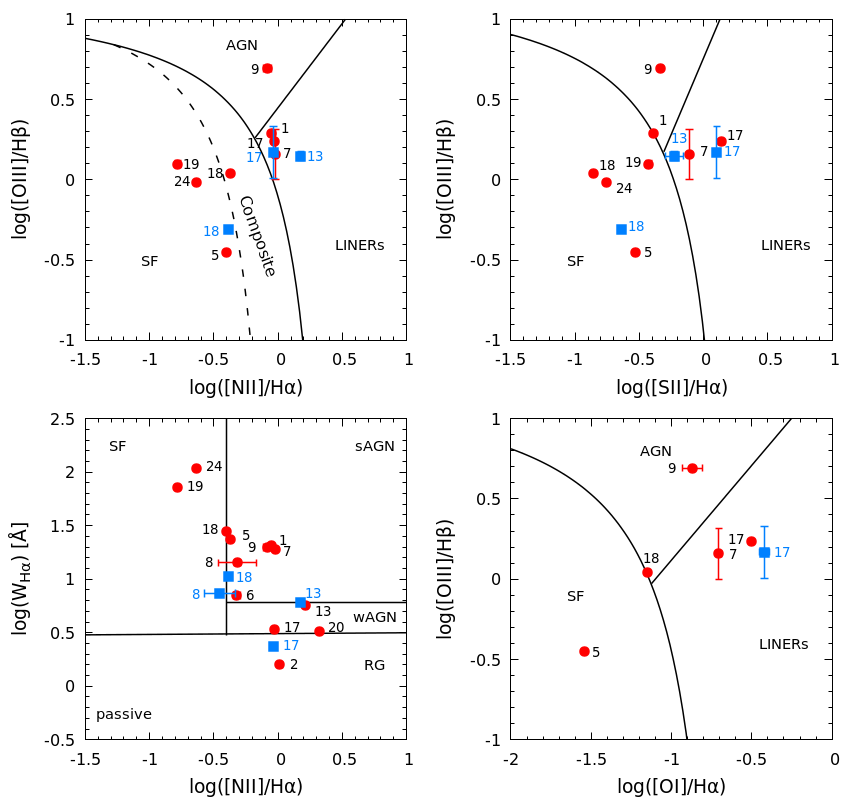}
\caption{Emission line diagnostic diagrams: $\rm [O\,III]\lambda5007/H_{\beta}$ vs three different line ratios \citep[BPT;][]{Baldwin, Kewley01, Kewley2006, Kauffmann} and $W_{H_{\alpha}}$ vs $\rm [N\,II]\lambda6584/H_{\alpha}$ \citep[WHAN;][]{CidFernandes2011}. The lines demarcate boundaries between sources classified as star-forming (SF) galaxies, AGNs, low ionization nuclear emission line regions
(LINERs), “Composite” sources, weak AGNs (wAGN), strong AGNs (sAGN), retired galaxies (RG) and passive galaxies. Classifications based on SDSS observations are indicated by blue squares, while red circles indicate the classifications based on Palomar and Keck observations. Numbers refer to source identifications in Table \ref{table:basic}.}
              \label{figure:bpt_diagram}%
    \end{figure*}

\subsection{Host galaxy properties}
The morphological classification of the host galaxies was performed based on a visual inspection of Pan-STARRS images. Galaxies were classified as spiral (S), elliptical (E), lenticular (L), or distorted (D). The images used for classification are presented in Figure \ref{figure:cut-outs}. 
Additionally, spectral analysis was conducted to characterize the host galaxies of our sources. The results indicate that more than half reside in massive elliptical galaxies,
where old stellar populations, aged $10^8$ years or older, dominate. However, in several cases, a significant contribution 
($>\,30\%$) from young stellar populations (younger than $10^7$ years) is evident, suggesting ongoing star formation. These sources are 024345$−$28, 064001$+$28, 164607$+$42, 180940$+$24, and 233058$+$10.

Next, using emission-line flux ratios, we diagnosed the nuclear activity by classifying the sources into different subclasses based on their position in the BPT and WHAN diagrams \citep{Baldwin, Kewley01, Kewley2006, CidFernandes2011}. 
The results of these classifications are presented in Table \ref{table:host_galaxies} and Figure \ref{figure:bpt_diagram}, with source labels corresponding to the IDs in Table \ref{table:basic}. Since the optical spectrum of 110239$-$06 is dominated by quasar emission, we exclude it from this analysis. Similarly, for several early-type galaxies, such an analysis was not possible due to the absence of emission lines in their spectra. 
This spectral analysis is based on our new observations conducted between 2021 and 2023, after the onset of radio activity. Additionally, six of our sources have spectroscopic data from SDSS, obtained between 2001 and 2008, likely before the radio brightening. For sources with detectable emission lines in these earlier spectra, we performed the same analysis to compare them with our new observations.

For two sources, this comparison suggests a shift from the star-forming galaxy region to either sAGN (095141$+$37, ID 8) or Composite objects (164607$+$42, ID 18), and changes within the group of weak AGNs in the case of source 155847$+$14 (ID 17). All of these shifts are likely caused by increased ionization of the emission lines. However, since we observe discrepancies in the spectroscopic classifications of these sources based on the BPT and WHAN diagrams, or large measurement uncertainties, this potential shift should be interpreted with caution.

\begin{figure*}[t!]
   \centering
   \includegraphics[scale=0.09]{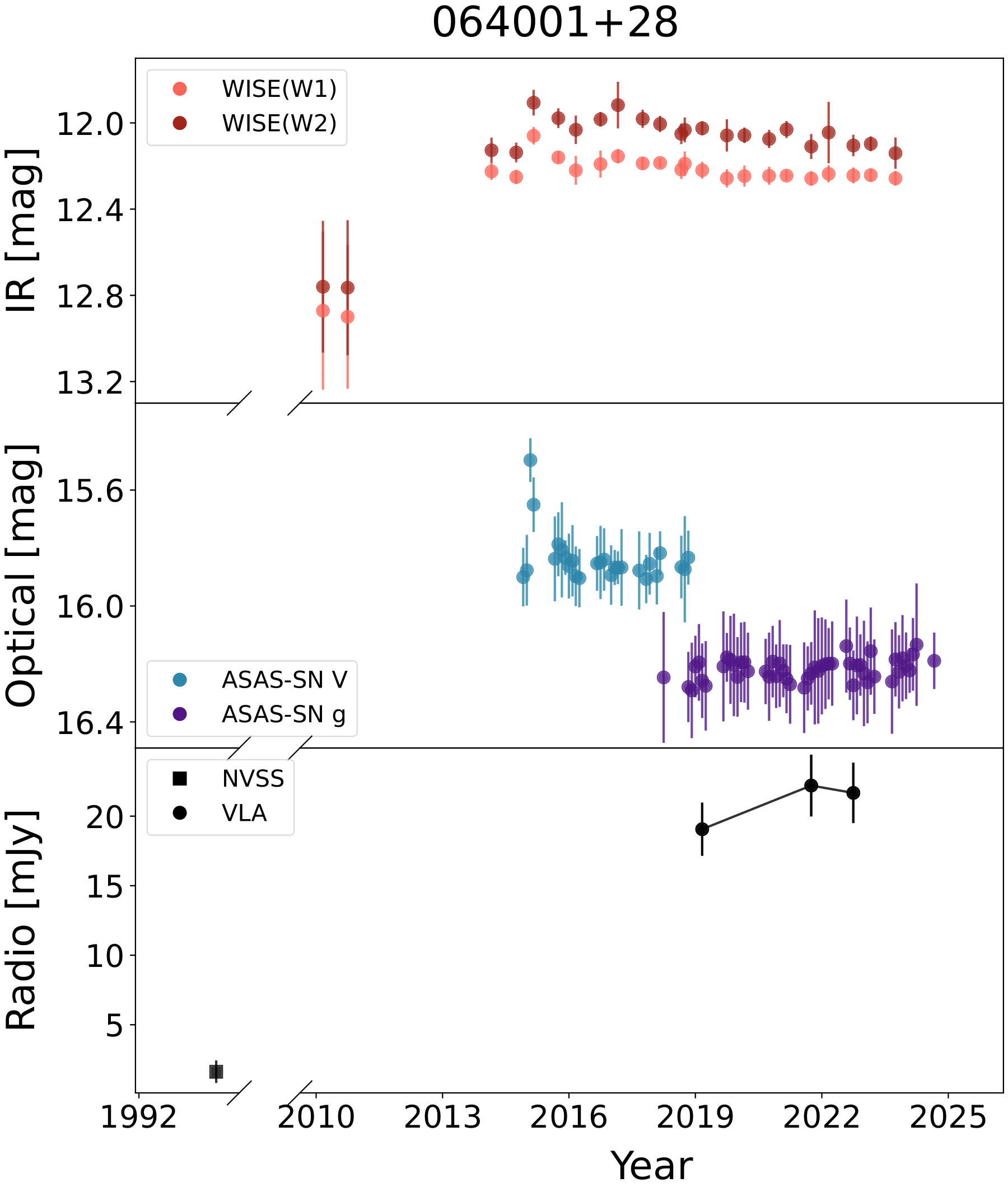}
   \includegraphics[scale=0.09]{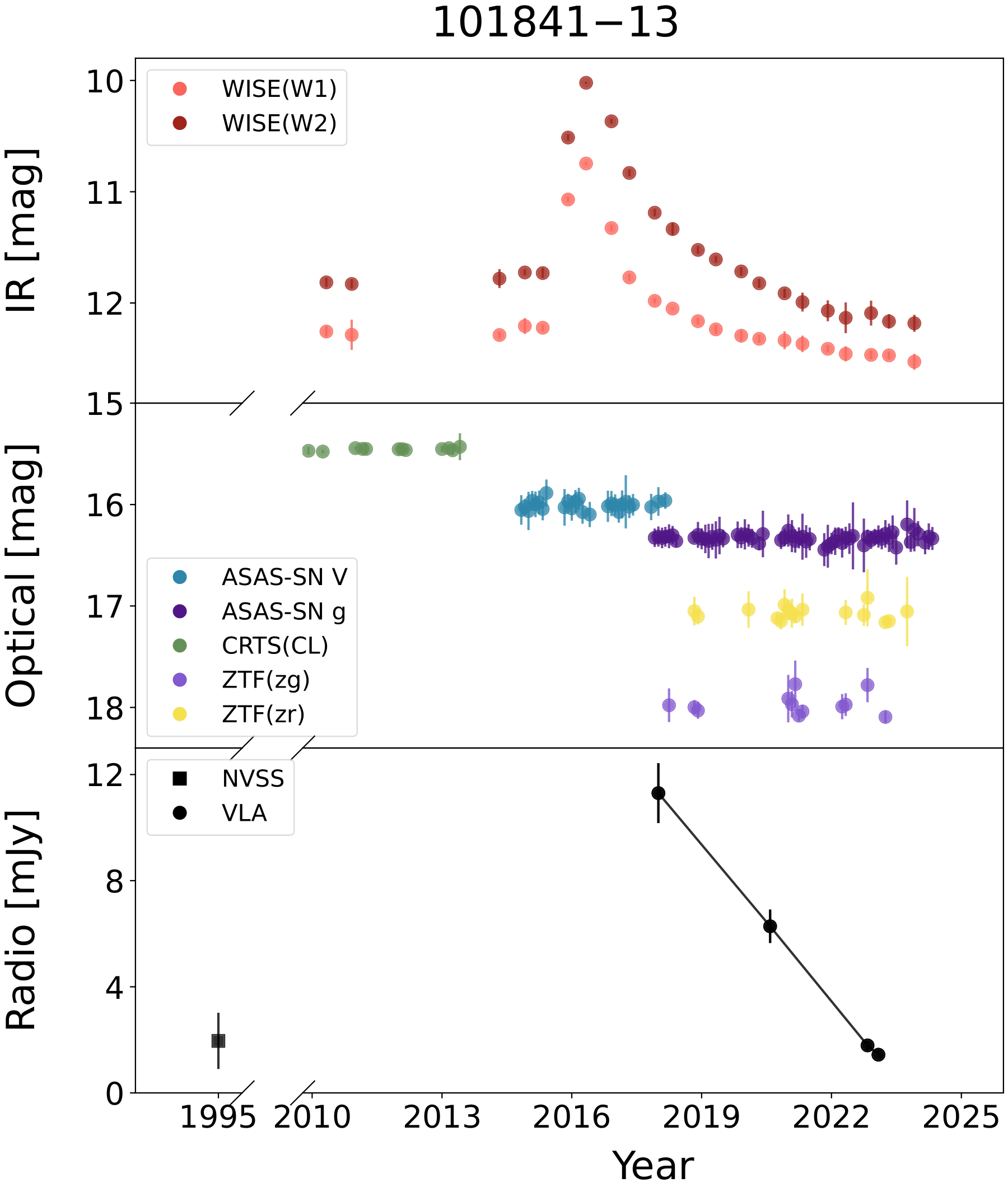}
   \includegraphics[scale=0.09]{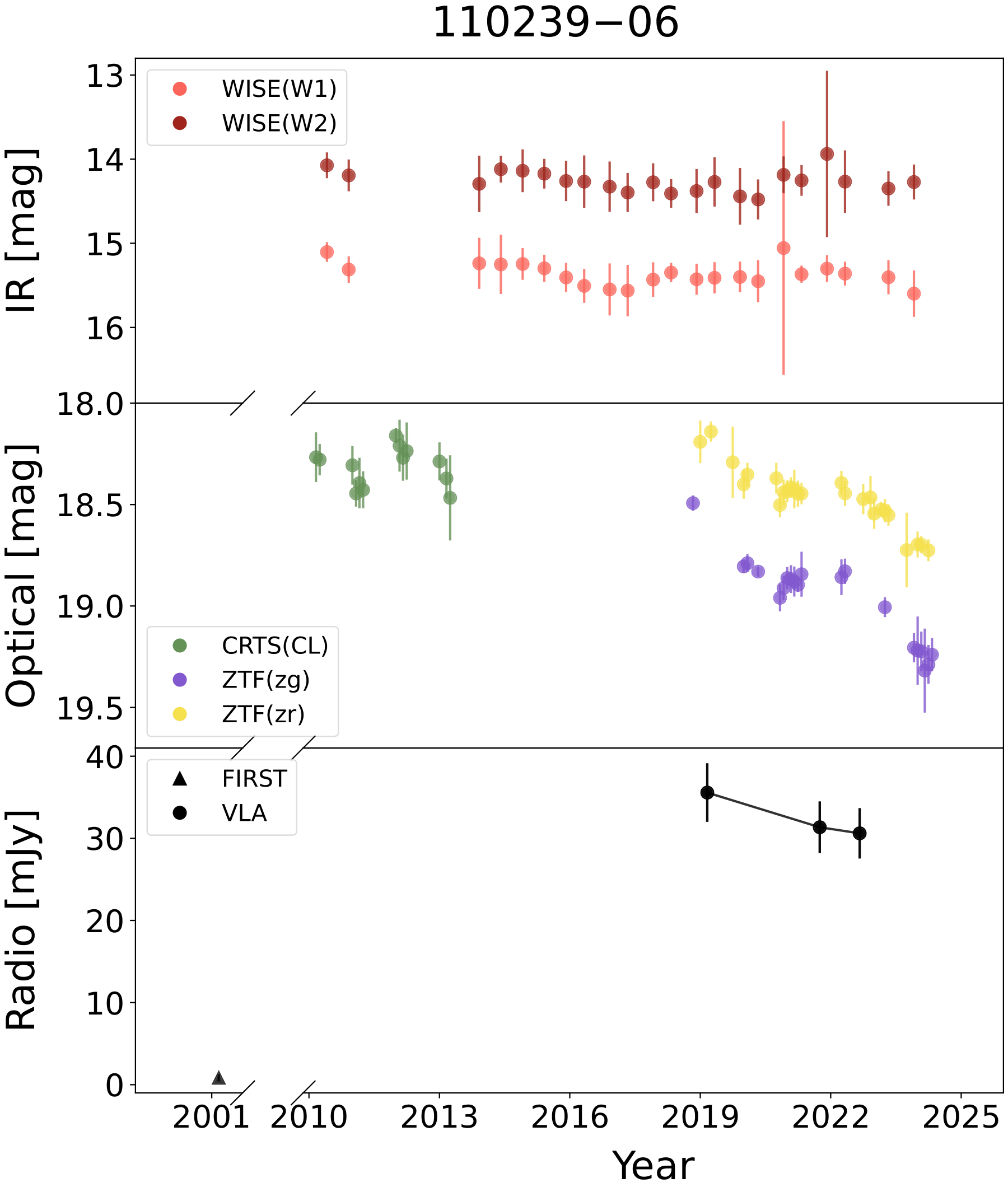}
   \includegraphics[scale=0.09]{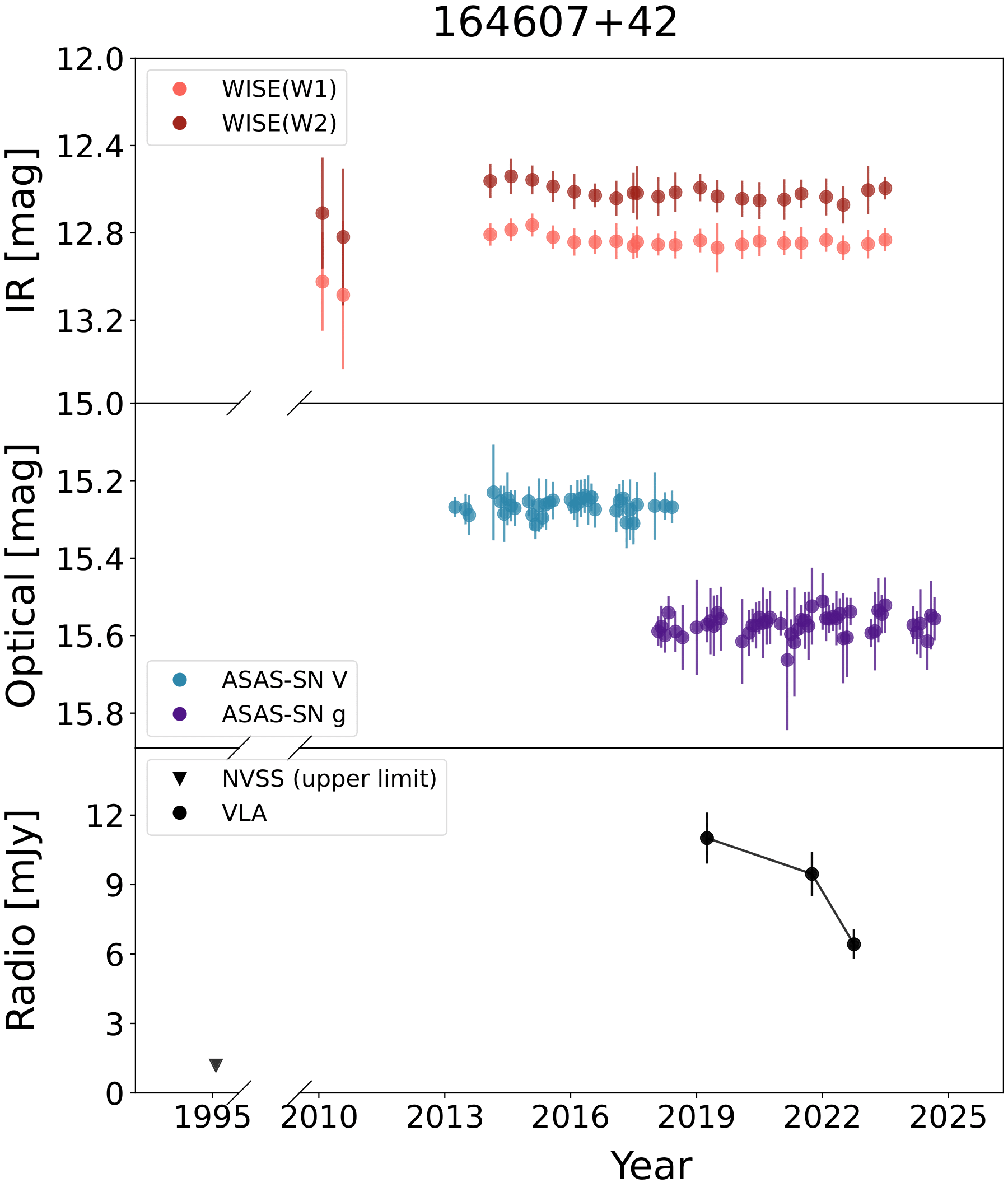}
   \includegraphics[scale=0.09]{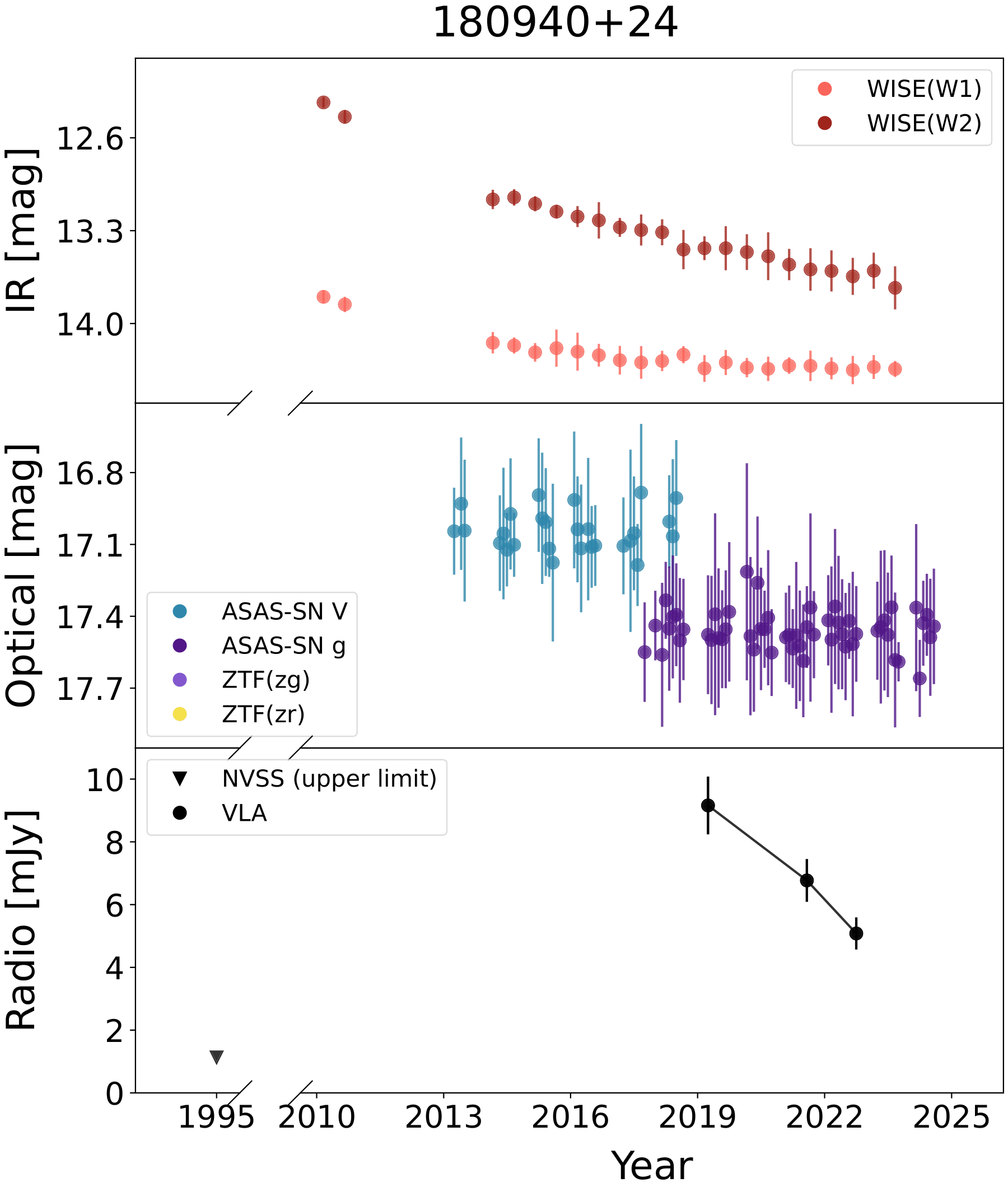}
   \includegraphics[scale=0.09]{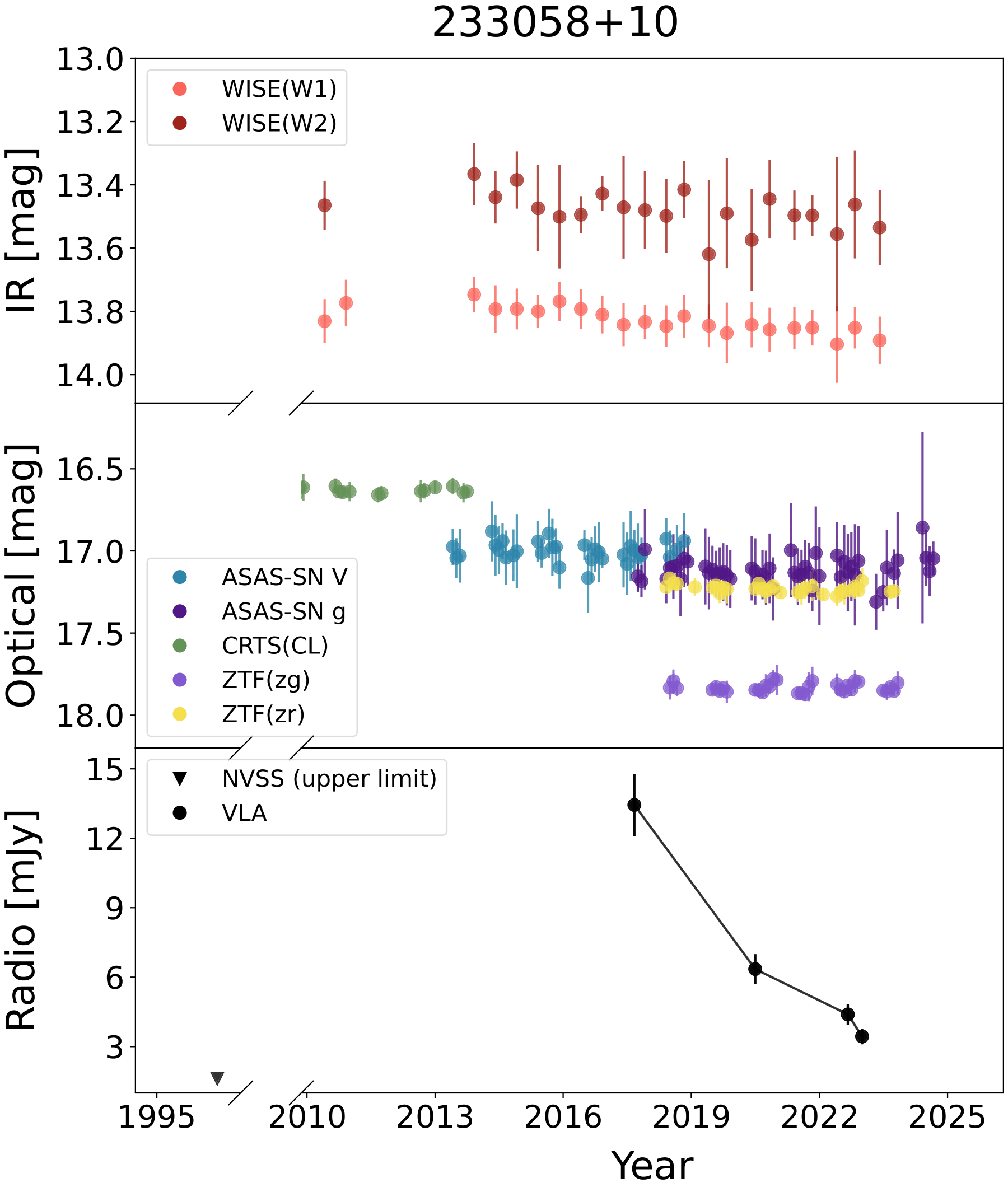}
   \caption{Light curves in the optical, infrared and radio (3 GHz) for selected objects. In case of optical and infrared curves the individual points represent monthly averaged values after $\rm 3\sigma$ clipping to reject outliers. For radio observations (NVSS, FIRST, VLA), the points correspond to single measurements. The VLA data come from the VLASS survey as well as dedicated observations, as reported in \citet{MKB2025}. }
   \label{figure:lightcurves}
\end{figure*}

The classification results reveal that most of the analyzed sources exhibit Seyfert or LINER ionization levels, indicating AGN activity. One object, 024609$+$34, is likely classified as retired galaxy with $\rm W_{H\alpha} <\,3\AA$. According to the discussion in \citet{CidFernandes2011}, this suggests that it is no longer forming stars and is ionized by hot evolved low-mass stars (HOLMES) instead. However, this does not necessarily imply that the central black hole is inactive, but rather that its ionizing power is weaker than that of HOLMES. The only sources that are unambiguously classified as star-forming galaxies in both the BPT and WHAN diagrams are 180940$+$24 (ID 19) and 233058$+$10 (ID 24). 
Objects that straddle the border between the populations of star-forming galaxies and AGNs are 064001$+$28 (ID 5) and 164607$+$42 (ID 18).   

We have estimated the SMBH masses of our sources using the $\rm M_{BH}-\sigma_{*}$ relation \citep{Kormendy}, as described in Section \ref{sec: physical_parameters}. The results are presented in Table \ref{table:host_galaxies}.
The derived SMBH mass range is quite broad, $\rm 10^5 - 10^9\,M_{\odot}$ which may indicate significant heterogeneity in our sample in terms of the nature of their transient radio emission. More than half our sources host very massive black holes ($\rm M_{BH} > 10^8\,M_{\odot}$). However, for 10 objects, the estimated black hole mass falls below this threshold, which also corresponds to the upper limit for the tidal disruption of a main-sequence star around a Schwarzschild black hole \citep{Gezari2021}.
Recently, \citet{Somal2024} suggested that radio-selected, optically bright TDEs tend to have even lower SMBH masses than those solely selected based on optical criteria. The average SMBH mass of radio-selected TDEs in their study is approximately $\rm 6.1 \times 10^6\,M_{\odot}$, which is consistent with the SMBH mass estimates for some of our objects.

To further characterize the SMBH activity, we estimated the Eddington ratio, $\rm \lambda_{Edd}=L_{bol}/L_{Edd}$, where $\rm L_{Edd}=1.3\times10^{38}\,(M_{BH}/M_{\odot})\,erg\,s^{-1}$. This ratio serves as a measure of accretion efficiency. 
We were able to estimate $\rm \lambda_{Edd}$ for a few sources, and the obtained values (Table \ref{table:host_galaxies}) indicate low accretion activity, with $\rm \lambda_{Edd} \approx 10^{-2}$. Such low values likely suggest a radiatively inefficient accretion regime, where most of the energy is transferred into radio-emitting jets \citep{Mingo}.
These jets are typically weak and characteristic of LINER-type objects \citep{Baldi2018, Kharb2021} or Fanaroff-Riley I (FR I) radio galaxies \citep{Fanaroff}, which aligns with our radio observations \citep{MKB2025}. However, not all objects in our sample show clear jet structures, which could be due to the limited spatial resolution of our observations.

\begin{figure*}[t!]
\sidecaption
   \includegraphics[width=12cm]{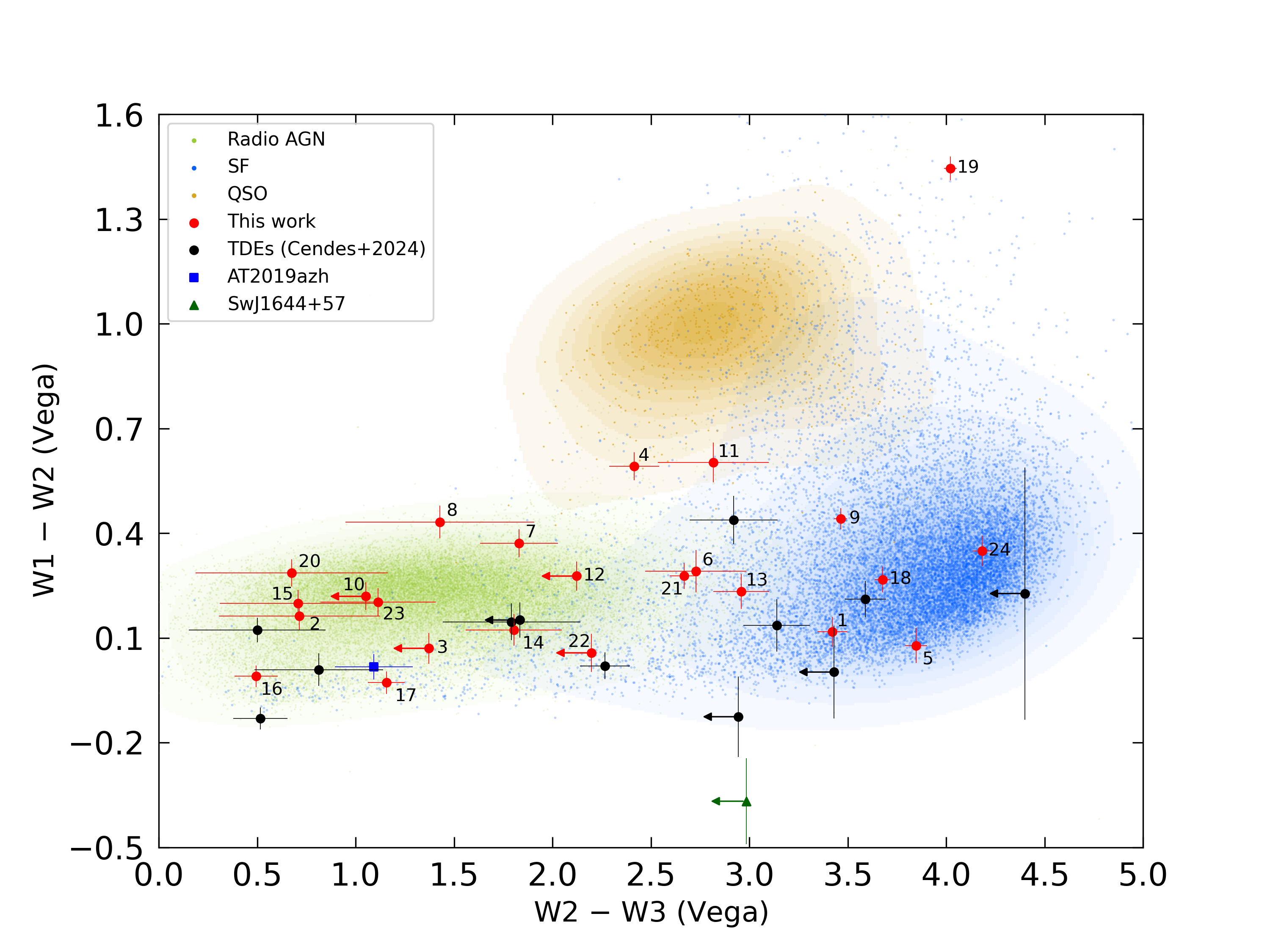}
\caption{WISE color–color plot for sources presented in this work (red dots). Numbers refer to source identifications in Table \ref{table:basic}. Additional points (black dots, blue square and green triangle) represent optically discovered TDEs as described in the text.
The colored density clouds represent the following objects: radio galaxies (green dots), star-forming galaxies (blue dots), and quasars (yellow dots). These data come from the ROGUE I \citep{Koziel2021} and ROGUE II (Kozieł-Wierzbowska, in prep.) catalogues. }
\label{figure:wise_color_color}
\end{figure*}

\subsection{Optical and infrared variability}
\label{sec:variability}
Analysis of long-term optical and infrared monitoring of our sources made with CRTS, ZTF, ASAS-SN, and WISE does not reveal significant variability or flares in the vast majority of them. The exceptions are four objects: 064001$+$28 (ID 5), 101841$−$13 (ID 9), 110239$−$06 (ID 11), and 180940$+$24 (ID 19), among which 110239$−$06 is the only quasar in our sample and exhibits typical AGN-like brightness fluctuations. The light curves of these objects are shown in Figure \ref{figure:lightcurves}, along with two additional sources, 164607$+$42 (ID 18) and 233058$+$10 (ID 24), which we refer to in the further discussion.

The most striking case of variability among our sources is the intense infrared flare observed in 101841$−$13 (ID 9), which was classified by \citet{Masterson2024} as a dust-obscured TDE. Its light curve shows a brightness increase of $\sim1.5$ and $\sim1.7$ magnitudes in the W1 and W2 bands, respectively, with the larger change in the redder W2 band suggesting an origin in hot dust emission. The shape of the infrared flare, characterized by a rapid rise followed by a slow, monotonic decline, is similar to optical TDEs \citep{vanVelzen2021}. 

However, our analysis shows that weak radio emission was already present in this object 23 years ago, as seen in the NVSS 1.4\,GHz observations \citep{MKB2025}. Additionally, archival spectroscopy from the 6dF survey \citep{Jones2004, Jones2009}, obtained in 2003 i.e. prior to the onset of the infrared and radio outburst, reveals significant [O\,III] and [N\,II] emission in 101841$−$13. Our analysis of the line ratios places this source in the AGN region of the BPT diagram.
It is important to note, however, that the 6dF spectrum of 101841$−$13 is not flux-calibrated, so the line ratios are estimated based on recorded counts and should therefore be treated with caution and considered as upper limits (see \citealt{MKB2025} for more details).
Nevertheless, our Palomar spectroscopic observations from 2022, conducted after the infrared flare had faded, clearly support the classification of this source as an AGN. Subsequent spectroscopy obtained one year later by \citet{Masterson2024} shows a decline in line luminosities and a shift of the source toward the Composite region of the BPT diagram. Taking all these observations into account, we suggest that different interpretations of the behavior of 101841$−$13 are possible: it may be the result of a dust-obscured TDE; it could reflect changes in the accretion rate onto the SMBH, similar to what is seen in changing-state AGNs; or alternatively, the event might still be consistent with a TDE occurring within an active AGN.

However, even if accretion is ongoing in this source, it appears to be at a very low level, $\rm \lambda_{Edd} << 10^{-2}$, and high-resolution VLBA imaging reveals only a weak, non-relativistic outflow responsible for the observed radio emission.

A similar infrared flare likely occurred in 180940$+$24 (ID 19). Its light curve shows a gradual decline in brightness in the W1 and W2 bands, with a large initial difference of approximately 1.5 mag between them, suggesting strong hot dust emission. However, the light curve captures only the final phase of the event, which likely took place before 2010. This initially large W1-W2 difference, derived from the 2010 measurements, makes this source an outlier on the WISE color-color diagram (Figure \ref{figure:wise_color_color}). However, over the next three years, this difference gradually decreases to about 0.6 mag, shifting the source into the region occupied by star-forming galaxies in Figure \ref{figure:wise_color_color}. Similar long-duration infrared variability has been discussed in the literature as dust-heated echoes of optical and X-ray TDEs \citep{Dou2016, Jiang2021}. For 180940$+$24, there is no optical monitoring data preceding the infrared flare that could confirm an associated optical outburst. However, as noted by \citet{Masterson2024}, many TDE events may lack detectable optical or X-ray variability due to strong dust obscuration within the host galaxy.

In the case of 064001$+$28 (ID 5), a significant change in infrared brightness of approximately 0.7 magnitudes is observed between 2010 and 2014. However, this variation is unlikely to be a flare or an infrared echo, and a gap in the observational data makes it impossible to confirm its exact nature. Nevertheless, since 2014, small infrared fluctuations are visible, with the most prominent one ($\sim0.2$ mag) occurring probably between October 2014 and March 2015. This variation may be related to an optical brightness change of $\sim0.4$ mag, observed in the ASAS-SN $V$-band monitoring between January and February 2015.
Due to the limited number of observations and their cadence, the origin of the observed variations remains uncertain. However, the radio activity of source 064001$+$28 has increased significantly in recent years, leading to the formation of a core–jet radio structure. According to our conservative interpretation, it does not show further significant radio variability. Moreover, radio observations suggest that 064001$+$28 may in fact be the central and newly reactivated component of a larger-scale radio structure currently in a fading phase \citep{MKB2025}. Such episodic jet activity is commonly attributed to unstable accretion onto the central black hole, for example due to internal processes in the accretion disk such as thermal–viscous instabilities \citep{Czerny}. In rare cases, it may also be triggered by a tidal disruption event \citep{Sullivan2024}.

For the sources 164607$+$42 (ID 18) and 233058$+$10 (ID 24), we do not detect any significant variations in either optical or infrared brightness.

As recently reported by \citep{Cendes2024}, approximately $\sim$40\% of optically discovered TDEs are detected in the radio hundreds to thousands of days after discovery. However, this emission is relatively weak, at the level of $\rm \sim10^{37} - 10^{39}\,erg\,s^{-1}$.
As demonstrated in our radio analysis, the vast majority of our sources exhibit significantly higher radio emission. Moreover, once initiated, this emission does not show a decline in the form of a radio flare \citep{MKB2025}. An exception is found in four sources: 101841$−$13 (ID 9), 164607$+$42 (ID 18), 180940$+$24 (ID 19) and 233058$+$10 (ID 24), two of which (101841$−$13 and 180940$+$24) also exhibit an additional infrared flare (Figure \ref{figure:lightcurves}).
If we assume that the infrared variation is a consequence of an UV flare, this would indeed suggest a delayed onset of the radio emission relative to a potential optical TDE — the timing of which we are unable to constrain. We can only roughly estimate the time delay between the infrared and radio flares.
For source 101841$-$13 (ID 9), the radio emission appears to follow relatively quickly, possibly within less than two years. In contrast, for source 180940$+$24 (ID 19), the radio emission seems to emerge only after approximately eight years. However, this estimate remains highly uncertain, as the precise onset of either the infrared or radio flare is unknown.

\begin{figure*}[th!]
   \sidecaption
   \includegraphics[width=12cm]{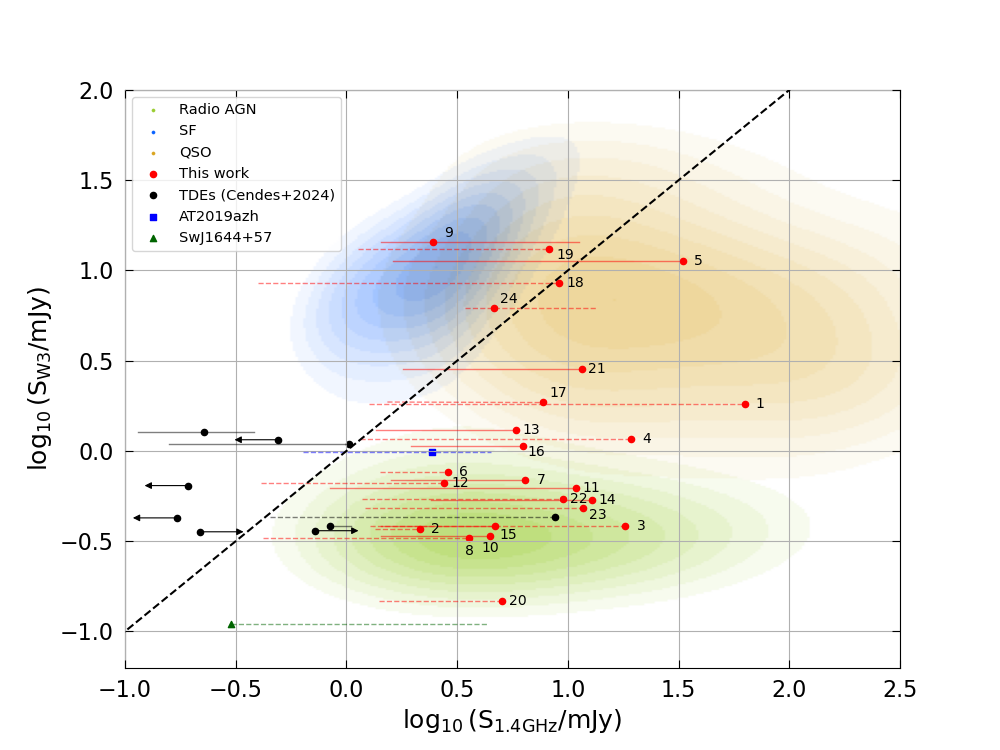}
\caption {Mid-infrared (W3) vs. radio (1.4\,GHz) flux density diagram for the sources presented in this work (red dots). Source numbers correspond to the identifications listed in Table \ref{table:basic}. Additional points—black dots, the blue square, and the green triangle—represent optically discovered TDEs. For clarity, error bars and upper limits are not shown in this figure. The solid and dashed lines indicate the evolution of radio flux density for individual sources, as described in detail in the text. To improve the readability of the plot, the main populations of objects are represented only by the contours of their density distributions: radio galaxies (green), star-forming galaxies (blue), and quasars (yellow). These reference data are taken from the ROGUE I \citep{Koziel2021} and ROGUE II (Kozieł-Wierzbowska, in prep.) catalogs. The main dashed line marks the relation $\rm S_{W3} = S_{1.4\,GHz}$, which serves to separate radio galaxies from star-forming galaxies based on the ROGUE I sample.}
              \label{figure:flux_wise_radio}
    \end{figure*}

\subsection{The {\it WISE} color-color plot}

In Figure \ref{figure:wise_color_color}, we present the distribution of our sources in the WISE color-color diagram (red dots), with source labels corresponding to the IDs in Table \ref{table:basic}. 
The color density clouds represent radio galaxies (radio AGN), star-forming (SF) galaxies, and quasars (QSO), identified through a cross-matching of the SDSS Seventh Data Release \citep[DR7;][]{Abazajian2009} with the First Images of the Radio Sky at Twenty cm survey \citep[FIRST;][]{White}, as described in \citet{Koziel2021}. For comparison, we also include a sample of optically discovered TDEs that exhibit late-time brightening in the radio band, taken from \citet{Cendes2024}, along with two well-known TDEs, SwJ1644$+$57 \citep{Bloom2011, Eftekhari2018, Cendes2021} and AT2019azh \citep{Goodwin2022}, both showing long-lived radio-bright emission. The criteria used to select these sources from the full TDE sample of \citet{Cendes2024} are described in the Appendix \ref{app:wise_and_radio}. A summary of the infrared properties of these TDEs and our sources is provided in Table \ref{table:wise_radio_measurements_TDEs} and Table \ref{table:wise_radio_measurements}.

By analyzing the distribution of both samples—TDEs and our radio transients—in this diagram, we find that although the TDEs appear scattered across the plot, the majority are located within regions occupied by the star-forming galaxy population. In our diagram, this population forms a large central concentration that extends in a long tail toward the domain of radio galaxies, and some of the TDEs also fall within this extended tail. In contrast, our radio transient sources exhibit a broader diversity of infrared properties, spanning multiple populations visible in the diagram, which further suggests considerable heterogeneity within this sample.
For our objects, which are predominantly galaxies, most lie below the W1–W2 = 0.75 threshold, a commonly used criterion to distinguish powerful AGNs from weaker AGNs and normal galaxies \citep{Stern2012}.
The most prominent outlier is 180940$+$24 (ID 19), where the unusually high W1–W2 value results from a long-lived, fading infrared flare rather than typical AGN activity (see discussion in section \ref{sec:variability}). The sources that clearly fall into the star-forming galaxy category are 024345$−$28 (ID 1), 064001$+$28 (ID 5), 101841$−$13 (ID 9), 164607$+$42 (ID 18), and 233058$+$10 (ID 24). Indeed, our stellar population analysis confirms that all these objects contain a significant fraction ($\sim 30\%$) of young stellar populations (age $< 10^7$ years), with 101841$−$13 (ID 9) exhibiting the lowest contribution ($< 17\%$). However, this purely infrared-based classification aligns with BPT/WHAN classifications only in the case of 233058$+$10 (ID 24).

\subsection{The mid-IR/radio diagram}

As the next step, we applied an additional diagnostic method to distinguish between galaxies whose radio emission is dominated by AGN activity and those in which it originates purely from star formation. Several attempts have been made in the past to separate these two populations using various criteria \citep{Best} and datasets. For instance, \citet{Sabater2019} employed deep radio surveys such as the LOFAR Two-Metre Sky Survey \citep[LoTSS;][]{Shimwell2019} for this purpose. Later, \citet{Koziel2021} proposed a simplified selection method that reduces the classification to just two observational parameters. This approach compares the radio flux density at 1.4\,GHz ($\rm S_{1.4\,GHz}$) with the mid-infrared WISE W3-band flux ($\rm S_{W3}$), which is primarily dominated by dust heated by ongoing star formation (see Figure~\ref{figure:flux_wise_radio}). Based on their study, \citet{Koziel2021} proposed a simple empirical division line, $\rm S_{W3} = S_{1.4\,GHz}$, which successfully classifies 99.5\% of extended radio sources as AGNs and correctly identifies approximately 99\% of star-forming (SF) galaxies, based on emission-line diagnostics. Very recently, the effectiveness of this W3/radio flux diagnostic in separating AGN and SF galaxies has been confirmed by \citet{Hardcastle2025} using the latest LoTSS observations at 144 MHz \citep{Shimwell2022}.
However, it is important to emphasize that this classification applies exclusively to galaxies and does not extend to quasars, which are also marked on the plot and include both radio-quiet and radio-loud objects. The origin of radio emission in radio-quiet quasars remains uncertain; many studies suggest it arises from a combination of star formation and AGN activity \citep{McCaffrey2022, Silpa2021}. Furthermore, in the case of quasars, it cannot be ruled out that, in addition to star formation, the AGN itself contributes to dust heating in the host galaxy, thereby influencing the W3-band luminosity. These factors make the classification of quasars into star formation- or AGN-dominated more complex and require additional criteria to be taken into account \citep{Hardcastle2025}.
The quasars shown in Figures~\ref{figure:wise_color_color} and~\ref{figure:flux_wise_radio} for comparison were also selected from the SDSS DR7 catalogue and will be discussed in detail in a forthcoming publication (ROGUE II, Kozieł-Wierzbowska et al., in prep.).

Therefore, we applied this classification exclusively to our sample of galaxies and show their distribution on the W3/radio diagnostic diagram in Figure \ref{figure:bpt_diagram}. The color scheme and symbols are the same as in Figure \ref{figure:wise_color_color}, and as in that figure, we also include optically discovered TDEs as a comparison sample.
Analyzing the distribution, we find that the vast majority of our sources lie in the region typically occupied by AGNs, which—together with their BPT/WHAN classifications—consistently indicates the presence of an active galactic nucleus in their host galaxies. Only three sources—101841$−$13 (ID 9), 180940$+$24 (ID 19), and 233058$+$10 (ID 24)—are located in the star-forming galaxy region, while source 164607$+$42 (ID 18) lies near the classification boundary (Figure \ref{figure:flux_wise_radio}). For 180940$+$24 (ID 19) and 233058$+$10 (ID 24), this classification is fully consistent with their BPT/WHAN classifications, as these are the only objects in our sample classified as star-forming in both diagnostic diagrams (Figure \ref{figure:bpt_diagram} and Table \ref{table:host_galaxies}).
As already mentioned, the classification of 101841$−$13 (ID 9) is ambiguous: it is classified as an AGN based on the BPT/WHAN diagrams, but its infrared colors are typical of star-forming galaxies. Nonetheless, there is evidence that the strength of the emission lines in this object is gradually declining, potentially moving it toward the star-forming galaxy regime \citep{Masterson2024}.
All four of these sources are considered TDE candidates in our study, based on their radio properties \citep{MKB2025}, and in the case of two of them—101841$−$13 (ID 9) and 180940$+$24 (ID 19)—also based on infrared variability. Their positions on the W3/radio diagram suggest high W3-band luminosities, likely due to intense star formation activity.
In contrast, the optically discovered TDEs included for comparison—if detected in the radio at all—typically exhibit much weaker radio emission, indicating a different origin than AGN-related processes. Nonetheless, their W3-band luminosities are comparable to those of radio galaxies, consistent with the properties of massive elliptical galaxies dominated by old stellar populations.

A key consideration is that the positions of our sources in Figure \ref{figure:flux_wise_radio} are based on their radio flux density measurements obtained with the VLA in 2021 and 2022—several years after the initial radio outburst—and therefore do not reflect the earlier or ongoing variability observed in many of these objects.
To account for the changes in radio brightness over time, we represent each source with a line on the plot, following the conventions described below.

\begin{itemize}
    
    \item A dashed line indicates that the starting point is based on an upper-limit estimate of the radio flux density (noise measurement), while a solid line indicates that it is based on a reliable detection. In both cases, the measurements were performed directly on the archival FIRST or NVSS images \citep{MKB2025}. 

    \item The point on each line represents the most recent radio flux measurement. If the point does not lie at the right end of the line, this indicates that its current brightness is lower than the highest measured radio flux density — i.e., a decline in radio brightness.

    \item For two sources, 101841$−$13 (ID 9) and 233058$+$10 (ID 24), we used 3 GHz flux density measurements from VLASS to better capture the sharp decline in radio brightness observed in recent years. This is because, at the time of our VLA observations at 1.4\,GHz, these objects had already nearly returned to their pre-outburst radio emission levels.

    \item We assumed that the W3-band brightness remained constant over the 2010–2023 period. While this may not be universally valid—since changes in SMBH accretion (e.g., due to a TDE or disk instabilities) can heat both the inner dust (traced by W1 and W2) and more distant, cooler dust emitting in W3—we find no significant infrared brightening in either W1 or W2 for most of our sources. In the two exceptions, 101841$−$13 (ID 9) and 180940$+$24 (ID 19), the observed radio variability occurs after the infrared flares have faded. Therefore, we consider the assumption of constant W3 brightness to be justified.

    \item We applied the same visualization rules for W3 and radio-band observations to the optically discovered TDEs. However, for many of these, we have only a single 1.4\,GHz flux density measurement, usually obtained during the peak or late stages of radio emission evolution. Nevertheless, multi-epoch flux density measurements at other frequencies allow us to infer the trend in radio brightness over time, which we indicate using arrows. In some cases (e.g., a few sources from \citealt{Cendes2024}), we show changes in radio brightness as lines, though these may represent incomplete data due to limited sensitivity or upper limits in the early epochs.

    \item  For the jetted TDE SwJ1644$+$57, radio brightness has been continuously monitored and the most recent measurements show a continued decline in brightness and non-relativistic expansion \citep{Cendes2021}. This source will likely remain visible in the radio for decades, and its position in Figure \ref{figure:flux_wise_radio} will gradually shift to the left. Similarly, TDE AT2019azh is also in a fading phase \citep{Goodwin2022} and may eventually fall below the upper limit for 1.4\,GHz radio emission indicated on the diagram.
    
\end{itemize}

As previously mentioned, the W3/radio diagram shown in Figure \ref{figure:flux_wise_radio} provides excellent separation between star-forming (SF) galaxies and AGNs, as it does not exclude AGNs that exhibit infrared colors similar to SF galaxies but show a significant radio excess beyond what is expected from star formation alone \citep{Koziel2021, Hardcastle2025}.
The positions of most of our sources and their radio variability tracks in the W3/radio diagram suggest that an AGN likely existed in these galaxies prior to the onset of the radio outburst and was the dominant source of their radio emission. For seven sources, this interpretation is directly supported by pre-burst measurements (indicated by solid lines in the diagram).
In contrast, for three sources—064001+28 (ID 5), 101841−13 (ID 9), and 195335−04 (ID 21)—their positions in the W3/radio diagram indicate that, prior to the outburst, their radio emission was likely driven by star formation activity. Upper limits on the radio flux densities for five additional objects—024345−28 (ID 1), 155847+14 (ID 17), 164607+42 (ID 18), 180940+24 (ID 19), and 233058+10 (ID 24)—also support this possibility.
Note that we do not discuss source 112940+39 (ID 12) in this context, as its location in the diagram is based on upper limits in both the radio flux and W3-band measurements.
In several of these sources, the increase in radio brightness was substantial enough to temporarily shift them into the AGN region of the diagram. However, as the radio flux declined rapidly over time, some of them began to move back—or have already returned—to their original positions associated with star formation-dominated emission.
A similar behavior is observed in optically detected TDEs: some of these events can produce radio emission as luminous as that of radio-loud AGNs, potentially placing them in the AGN region of the W3/radio diagram for years or even decades. Some of them may be potential candidates for proto-radio sources \citep{Readhead2024, Sullivan2024}.

\subsection{Origin of the brightness variability}
\label{sec:origin}
All the sources discussed in this paper were discovered based on their significant brightening in the radio band, as shown in Figure \ref{figure:lightcurves} for selected objects. Our analysis of their radio properties and evolution after the radio transient event indicates a rather stable and persistent change, which could continue and lead to further development of radio structures over the next decades or longer. Such a change is most likely caused by intrinsic variations in accretion properties, similar to what is observed in changing-state AGNs \citep{MKB2025}. However, a small fraction of sources, less than 17\%, exhibit more pronounced changes in radio flux density, characterized by a declining trend, which can be classified as radio flares. These sources—101841$−$13 (ID 9), 164607$+$42 (ID 18), 180940$+$24 (ID 19), and 233058$+$10 (ID 24)—cannot be ruled out as potential TDE candidates responsible for the observed variability. Additionally, in the cases of 101841$−$13 (ID 9) and 180940$+$24 (ID 19), the radio variability was preceded by an infrared flare, which typically occurs after an optical TDE flare and is interpreted as nuclear dust reprocessing the UV/optical burst \citep{vanVelzen2016}. However, none of these sources exhibits significant optical variability. This could be due to heavy dust obscuration, preventing the detection of optical changes—suggesting that dust-obscured TDEs might be more common than previously thought \citep{Masterson2024}. On the other hand, changing-state AGNs also display infrared variability as a result of intrinsic changes in the AGN accretion rate \citep{stern2018,Yang2019}. 
Moreover, AGN activity and TDEs are not mutually exclusive phenomena and can occur within the same galaxy. However, numerical simulations show that a TDE in an AGN behaves very differently from one in a quiescent galaxy. The interaction between the stellar debris stream and the pre-existing accretion disk leads to strong shocks, super-Eddington mass inflow, and complex, often non-thermal radiation. As a result, such TDEs may appear very different from “classical” TDEs and can be challenging to identify observationally \citep{Chan2019, Cannizzaro2022}. Finally, there are numerous cases where a clear-cut interpretation is not possible, and the observed flare could be attributed either to a TDE or to an episode of enhanced AGN activity \citep{Merloni2015, Goodwin2024, Somal2023}.

Examining the host galaxies of our TDE candidates, we find that they are classified as star-forming or exhibit a significant presence of young stellar populations, indicating ongoing star formation activity.
This is consistent with recent results by \citet{Somal2024}, who suggested that radio-selected TDEs may preferentially occur in more actively star-forming host galaxies than optically selected TDEs, and also in galaxies with more recent or enhanced AGN activity.

We also note that our infrared–radio analysis highlights four objects that we suspect may be TDEs. Their positions on the W3/radio diagram indicate stronger W3 emission than that of the remaining sources in our sample, as well as optically discovered TDEs (Figure \ref{figure:flux_wise_radio}).
The strong increase in radio brightness caused some of these sources to temporarily shift into the AGN region of the diagram—similarly to what is observed for radio-luminous, optically detected TDEs.
At present, however, for all four sources—101841$−$13 (ID 9), 164607$+$42 (ID 18), 180940$+$24 (ID 19), and 233058$+$10 (ID 24)—the trend in radio brightness is decreasing, and for 101841$−$13, the radio luminosity has nearly returned to its pre-outburst level.

In the context of these changes, which caused some of the sources to cross the star formation/AGN boundary, we also note the behavior of four additional sources: 064001$+$28 (ID 5), 195335$-$04 (ID 21), 024345$−$28 (ID 1), and 155847$+$14 (ID 17), although the pre-burst positions of the last two are based on upper limits. After strong radio brightening, which also shifted them into the AGN region, their radio fluxes have so far shown no significant variability.
Moreover, the radio morphology of sources 064001$+$28 (ID 5) and 024345$−$28 (ID 1) consists of more than one component and has been preliminarily classified as "core-jet" and "double-lobed," respectively \citep{MKB2025}. We speculate that single-epoch radio observations of such objects could easily result in their classification as young radio sources of the GPS/compact symmetric objects (CSO) type—which they may already be, or may become. Therefore, long-term monitoring (spanning at least several years) of the radio evolution of our sources remains essential for better understanding the nature of the phenomena that occurred in these galaxies.

Recent work by \citet{Readhead2024} and \citet{Sullivan2024} suggests that some compact radio sources—specifically a subclass of CSOs—may be triggered by TDEs involving giant-branch stars on massive SMBHs. However, even much shorter TDEs involving main-sequence stars can reproduce similar radio activity. For a TDE to develop into a long-lived and stable radio source, the accretion must be long-lasting, sub-Eddington, and accompanied by preferential jet-launching conditions \citep{Sullivan2024}.
The TDE Swift J1644+57 (marked with a green triangle in Figure \ref{figure:flux_wise_radio}) serves as a counter-example. It was a short, super-Eddington accretion episode with an immediately launched jet, and after about 10 years of activity, its radio emission has returned to its pre-flare levels.
Nevertheless, long-duration (>1 yr), slowly evolving jetted TDEs can be viable proto-CSO candidates \citep{Sullivan2024}.
It is possible that some of the sources listed above could also be candidates for such phenomena. However, this interpretation is weakened by the lack of significant variability in the optical and infrared domains for most of these sources.

As the preceding discussion demonstrates, our sample of radio transient sources is not homogeneous in terms of their optical and infrared properties, nor in their radio variability behavior. When examining Figure \ref{figure:flux_wise_radio}, we can distinguish at least two distinct groups within the sample.
The first group is characterized by higher luminosities in the W3 band and a significant presence of young stellar populations in their host galaxies. Several of these sources also show noticeable variability in the infrared, which may indicate the occurrence of a TDE. These are objects that likely originally resided in the region occupied by star-forming galaxies, but the increase in their radio brightness temporarily shifted them into the AGN region.
The second group consists of sources located in the area typical for radio galaxies. While their radio brightening shifts their position within that region, they do not exhibit any significant variability in the optical or infrared bands that would suggest an external trigger for the radio outburst. These characteristics distinguish them not only from the first group but also from optically selected TDEs, whose behavior more closely resembles that of the sources in the first group.
Based on these findings, we hypothesize that the W3/radio diagnostic diagram (Figure \ref{figure:flux_wise_radio}) may serve as a useful tool for distinguishing between radio transients triggered by TDEs (where $\rm S_{W3} > S_{1.4\,GHz}$) and those resulting from intrinsic AGN variability (where $\rm S_{W3} < S_{1.4\,GHz}$).
Applying this interpretation to our sample suggests that the radio transients in eight sources—those with IDs 1, 5, 9, 17, 18, 19, 21, and 24—are likely triggered by TDEs. Among these, the sources with IDs 1, 5, 17, and 21 had not previously been considered TDE candidates based on their radio properties alone, as they exhibit different radio characteristics \citep{MKB2025}.
This reinforces the idea that even within this small group, TDE candidates are not homogeneous, and a definitive interpretation of the observed behavior will require further monitoring. Ultimately, deep radio observations of optically or infrared-detected TDEs, combined with continued monitoring and targeted searches for a broader population of radio transients, are essential for directly testing the hypothesis we propose.

There are also additional caveats that must be acknowledged regarding the interpretation and validation of the W3/radio diagnostic diagram.
First, while we compare the locations and evolutionary tracks of our radio transients with those of optically discovered TDEs, it is important to note that the two samples are selected in fundamentally different ways: optically selected TDEs are, by construction, chosen to avoid AGN activity (e.g., W1–W2 $\leq$ 0.7) and are generally much fainter in the radio than the sources in our sample. In our context, TDEs are therefore not used as a statistically comparable population, but as a qualitative reference to explore whether the diagram can help distinguish intrinsic AGN variability from possible TDE-driven flares.

Second, although the starting points of our radio variability tracks often lie below the nominal flux density limits of the FIRST and NVSS catalogues—because they are based on direct measurements from archival radio maps—our sample is still ultimately constrained by the intrinsic sensitivity limits of these surveys. This means that, in cases where no prior radio emission is detected, the starting point of the variability track is inherently limited by the survey’s detection threshold.

Third, redshift-related biases may influence the detectability of infrared variability. Sources located in the star-forming galaxy region of the diagram in our sample tend to be at lower redshifts, where the same luminosity change translates to a larger observed flux change, potentially making IR variability easier to detect compared to higher-redshift AGNs.

Moreover, the applicability of the W3–radio relation at lower flux densities remains uncertain. The absence of sources below certain flux thresholds on the diagram is primarily a consequence of the sensitivity limits of the W3 and 1.4 GHz observations. Deeper data would be needed to test whether the relation holds in this regime, or whether other populations of transients emerge there.

Furthermore, it remains unclear how TDEs occurring in AGNs would behave on the W3/radio diagram. While such events are plausible, there is no definitive signature that would allow us to distinguish a TDE in an AGN from other types of TDEs or from AGN-related flares \citep{Chan2019}. Similarly, intrinsic variability in AGNs is still not fully understood. It is difficult to predict and likely differs among AGNs. 

Undoubtedly, further evaluation of the W3/radio diagram as a classification tool for these phenomena needs to be the subject of future studies.

\section{Conclusions}
In this article, we present an analysis of the optical and infrared properties of 24 objects discovered based on their transient radio emission. Our previous study of their radio properties showed that they resemble a class of young AGNs known as GPS radio sources but are characterized by low-power radio jets, some of which are likely non-relativistic. We suggested that, in most cases, the cause of the radio burst activity in these objects is likely intrinsic changes in the accretion process.
Here, we continue our investigation, focusing primarily on the properties of their host galaxies and central black holes. Our findings are as follows:

\begin{itemize}

    \item Most of our objects reside in massive elliptical galaxies dominated by old stellar populations. Many of them lack visible emission lines in their spectra, and among those that do, the vast majority exhibit ionization levels typical of Seyferts or LINERs, indicating AGN activity.

    \item The majority of sources have black hole masses typical of radio-loud AGNs, $\rm >10^7\,M_{\odot}$. However, their estimated Eddington ratios, $\rm \lambda_{Edd} < 10^{-2}$, indicate low accretion activity.

    \item The vast majority of objects do not exhibit significant variability outside the radio spectrum. Only in the case of two sources, 101841$−$13 and 180940$+$24, can we confidently state that they also experienced infrared flares, which can be interpreted as nuclear dust reprocessing a possible UV/optical TDE flare.

    \item Based on the $\rm W3 - 1.4\,GHz$ diagnostic diagram, we conclude that the infrared luminosities and radio behavior of the majority of our sources are typical of radio galaxies and indicate the presence of an AGN. However, for perhaps as many as one-third of our sources, the evolution of their radio brightness on the W3/radio diagram—resembling the radio variability observed in optically discovered TDEs with radio emission—may suggest the possibility of a TDE. We further hypothesize that the W3/radio diagram could serve as a useful tool for distinguishing between radio transients caused by TDEs and those originating from intrinsic AGN variability.
    
    \item The objects in which we suspect that their variability may be the result of a TDE are either pure star-forming galaxies or have a very significant ongoing star-forming process similar to radio-selected optically-detected TDEs.

\end{itemize}

\begin{acknowledgements}
The Pan-STARRS1 Surveys (PS1) have been made possible through contributions of the Institute for Astronomy, the University of Hawaii, the Pan-STARRS Project Office, the Max-Planck Society and its participating institutes, the Max Planck Institute for Astronomy, Heidelberg and the Max Planck Institute for Extraterrestrial Physics, Garching, The Johns Hopkins University, Durham University, the University of Edinburgh, Queen's University Belfast, the Harvard-Smithsonian Center for Astrophysics, the Las Cumbres Observatory Global Telescope Network Incorporated, the National Central University of Taiwan, the Space Telescope Science Institute, the National Aeronautics and Space Administration under Grant No. NNX08AR22G issued through the Planetary Science Division of the NASA Science Mission Directorate, the National Science Foundation under Grant No. AST-1238877, the University of Maryland, and Eotvos Lorand University (ELTE).

The CSS survey is funded by the National Aeronautics and Space Administration under Grant No. NNG05GF22G issued through the Science Mission Directorate Near-Earth Objects Observations Program. The CRTS survey is supported by the U.S.~National Science Foundation under
grants AST-0909182.

ZTF is supported by the National Science Foundation under Grant No. AST-1440341 and AST-2034437 and a collaboration including current partners Caltech, IPAC, the Oskar Klein Center at Stockholm University, the University of Maryland, University of California, Berkeley , the University of Wisconsin at Milwaukee, University of Warwick, Ruhr University, Cornell University, Northwestern University and Drexel University. Operations are conducted by COO, IPAC, and UW.

This publication makes use of data products from the Wide-field Infrared Survey Explorer, which is a joint project of the University of California, Los Angeles, and the Jet Propulsion Laboratory/California Institute of Technology, and NEOWISE, which is a project of the Jet Propulsion Laboratory/California Institute of Technology. WISE and NEOWISE are funded by the National Aeronautics and Space Administration.

We acknowledge the ASAS-SN team for providing public access to the Sky Patrol data. ASAS-SN is supported by the Gordon and Betty Moore Foundation through grant GBMF5490, the NSF grant AST-1515927, the Center for Cosmology and AstroParticle Physics at The Ohio State University, and the Mt. Cuba Astronomical Foundation. ASAS-SN also benefits from the participation of more than 10,000 volunteers via the Citizen ASAS-SN project, which is hosted by the Zooniverse platform.

We would like to thank Paweł Zieliński, Łukasz Wyrzykowski and the BHTOM team for the discussion and help. BHTOM is based on the open-source TOM Toolkit by LCO and has been developed with funding from the OPTICON-RadioNet Pilot (ORP) of the European Union's Horizon 2020 research and innovation programme under grant agreement No 101004719 (2021-2025). This project has received funding the European Union's Horizon Europe Research and Innovation programme ACME under grant agreement No 101131928 (2024-2028). 

MKB acknowledges support from the 'National Science Centre, Poland' under grant no. 2017/26/E/ST9/00216.

DKW acknowledges support from the 'National Science Centre, Poland' under grant no. 2021/43/B/ST9/03246.

\end{acknowledgements}

\bibliographystyle{aa}
\bibliography{ms.bib}

\begin{appendix} 
\section{Additional notes on individual sources}
\label{app:notes}

\hspace{1em}
{\bf 024345$-$28}.
Spiral galaxy, with majority of old populations but with a significant fraction ($\sim$30\%) of young stellar populations. 

{\bf 024609$+$34}.
Massive elliptical galaxy with old stellar populations only.

{\bf 031115$+$08}.
Spiral or lenticular edge-on galaxy with a burst of very young populations but low extinction. 
There are no emission lines detected in the spectrum, maybe because of the orientation of the galaxy and the galaxy disk obscuration. 

{\bf 053509$+$83}.
Very low S/N spectrum, hence fit is not reliable. Only old stellar population was found, with very low extinction. 
No emission lines are detected. 

{\bf 064001$+$28}.
Low S/N spectrum of lenticular/spiral galaxy with about 40$\%$ of stellar populations with age less than $10^7$ years, and significant extinction. 

{\bf 070837$+$32}.
Massive, elliptical galaxy with majority of old stellar populations ($\rm >10^8\,years$). 

{\bf 071829$+$59}.
Massive, elliptical galaxy with majority old stellar populations and small fraction of young stars. High value of the velocity dispersion indicates very massive black hole. 
 
{\bf 095141$+$37}.
Elliptical galaxy with old stellar populations and low extinction. 

{\bf 101841$-$13}.
Spiral galaxy with a possible dwarf interacting neighbour and significant amount of young stellar populations. 

{\bf 105035$-$07}.
Early type galaxy with a faint tail which may indicate past interaction. 
Galaxy hosts massive black hole. Gap in the spectrum at the position of the [O\,III] lines and the telluric line near the position of $\rm H\alpha$ and [N\,II] lines prevents emission line classification. However, presence of [O\,II] line may suggest the presence of a central AGN.

{\bf 110239$-$06}.
Quasar with strong C\,IV, C\,III] and Mg\,II lines, massive black hole and high Eddington ratio.

{\bf 112940$+$39}.
Elliptical galaxy with a possible interaction, old stellar populations and a small fraction of young stars ($\sim11\%$). In the bulge, as probed by the SDSS data, the contribution of young stellar populations is much smaller while the even higher velocity dispersion indicates a massive black hole. There are no emission lines detected with a sufficient S/N to classify the galaxy. 

{\bf 114101$+$10}.
Disc galaxy, most likely lenticular, with intermediate and old stellar populations.
The low velocity dispersion, close to the instrumental resolution, suggests a low-mass black hole, as indicated by both the Palomar and SDSS spectra.

{\bf 121619$+$12}.
Lenticular galaxy with a possible neighbour. Stellar populations 
dominated by old stars with ages $\rm >10^8\,years$ as indicated by both spectra.
There are no emission lines detected with a sufficient S/N to classify the galaxy. 

{\bf 130400$-$11}.
Massive, elliptical galaxy with old and intermediate stellar populations. Galaxy hosts massive black hole. There are no emission lines visible in the spectrum suitable for the optical classification. 

{\bf 150415$+$28}.
Massive, elliptical galaxy with old stellar populations and small ($\sim5\%$) fraction of young stars. Galaxy hosts massive black hole. No emission lines are detected.

{\bf 155847$+$14}.
Massive, elliptical galaxy with old stellar populations and low extinction. 

{\bf 164607$+$42}.
Spiral galaxy with a bar. Keck spectrum shows periodic bursts of star-formation. In the bulge, probed by the SDSS observations, the contribution of young stellar populations is smaller, but they are still present.

{\bf 180940$+$24}.
Low mass, possibly spiral galaxy with periodic episodes of star-formation and significant amount of populations with age $10^6$ years.

{\bf 183415$+$61}.
Elongated, possibly elliptical galaxy with old stellar populations only. Very massive.

{\bf 195335$-$04}.
Elliptical galaxy with old stellar populations and small ($<15\%$) addition of young stars. 
No emission lines are detected.

{\bf 203909$-$30}.
Probably elliptical galaxy with majority of old stellar populations with age $\rm >10^8\,years$. No emission lines are detected.

{\bf 223933$-$22}.
Elliptical galaxy with majority of old stellar populations with age $\rm >10^8\,years$. 
No emission lines are detected.

{\bf 233058$+$10}.
Probably elliptical galaxy with periodic episodes of star-formation and significant amount ($\sim40\%$) of young stars.

\newpage
\onecolumn
\section{Emission line measurements}
\label{app:emission_lines}

\begin{table*}[h]
\footnotesize
\caption{Emission line measurements }
\centering
\begin{tabular}{c r r r r r r r r r}
\hline\hline
\noalign{\smallskip}
Name & 
\multicolumn{1}{c}{Epoch} &
\multicolumn{1}{c}{$\rm H\beta$}& 
\multicolumn{1}{c}{[O\,III]} & 
\multicolumn{1}{c}{[O\,I]} & 
\multicolumn{1}{c}{$\rm H\alpha$}& 
\multicolumn{1}{c}{[N\,II]}& 
\multicolumn{1}{c}{[S\,II]}& 
\multicolumn{1}{c}{[S\,II]}&
\multicolumn{1}{c}{$\rm W_{H\alpha}$}\\
   & &
\multicolumn{1}{c}{$\lambda4861$}& 
\multicolumn{1}{c}{$\lambda5007$}& 
\multicolumn{1}{c}{$\lambda6300$}& 
\multicolumn{1}{c}{$\lambda6563$}& 
\multicolumn{1}{c}{$\lambda6584$}& 
\multicolumn{1}{c}{$\lambda6717$}& 
\multicolumn{1}{c}{$\lambda6731$}& 
\multicolumn{1}{c}{[$\AA$]}\\
\noalign{\smallskip}
\hline
\noalign{\smallskip}
024345$-$28 & Sep 2021& 2.97$\pm$0.07& 5.78$\pm$0.12& \multicolumn{1}{c}{$-$}&19.83$\pm$0.03& 17.57$\pm$0.07 & 4.50$\pm$0.02 & 3.52$\pm$0.06 & 20.70$\pm$0.01\\
024609$+$34 & Sep 2021& \multicolumn{1}{c}{$-$}&\multicolumn{1}{c}{$-$} & \multicolumn{1}{c}{$-$}& 5.72$\pm$0.17& 5.91$\pm$0.11& \multicolumn{1}{c}{$-$}& \multicolumn{1}{c}{$-$}& 1.59$\pm$0.03 \\
064001$+$28 & Sep 2021& 11.37$\pm$0.02 & 4.02$\pm$0.03 & 1.77$\pm$0.04 &61.66$\pm$0.06 & 24.39$\pm$0.17 & 10.29$\pm$0.65 & 7.92$\pm$0.75 & 27.84$\pm$0.01\\ 
070837$+$32            & Jan 2023& 
1.97$\pm$0.11 &\multicolumn{1}{c}{$-$}&\multicolumn{1}{c}{$-$}&4.46$\pm$0.05 & 2.13$\pm$0.10&\multicolumn{1}{c}{$-$} &\multicolumn{1}{c}{$-$} & 7.04$\pm$0.04\\ 
071829$+$59            & Jan 2023& 8.70$\pm$3.00& 12.53$\pm$0.64& 3.68$\pm$0.11 & 18.64$\pm$0.04& 17.80$\pm$0.05 & 9.71$\pm$0.13 & 4.79$\pm$0.09 & 18.86$\pm$0.01\\ 
095141$+$37 & Dec 2003  & \multicolumn{1}{c}{$-$} & 2.19$\pm$0.40 &\multicolumn{1}{c}{$-$} &5.42$\pm$0.33 & 1.90$\pm$0.10 & 2.54$\pm$0.08 & 1.67$\pm$0.09 & 7.36$\pm$0.33\\ 
            & Feb 2022& 
\multicolumn{1}{c}{$-$} & \multicolumn{1}{c}{$-$} &\multicolumn{1}{c}{$-$}&3.77$\pm$0.37& 1.82$\pm$0.12 & 2.02$\pm$0.06 & 2.14$\pm$0.08 & 14.20$\pm$0.78\\
101841$-$13 & Feb 2022&  4.09$\pm$0.11  & 20.14$\pm$0.06 &3.80$\pm$0.18 & 28.08$\pm$0.19 & 23.20$\pm$0.24 & 7.23$\pm$0.31 & 5.81$\pm$0.05 & 19.74$\pm$0.06\\  
114101$+$10 & May 2003 & 1.73$\pm$0.06 & 2.43$\pm0.06$ & \multicolumn{1}{c}{$-$} &5.70$\pm$0.06 & 8.57$\pm$0.07 & 1.92$\pm$0.01 & 1.45$\pm$0.02 & 5.99$\pm$0.04\\ 
            & Jan 2023& \multicolumn{1}{c}{$-$} & \multicolumn{1}{c}{$-$} &\multicolumn{1}{c}{$-$} &2.02$\pm$0.04 & 3.31$\pm$0.03 & 0.99$\pm$0.02 & 0.91$\pm$0.02 & 5.64$\pm$0.05\\
121619$+$12 & May 2004 & \multicolumn{1}{c}{$-$} & 4.27$\pm$ 0.04 &\multicolumn{1}{c}{$-$}&\multicolumn{1}{c}{$-$} &\multicolumn{1}{c}{$-$} & \multicolumn{1}{c}{$-$} &\multicolumn{1}{c}{$-$} &\multicolumn{1}{c}{$-$}\\ 
155847$+$14 & Apr 2008 & 8.09$\pm$3.00 & 11.90$\pm$0.14 &6.69$\pm$0.34 & 17.71$\pm$ 0.03 & 16.23$\pm$0.03 & 12.25$\pm$0.04 & 10.20$\pm$0.08 & 2.35$\pm$0.01\\
            & Sep 2021 & 7.14$\pm$0.09 & 12.32$\pm$0.09 & 7.52$\pm$0.16 & 23.77$\pm$0.03 & 22.41$\pm$0.02 & 18.19$\pm$0.25 & 14.56$\pm$0.05 & 3.42$\pm$0.02\\
164607$+$42 & Jun 2001 & 5.70$\pm$0.01 & 2.78$\pm$0.01 & \multicolumn{1}{c}{$-$} & 19.25$\pm$0.01 & 7.84$\pm$0.01 & 2.38$\pm$0.02 & 2.00$\pm$0.01 & 10.56$\pm$0.01\\ 
            & Sep 2021& 9.21$\pm$0.04 & 10.11$\pm$0.08 &2.19$\pm$0.02 &31.03$\pm$0.16 & 13.13$\pm$0.20 & 2.19$\pm$0.18 & 2.12$\pm$0.19 &23.36$\pm$0.06\\
180940$+$24 & Sep 2021&  7.51$\pm$0.01  & 9.40$\pm$0.02 & \multicolumn{1}{c}{$-$} &37.25$\pm$0.07 & 6.17$\pm$0.04 & 8.11$\pm$0.17 & 5.80$\pm$0.17 & 71.59$\pm$0.01\\
183415$+$61 & Jul 2021    & \multicolumn{1}{c}{$-$}& 0.89$\pm$0.03 &\multicolumn{1}{c}{$-$} &1.41$\pm$0.16 &2.93$\pm$0.09 & \multicolumn{1}{c}{$-$}& \multicolumn{1}{c}{$-$}& 3.26$\pm$0.22\\
195335$-$04 & Sep 2021& 
\multicolumn{1}{c}{$-$}& 16.28$\pm$0.21 &\multicolumn{1}{c}{$-$}& \multicolumn{1}{c}{$-$}& \multicolumn{1}{c}{$-$}&\multicolumn{1}{c}{$-$} &\multicolumn{1}{c}{$-$} &\multicolumn{1}{c}{$-$}\\ 
233058$+$10 & Sep 2021& 33.18$\pm$0.08 & 32.04$\pm$0.11 &\multicolumn{1}{c}{$-$} &174.09$\pm$0.23 & 40.33$\pm$0.37 & 17.23$\pm$0.36 & 13.29$\pm$0.30 & 109.15$\pm$0.06\\ 
\noalign{\smallskip}
\hline\hline
\noalign{\smallskip}
Name & 
\multicolumn{1}{c}{Epoch} &
\multicolumn{1}{c}{C\,IV} &
\multicolumn{1}{c}{C\,III]} &
\multicolumn{1}{c}{Mg\,II} & 
\multicolumn{1}{c}{Mg\,II\,FWHM} & & \\
   & &
\multicolumn{1}{c}{$\lambda1549$}&
\multicolumn{1}{c}{$\lambda1909$}& 
\multicolumn{1}{c}{$\lambda2800$} & 
\multicolumn{1}{c}{$\rm (km\,s^{-1})$}& & \\
\hline
\noalign{\smallskip}
110239$-$06 & Jan 2023& 153.40$\pm$36.60 &65.20$\pm$7.50& 37.40$\pm$2.45&\multicolumn{1}{c}{5166$\pm$91} & & &\\ 
\noalign{\smallskip}
\hline
\end{tabular}
\tablefoot{Line fluxes are in units of $\rm 10^{-16}\,erg\,s^{-1}\,cm^{-2}$. The fluxes are not corrected for reddening in the host galaxies. 
} 
\label{table:emission_line_measurements}
\end{table*}

\section{WISE and radio data}
\label{app:wise_and_radio}

Flux density measurements of our sources from follow-up VLA observations \citep{MKB2025}, along with their infrared colors from the ALLWISE catalog, are presented in Table~\ref{table:wise_radio_measurements}. Table~\ref{table:wise_radio_measurements_TDEs} lists infrared colors from the ALLWISE catalog and radio flux densities from \citet{Cendes2024} for optically discovered TDEs, which are shown in Figures~\ref{figure:wise_color_color} and~\ref{figure:flux_wise_radio}. This table also includes data for two additional TDEs: AT2019azh and SwJ1644+57.

From the TDEs discussed in \citet{Cendes2024}, we selected only those with radio measurements at frequencies close to 1.4\,GHz (specifically 1.25, 1.36, and 1.75\,GHz). We excluded sources classified as "TDEs with Ambiguous or Host/AGN Radio Emission" in that work. For several objects, we used additional literature data to plot the lines representing their radio flux evolution:

\begin{itemize}

    \item For AT2018hyz, we used data from \citet{Cendes2022}; we estimated the upper limit on the flux density at 1.4\,GHz ($<$0.45\,mJy) based on FIRST radio observations.

    \item For AT2019azh, we adopted radio data from \citet{Goodwin2022}. The 1.4\,GHz flux density upper limit was estimated by fitting a power-law spectrum to the VLA data from the first observing epoch presented in that work.

    \item For SwJ1644+57, we used data from \citet{Cendes2021}, and we adopted an upper limit of $<$0.3\,mJy at 1.4\,GHz based on their analysis.
\end{itemize}

Additionally, for better visual clarity in Figure~\ref{figure:flux_wise_radio}, the W3 flux values of sources 155847$+$14 (ID 17), 180940$+$24 (ID 19), and AT2019eve were slightly adjusted to avoid overlapping with neighboring points. However, the values reported in the table correspond to the actual measured W3 fluxes.

\noindent
\begin{center} 
\begin{sidewaystable*}[htbp]
\centering
\footnotesize
\caption{WISE infrared data and VLA radio data of our radio transient sources}
\begin{tabular}{c c r r r r r r c r r}
\hline\hline
\noalign{\smallskip}
ID & Name & 
\multicolumn{1}{c}{W1} &
\multicolumn{1}{c}{W2}& 
\multicolumn{1}{c}{W3} & 
\multicolumn{1}{c}{W1 - W2}& 
\multicolumn{1}{c}{W2 - W3}& 
\multicolumn{1}{c}{$\rm S_{1.4GHz}$}& 
\multicolumn{1}{c}{$\rm log\,S_{1.4GHz}$}&
\multicolumn{1}{c}{$\rm S_{W3}$}& 
\multicolumn{1}{c}{$\rm log\,S_{W3}$}
\\
&  &
\multicolumn{1}{c}{(mag)} &
\multicolumn{1}{c}{(mag)} &
\multicolumn{1}{c}{(mag)} &
\multicolumn{1}{c}{(mag)} &
\multicolumn{1}{c}{(mag)} &
\multicolumn{1}{c}{(mJy)} & &
\multicolumn{1}{c}{(mJy)} & \\
\noalign{\smallskip}
\hline
\noalign{\smallskip}
1 &024345$-$28	&	14.143 $\pm$ 0.026			&	14.026	$\pm$	0.034	&	10.604	$\pm$	0.068	&	0.117	$\pm$	0.043	&	3.422	$\pm$	0.076	&	62.91	$\pm$	1.96	&	$1.80^{+0.01}_{-0.01}$	&	1.82	$\pm$	0.11	&	$0.26^{+0.03}_{-0.03}$\\
\noalign{\smallskip}
2 &024609$+$34	&	13.212 $\pm$ 0.024			&	13.050	$\pm$	0.030	&	12.335	$\pm$	0.408	&	0.162	$\pm$	0.038	&	0.715	$\pm$	0.409	&	2.14	$\pm$	0.45	&	$0.33^{+0.08}_{-0.10}$	&	0.37	$\pm$	0.14	&	$-0.43^{+0.14}_{-0.20}$	\\
\noalign{\smallskip}
3 &031115$+$08	&	13.727	$\pm$	0.026	&	13.657	$\pm$	0.036	&	$>$12.286			&	0.070	$\pm$	0.044	&	$<$1.371			&	18.21	$\pm$	0.69	&	$1.26^{+0.02}_{-0.02}$&	$<$0.39			&	$<$-0.41	\\
\noalign{\smallskip}
4 &053509$+$83	&	14.098	$\pm$	0.027	&	13.507	$\pm$	0.029	&	11.092	$\pm$	0.124	&	0.591	$\pm$	0.040	&	2.415	$\pm$	0.127	&	19.29	$\pm$	0.90	&	$1.29^{+0.02}_{-0.02}$&	1.16	$\pm$	0.13	&	$0.06^{+0.05}_{-0.05}$	\\
\noalign{\smallskip}
5 &064001$+$28	&	12.546	$\pm$	0.033	&	12.468	$\pm$	0.038	&	8.620	$\pm$	0.041	&	0.078	$\pm$	0.050	&	3.848	$\pm$	0.056	&	32.99	$\pm$	0.64	&	$1.52^{+0.01}_{-0.01}$&	11.29	$\pm$	0.43	&	$1.05^{+0.02}_{-0.02}$	\\
\noalign{\smallskip}
6 &070837$+$32	&	14.569	$\pm$	0.032	&	14.278	$\pm$	0.052	&	11.550	$\pm$	0.251	&	0.291	$\pm$	0.061	&	2.728	$\pm$	0.256	&	2.87	$\pm$	0.36	&	$0.46^{+0.05}_{-0.06}$&	0.76	$\pm$	0.18	&	$-0.12^{+0.09}_{-0.11}$\\
\noalign{\smallskip}
7 &071829$+$59	&	13.850	$\pm$	0.026	&	13.479	$\pm$	0.031	&	11.649	$\pm$	0.196	&	0.371	$\pm$	0.040	&	1.830	$\pm$	0.198	&	6.43	$\pm$	0.93	&	$0.81^{+0.06}_{-0.07}$&	0.69	$\pm$	0.13	&	$-0.16^{+0.07}_{-0.09}$	\\
\noalign{\smallskip}
8 &095141$+$37	&	14.320	$\pm$	0.028	&	13.888	$\pm$	0.038	&	12.460	$\pm$	0.478	&	0.432	$\pm$	0.047	&	1.428	$\pm$	0.480	&	3.58	$\pm$	0.96	&	$0.55^{+0.10}_{-0.14}$&	0.33	$\pm$	0.14	&	$-0.48^{+0.16}_{-0.25}$\\
\noalign{\smallskip}
9 &101841$-$13	&	12.262	$\pm$	0.022	&	11.821	$\pm$	0.021	&	8.357	$\pm$	0.025	&	0.441	$\pm$	0.030	&	3.464	$\pm$	0.033	&	2.46	$\pm$	0.75	&	$0.39^{+0.12}_{-0.16}$&	14.38	$\pm$	0.33	&	$1.16^{+0.01}_{-0.01}$\\
\noalign{\smallskip}
10 &105035$-$07	&	13.697	$\pm$	0.026	&	13.478	$\pm$	0.031	&	$>$12.427			&	0.219	$\pm$	0.040	&	$<$1.051			&	4.47	$\pm$	0.30	&	$0.65	^{+0.03}_{-0.03}$&	$<$0.34			&	$<$-0.47	\\
\noalign{\smallskip}
11 &110239$-$06	&	15.190	$\pm$	0.038	&	14.588	$\pm$	0.042	&	11.771	$\pm$	0.278	&	0.602	$\pm$	0.057	&	2.817	$\pm$	0.281	&	10.90	$\pm$	2.25	&	$1.04^{+0.08}_{-0.10}$&	0.62	$\pm$	0.16	&	$-0.21^{+0.10}_{-0.13}$\\
\noalign{\smallskip}
12 &112940$+$39	&	14.094	$\pm$	0.026	&	13.817	$\pm$	0.033	&	$>$11.694			&	0.277	$\pm$	0.042	&	$<$2.123			&	2.77	$\pm$	0.08	&	$0.44	^{+0.01}_{-0.01}$&	$<$0.67			&	$<$-0.18	\\
\noalign{\smallskip}
13 &114101$+$10	&	14.157	$\pm$	0.028	&	13.924	$\pm$	0.041	&	10.965	$\pm$	0.136	&	0.233	$\pm$	0.050	&	2.959	$\pm$	0.142	&	5.83	$\pm$	1.29	&	$0.77^{+0.09}_{-0.11}$	&	1.30	$\pm$	0.16	&	$0.11^{+0.05}_{-0.06}$\\
\noalign{\smallskip}
14 &121619$+$12	&	13.853	$\pm$	0.027	&	13.730	$\pm$	0.036	&	11.926	$\pm$	0.241	&	0.123	$\pm$	0.045	&	1.804	$\pm$	0.244	&	12.83	$\pm$	0.95	&	$1.11^{+0.03}_{-0.03}$&	0.54	$\pm$	0.12	&	$-0.27^{+0.09}_{-0.11}$\\
\noalign{\smallskip}
15 &130400$-$11	&	13.192	$\pm$	0.025	&	12.993	$\pm$	0.028	&	12.286	$\pm$	0.396	&	0.199	$\pm$	0.038	&	0.707	$\pm$	0.397	&	4.71	$\pm$	1.08	&	$0.67^{+0.09}_{-0.11}$&	0.39	$\pm$	0.14	&	$-0.41^{+0.14}_{-0.20}$	\\
\noalign{\smallskip}
16 &150415$+$28	&	11.673	$\pm$	0.023	&	11.683	$\pm$	0.021	&	11.188	$\pm$	0.108	&	-0.010	$\pm$	0.031	&	0.495	$\pm$	0.110	&	6.30	$\pm$	0.91	&	$0.80^{+0.06}_{-0.07}$&	1.06	$\pm$	0.11	&	$0.03^{+0.04}_{-0.05}$\\
\noalign{\smallskip}
17 &155847$+$14	&	11.709	$\pm$	0.023	&	11.737	$\pm$	0.022	&	10.581	$\pm$	0.090	&	-0.028	$\pm$	0.032	&	1.156	$\pm$	0.093	&	7.74	$\pm$	1.55	&	$0.89^{+0.08}_{-0.10}$&	1.85	$\pm$	0.15	&	$0.27^{+0.03}_{-0.04}$\\
\noalign{\smallskip}
18 &164607$+$42	&	12.866	$\pm$	0.025	&	12.599	$\pm$	0.027	&	8.921	$\pm$	0.031	&	0.267	$\pm$	0.037	&	3.678	$\pm$	0.041	&	9.16	$\pm$	0.97	&	$0.96^{+0.04}_{-0.05}$&	8.56	$\pm$	0.24	&	$0.93^{+0.01}_{-0.01}$\\
\noalign{\smallskip}
19 &180940$+$24	&	13.825	$\pm$	0.025	&	12.380	$\pm$	0.023	&	8.358	$\pm$	0.022	&	1.445	$\pm$	0.034	&	4.022	$\pm$	0.032	&	8.25	$\pm$	1.55	&	$0.92^{+0.07}_{-0.09}$&	14.37	$\pm$	0.29	&	$1.16^{+0.01}_{-0.01}$\\
\noalign{\smallskip}
20 &183415$+$61	&	14.295	$\pm$	0.025	&	14.009	$\pm$	0.030	&	13.333	$\pm$	0.487	&	0.286	$\pm$	0.039	&	0.676	$\pm$	0.488	&	5.06	$\pm$	0.58	&	$0.70^{+0.05}_{-0.05}$&	0.15	$\pm$	0.07	&	$-0.83^{+0.16}_{-0.26}$\\
\noalign{\smallskip}
21 &195335$-$04	&	13.056	$\pm$	0.024	&	12.779	$\pm$	0.028	&	10.111	$\pm$	0.066	&	0.277	$\pm$	0.037	&	2.668	$\pm$	0.072	&	11.56	$\pm$	3.11	&	$1.06^{+0.10}_{-0.14}$&	2.86	$\pm$	0.17	&	$0.46^{+0.03}_{-0.03}$\\
\noalign{\smallskip}
22 &203909$-$30	&	14.176	$\pm$	0.029	&	14.119	$\pm$	0.047	&	$>$11.922			&	0.057	$\pm$	0.055	&	$<$2.197			&	9.55	$\pm$	1.82	&	$0.98^{+0.08}_{-0.09}$&	$<$0.54			&	$<$-0.27	\\
\noalign{\smallskip}
23 &223933$-$22	&	13.354	$\pm$	0.025	&	13.152	$\pm$	0.031	&	12.038	$\pm$	0.290	&	0.202	$\pm$	0.040	&	1.114	$\pm$	0.292	&	11.68	$\pm$	2.70	&	$1.07^{+0.09}_{-0.11}$&	0.48	$\pm$	0.13	&	$-0.31^{+0.10}_{-0.13}$	\\
\noalign{\smallskip}
24 &233058$+$10	&	13.806	$\pm$	0.026	&	13.457	$\pm$	0.036	&	9.274	$\pm$	0.034	&	0.349	$\pm$	0.044	&	4.183	$\pm$	0.050	&	4.63	$\pm$	0.87	&	$0.67^{+0.07}_{-0.09}$&	6.18	$\pm$	0.19	&	$0.79^{+0.01}_{-0.01}$	\\

\noalign{\smallskip}
\hline
\end{tabular}
\tablefoot{WISE colors in Vega magnitudes \citep{Wright} are taken from ALLWISE catalog. These data are used in Figure \ref{figure:wise_color_color}. “<” and ">" mark an upper and lower limits, respectively.}
\label{table:wise_radio_measurements}
\end{sidewaystable*}
\end{center}

\noindent
\begin{center} 
\begin{sidewaystable*}[htbp]
\centering
\footnotesize
\caption{WISE infrared data and radio data for optically discovered TDEs}
\begin{tabular}{l r r r r r r r r r}
\hline\hline
\noalign{\smallskip}
\multicolumn{1}{c}{Name} & 
\multicolumn{1}{c}{W1} &
\multicolumn{1}{c}{W2}& 
\multicolumn{1}{c}{W3} & 
\multicolumn{1}{c}{W1 - W2}& 
\multicolumn{1}{c}{W2 - W3}& 
\multicolumn{1}{c}{$\rm S_{1.4GHz}$}& 
\multicolumn{1}{c}{$\rm log\,S_{1.4GHz}$}&
\multicolumn{1}{c}{$\rm S_{W3}$}& 
\multicolumn{1}{c}{$\rm log\,S_{W3}$}
\\
&
\multicolumn{1}{c}{(mag)} &
\multicolumn{1}{c}{(mag)} &
\multicolumn{1}{c}{(mag)} &
\multicolumn{1}{c}{(mag)} &
\multicolumn{1}{c}{(mag)} &
\multicolumn{1}{c}{(mJy)} & &
\multicolumn{1}{c}{(mJy)} & \\
\noalign{\smallskip}
\hline
\noalign{\smallskip}
iPTF16fnl	&	13.060	$\pm$	0.031	&	13.050	$\pm$	0.035	&	12.238	$\pm$	0.325	&	0.010	$\pm$	0.047	&	0.812	$\pm$	0.327	&	\multicolumn{1}{c}{$-$}	&		\multicolumn{1}{c}{$-$}	&	\multicolumn{1}{c}{$-$}	&	\multicolumn{1}{c}{$-$}	\\
\noalign{\smallskip}																				AT2018hco	&	12.801	$\pm$	0.023	&	12.679	$\pm$	0.027	&	12.179	$\pm$	0.346	&	0.122	$\pm$	0.035	&	0.500	$\pm$	0.347	&	0.171	$\pm$	0.025	&	$	-0.77	^{	+0.06	}_{-0.07	}	$	&	0.426	$\pm$	0.136	&	$	-0.37	^{+0.12	}_{-0.17}	$	\\
\noalign{\smallskip}																			PS16dtm	&	15.089	$\pm$	0.033	&	14.651	$\pm$	0.061	&	11.731	$\pm$	0.216	&	0.438	$\pm$	0.069	&	2.920	$\pm$	0.224	&	0.193	$\pm$	0.047	&	$	-0.71	^{	+0.09	}_{-0.12	}	$	&	0.643	$\pm$	0.128	&	$	-0.19	^{+0.08	}_{-0.10}	$	\\
\noalign{\smallskip}																				
AT2018zr	&	15.191	$\pm$	0.038	&	15.317	$\pm$	0.109	&	$>$12.372&	-0.126	$\pm$	0.115	&	$<$2.945&	0.218	$\pm$	0.041	&	$	-0.66	^{	+0.07	}_{-0.09	}	$	&	$<$0.356&	$<$-0.45\\
\noalign{\smallskip}																				
AT2018hyz	&	14.152	$\pm$	0.028	&	14.001	$\pm$	0.042	&	$>$12.167&	0.151	$\pm$	0.050	&	$<$1.834&	8.740	$\pm$	0.093	&	$	0.94	^{+0.01	}_{-0.00}	$	&	$<$0.430&	$<$-0.37\\
\noalign{\smallskip}																				
ASASSN-14ae	&	14.296	$\pm$	0.027	&	14.150	$\pm$	0.046	&	12.358	$\pm$	0.346	&	0.146	$\pm$	0.053	&	1.792	$\pm$	0.349	&	0.720	$\pm$	0.041	&	$	-0.14	^{	+0.02	}_{-0.02	}	$	&	0.361	$\pm$	0.115	&	$	-0.44	^{+0.12	}_{-0.17	}	$	\\
\noalign{\smallskip}																				
AT2019eve	&	16.976	$\pm$	0.117	&	16.749	$\pm$	0.341	&	$>$12.350&	0.227	$\pm$	0.360	&	$<$4.399&	0.846	$\pm$	0.175	&	$	-0.07	^{+0.08	}_{-0.10	}	$	&	$<$0.364&	$<$-0.44\\
\noalign{\smallskip}																				
AT2019ehz	&	16.027	$\pm$	0.044	&	16.025	$\pm$	0.126	&	$>$12.594&	0.002	$\pm$	0.133	&	$<$3.431&	\multicolumn{1}{c}{$-$}	&	\multicolumn{1}{c}{$-$}	&	\multicolumn{1}{c}{$-$}	&	\multicolumn{1}{c}{$-$}	\\
\noalign{\smallskip}																				
AT2020neh	&	14.909	$\pm$	0.033	&	14.773	$\pm$	0.067	&	11.634	$\pm$	0.155	&	0.136	$\pm$	0.075	&	3.139	$\pm$	0.169	&	\multicolumn{1}{c}{$-$}	&	\multicolumn{1}{c}{$-$}	&	\multicolumn{1}{c}{$-$}	&	\multicolumn{1}{c}{$-$}	\\
\noalign{\smallskip}																				
AT2018dyb	&	11.545	$\pm$	0.023	&	11.676	$\pm$	0.022	&	11.160	$\pm$	0.135	&	-0.131	$\pm$	0.032	&	0.516	$\pm$	0.137	&	1.031	$\pm$	0.068	&	$	0.01	^{	+0.03	}_{-0.03	}	$	&	1.089	$\pm$	0.135	&	$	0.04	^{+0.05	}_{-0.06	}	$	\\
\noalign{\smallskip}																				
AT2019teq	&	14.899	$\pm$	0.029	&	14.688	$\pm$	0.043	&	11.098	$\pm$	0.096	&	0.211	$\pm$	0.052	&	3.590	$\pm$	0.105	&	0.492	$\pm$	0.088	&	$	-0.31	^{	+0.07}_{-0.08}	$	&	1.152	$\pm$	0.102	&	$	0.06^{+0.04	}_{-0.04}	$	\\
\noalign{\smallskip}																				
AT2019dsg	&	13.272	$\pm$	0.024	&	13.252	$\pm$	0.029	&	10.986	$\pm$	0.124	&	0.020	$\pm$	0.037	&	2.266	$\pm$	0.127	&	0.228	$\pm$	0.026	&	$	-0.64	^{	+0.05}_{-0.05}	$	&	1.277	$\pm$	0.146	&	$	0.11^{+0.05}_{-0.05}	$	\\
\noalign{\smallskip}
\hline
\noalign{\smallskip}
AT2019azh	&	12.372	$\pm$	0.025	&	12.355	$\pm$	0.026	&	11.263	$\pm$	0.196	&	0.017	$\pm$	0.036	&	1.092	$\pm$	0.198	&	2.44	$\pm$	0.10	&	$	0.387	^{	+0.02}_{-0.02}	$	&	0.990	$\pm$	0.176	&	$	0.00	^{+0.07	}_{-0.09}	$	\\
\noalign{\smallskip}
SwJ1644+57	&	16.271	$\pm$	0.036	&	16.639	$\pm$	0.118	&	$>$13.653&	-0.368	$\pm$	0.123	&	$<$2.986	&	$<$0.3	&$<$-0.523&	$<$0.11		&	$<$-0.96							\\
\noalign{\smallskip}
\hline
\end{tabular}
\tablefoot{WISE colors in Vega magnitudes are taken from ALLWISE catalog. The radio data are taken from \citet{Cendes2021,Cendes2022,Goodwin2022,Cendes2024}. These data are used in Figure \ref{figure:wise_color_color}. “<” and ">" mark an upper and lower limits, respectively.}
\label{table:wise_radio_measurements_TDEs}
\end{sidewaystable*}
\end{center}

\clearpage
\section{Optical spectra}
\label{app:opt_spec}

\begin{figure}[h!]
   \centering
   \includegraphics[width=0.45\textwidth]{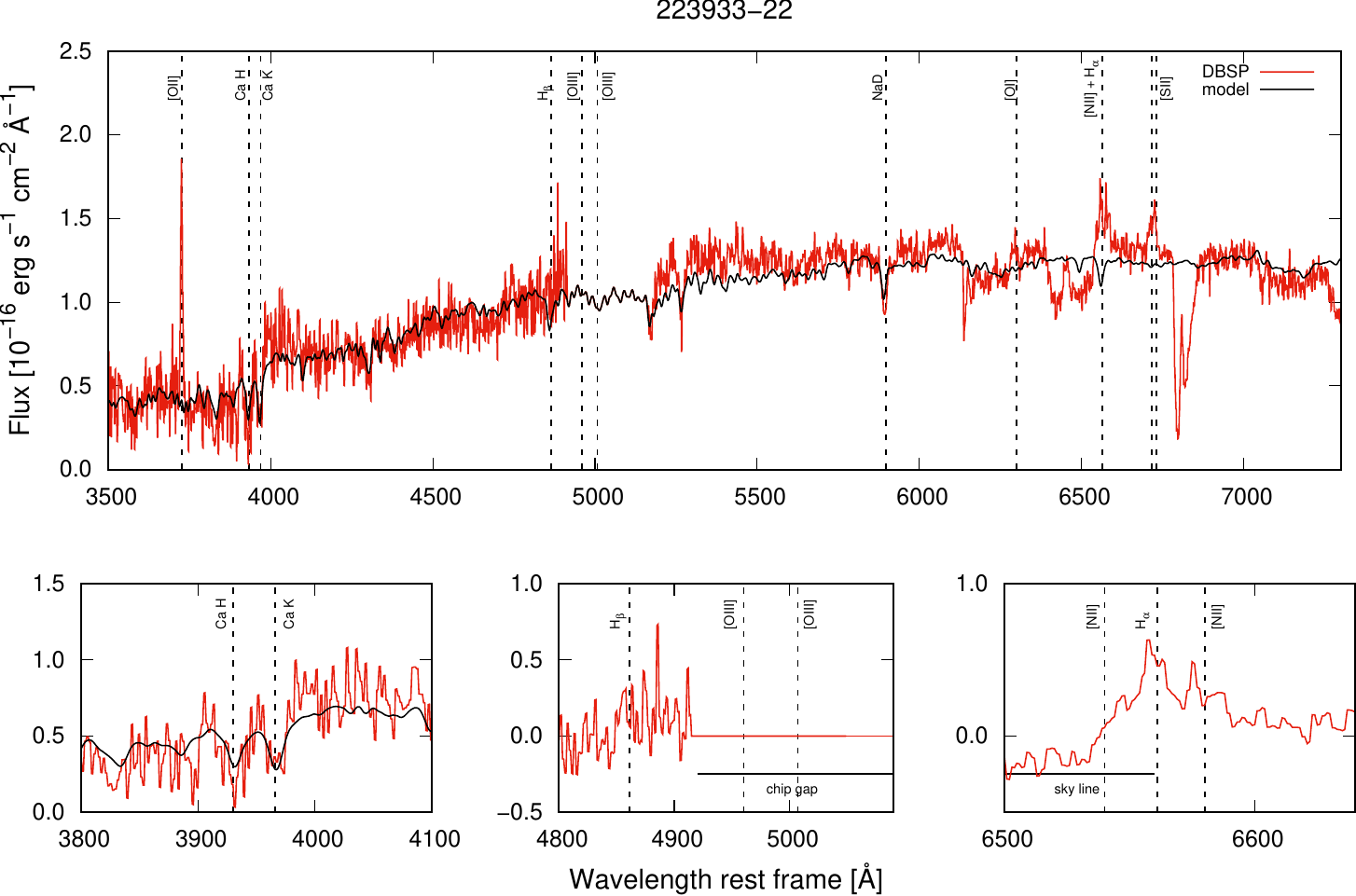}
   \includegraphics[width=0.45\textwidth]{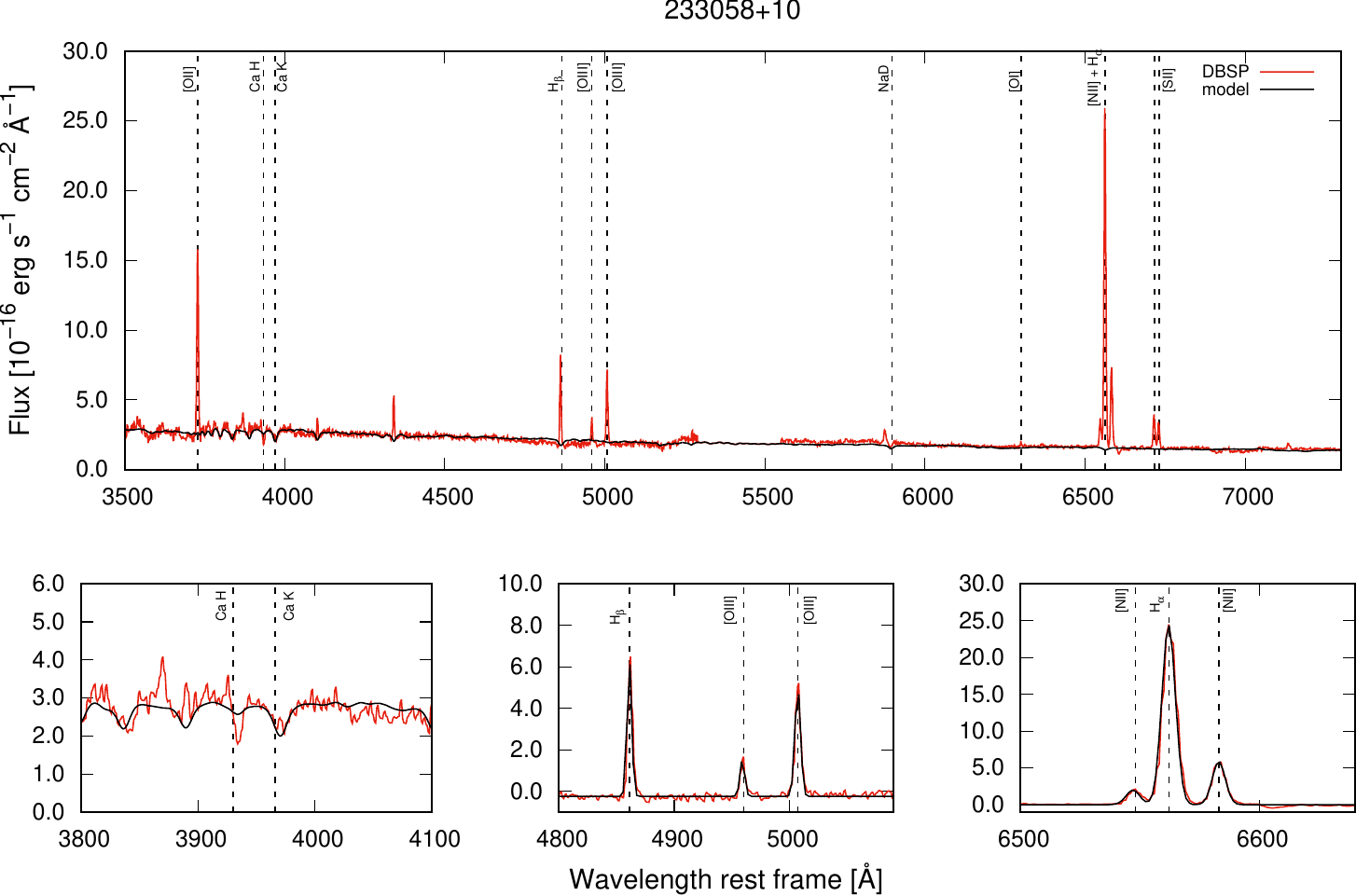}
\caption{Keck (LRIS) and Palomar (DBSP) optical spectra of all sources from our sample.} 
\label{figure:optical_spectra}
    \end{figure}

In Figure \ref{figure:optical_spectra}, we present optical spectra obtained with Keck (LRIS) and Palomar (DBSP), showing characteristic emission and absorption lines (in red), along with the stellar population model fitted using STARLIGHT (in black).
The positions of the most common emission and absorption lines are labelled and indicated by dashed vertical lines.

The lower panels display, from left to right: the observed (red) and modeled (black) Ca\,H\&K absorption lines; the H$\beta$ and [O\,III] emission lines after subtraction of the stellar continuum (red), along with the fitted Gaussian models (black); and the [N\,II] and H$\alpha$ lines with their corresponding Gaussian fits. Breaks in the observed spectra indicate gaps between CCD chips. Where applicable, the positions of strong sky lines are also marked in the lower panels.

\begin{figure}[h!]
   \centering
   \includegraphics[width=0.45\textwidth]{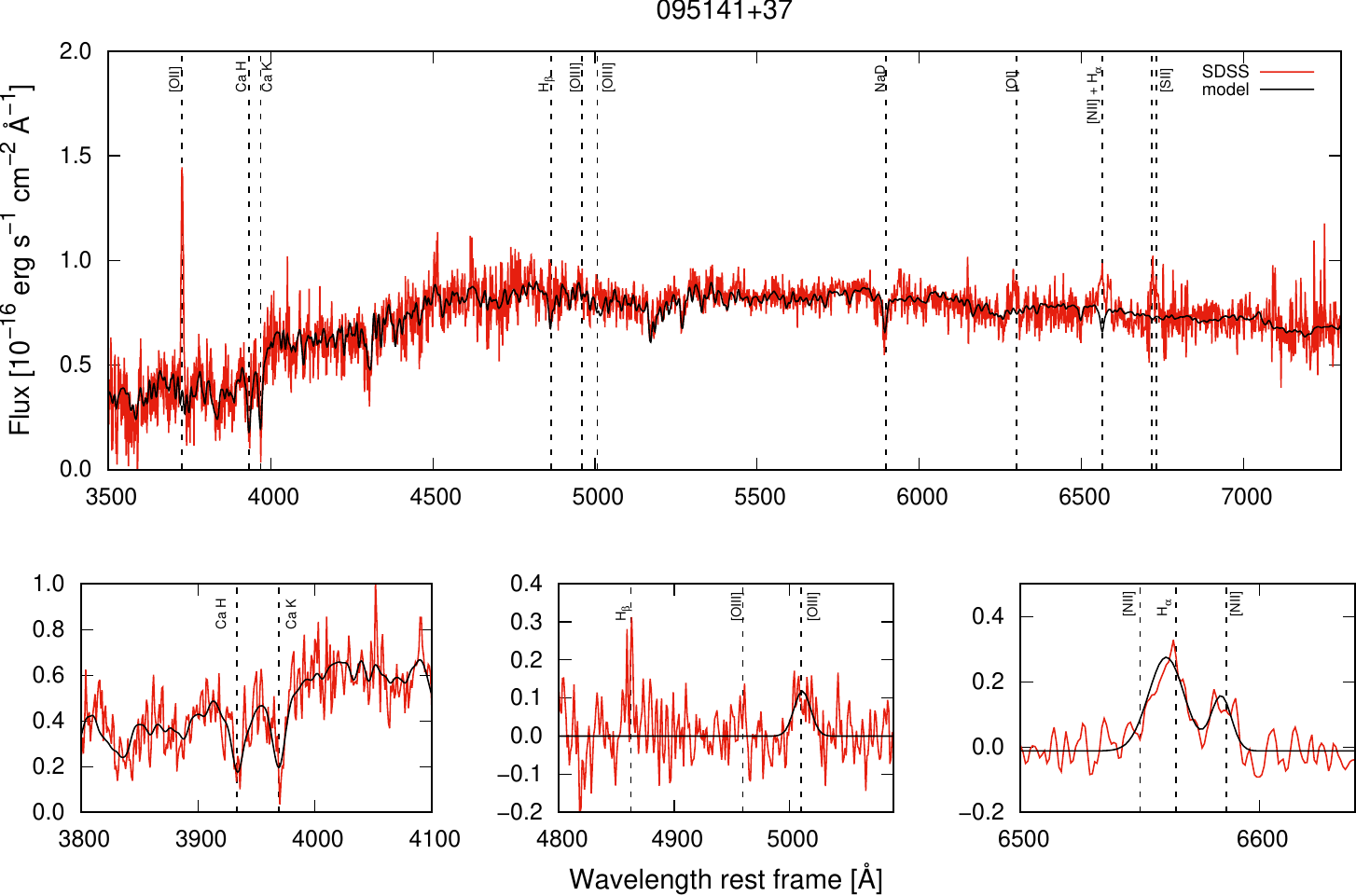}
   \includegraphics[width=0.45\textwidth]{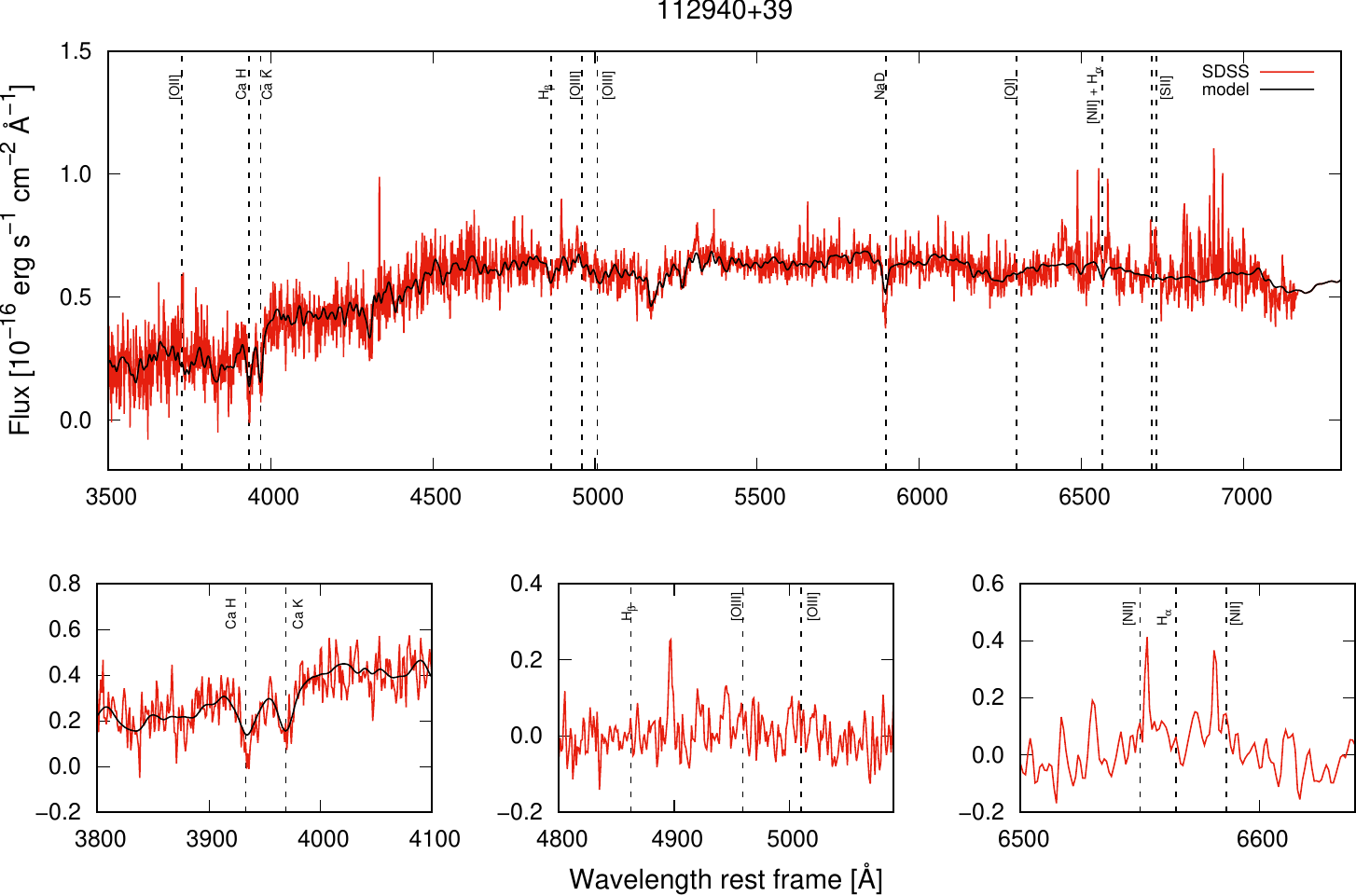}
   \includegraphics[width=0.45\textwidth]{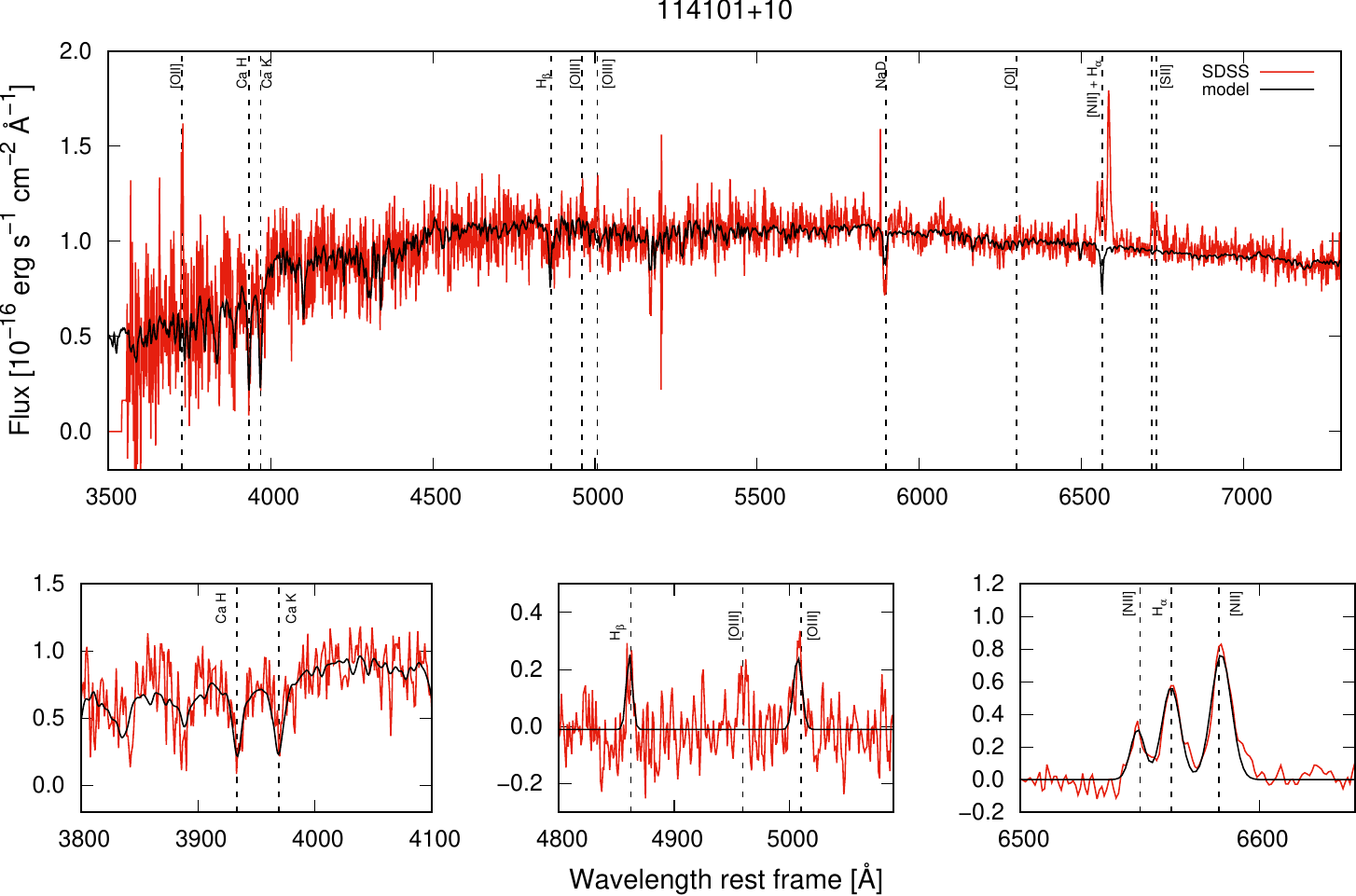}
   \includegraphics[width=0.45\textwidth]{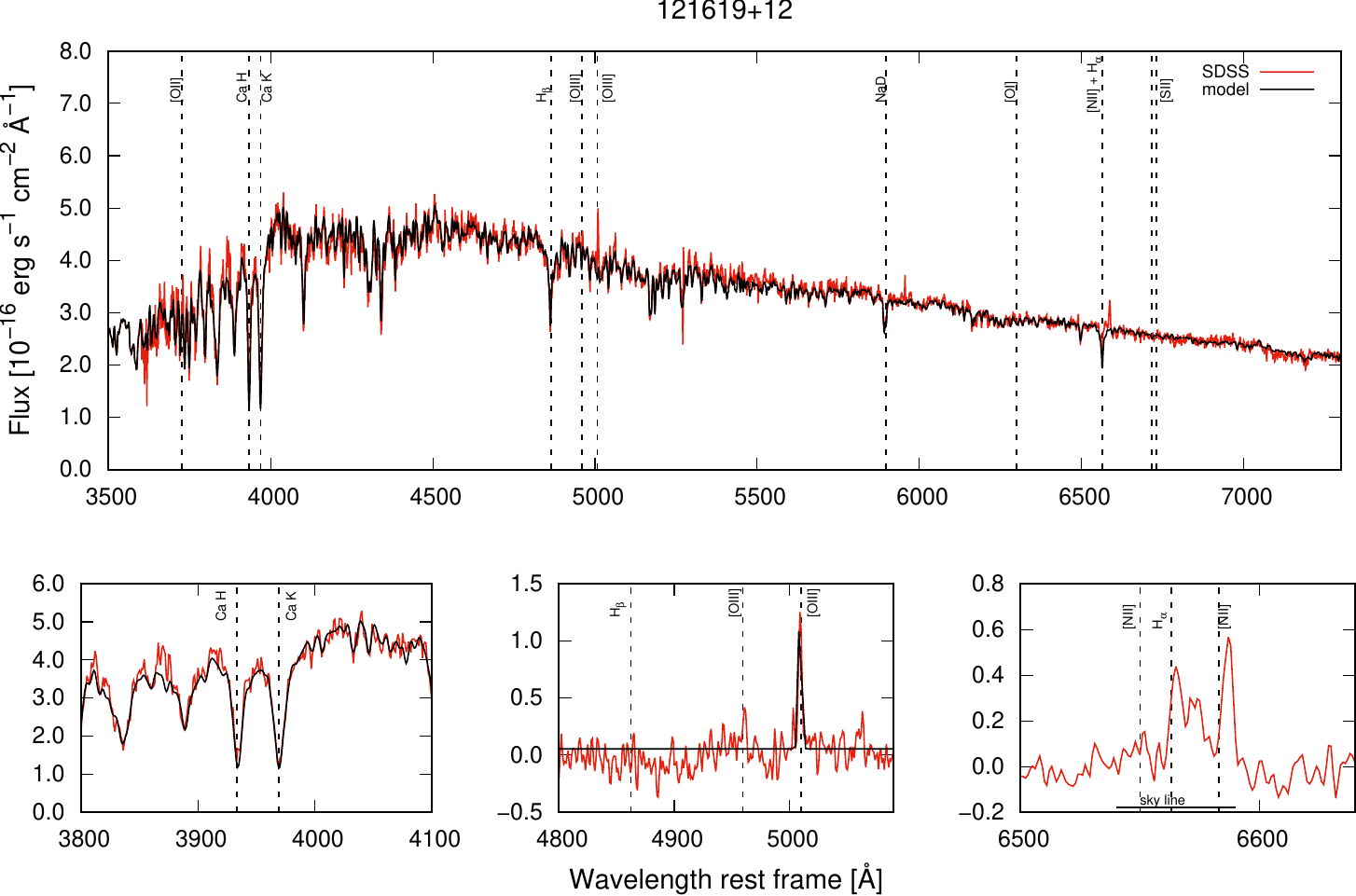}
   \includegraphics[width=0.45\textwidth]{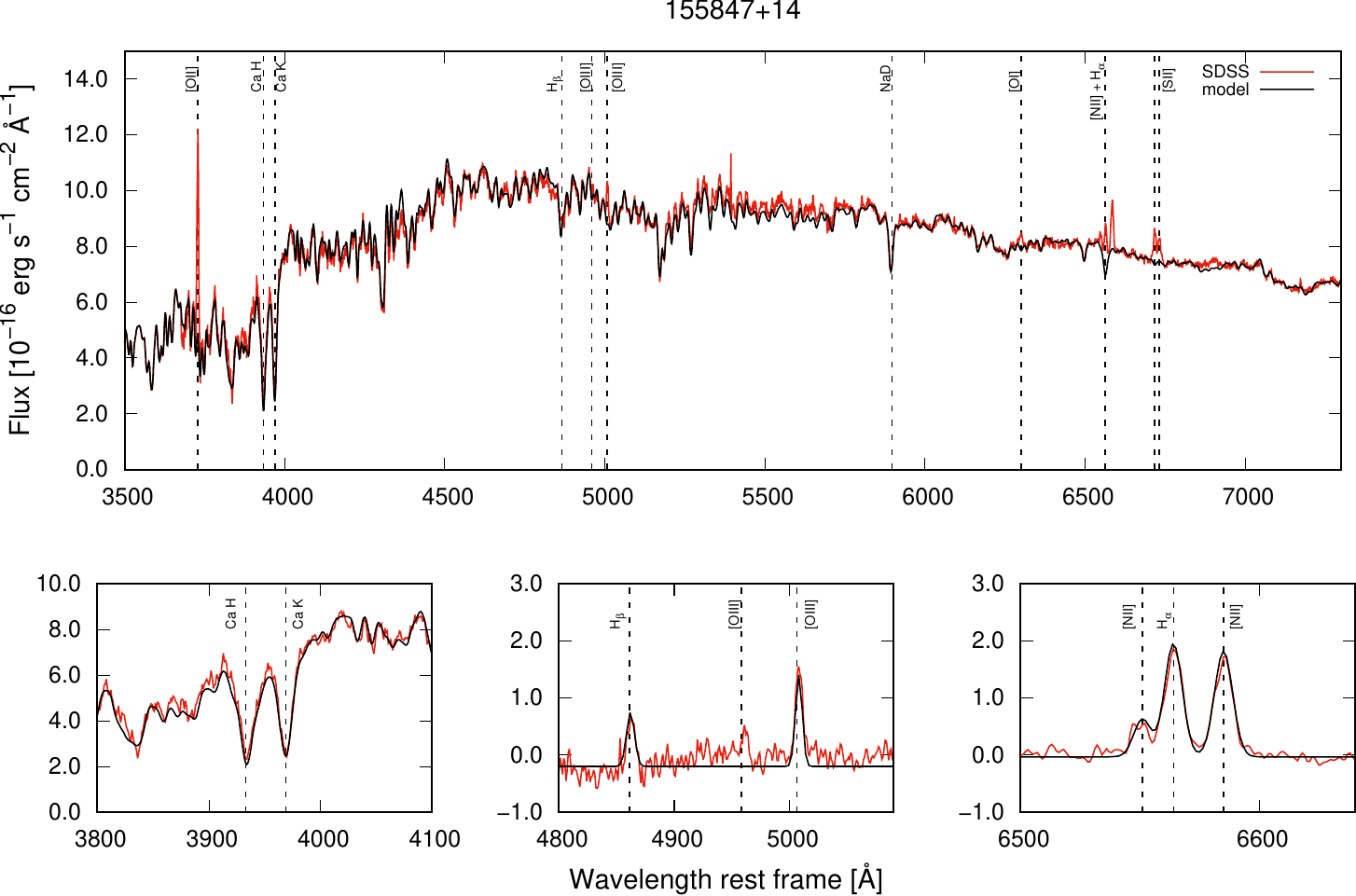}
   \includegraphics[width=0.45\textwidth]{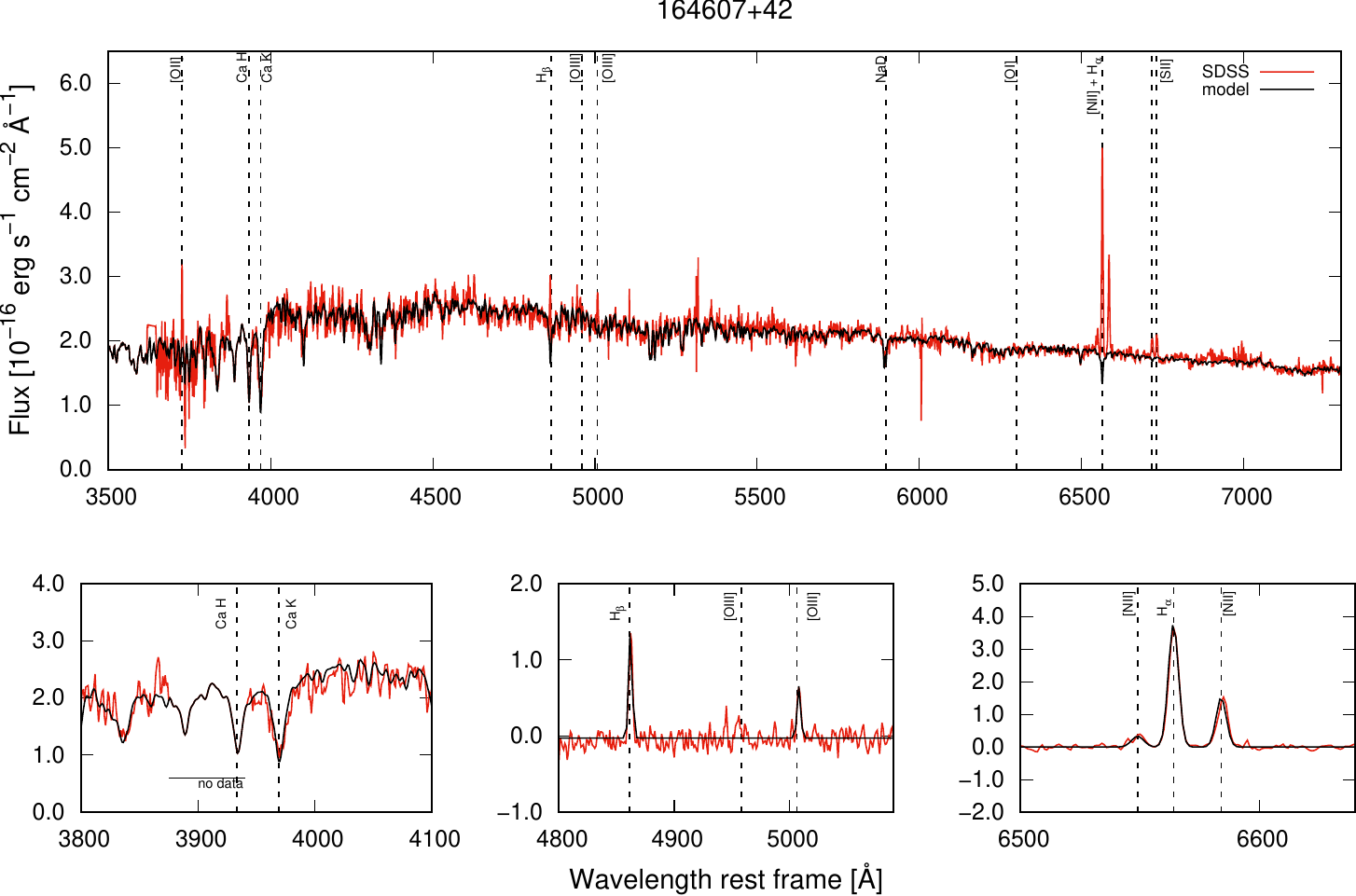}
\caption{SDSS optical spectra of six sources from our sample.}
\label{figure:optical_sdss_spectra}
    \end{figure}

The SDSS spectra are shown in Figure \ref{figure:optical_sdss_spectra}, following the same labeling and formatting scheme.

\begin{figure}[h!]
   \centering
   \includegraphics[width=0.45\textwidth]{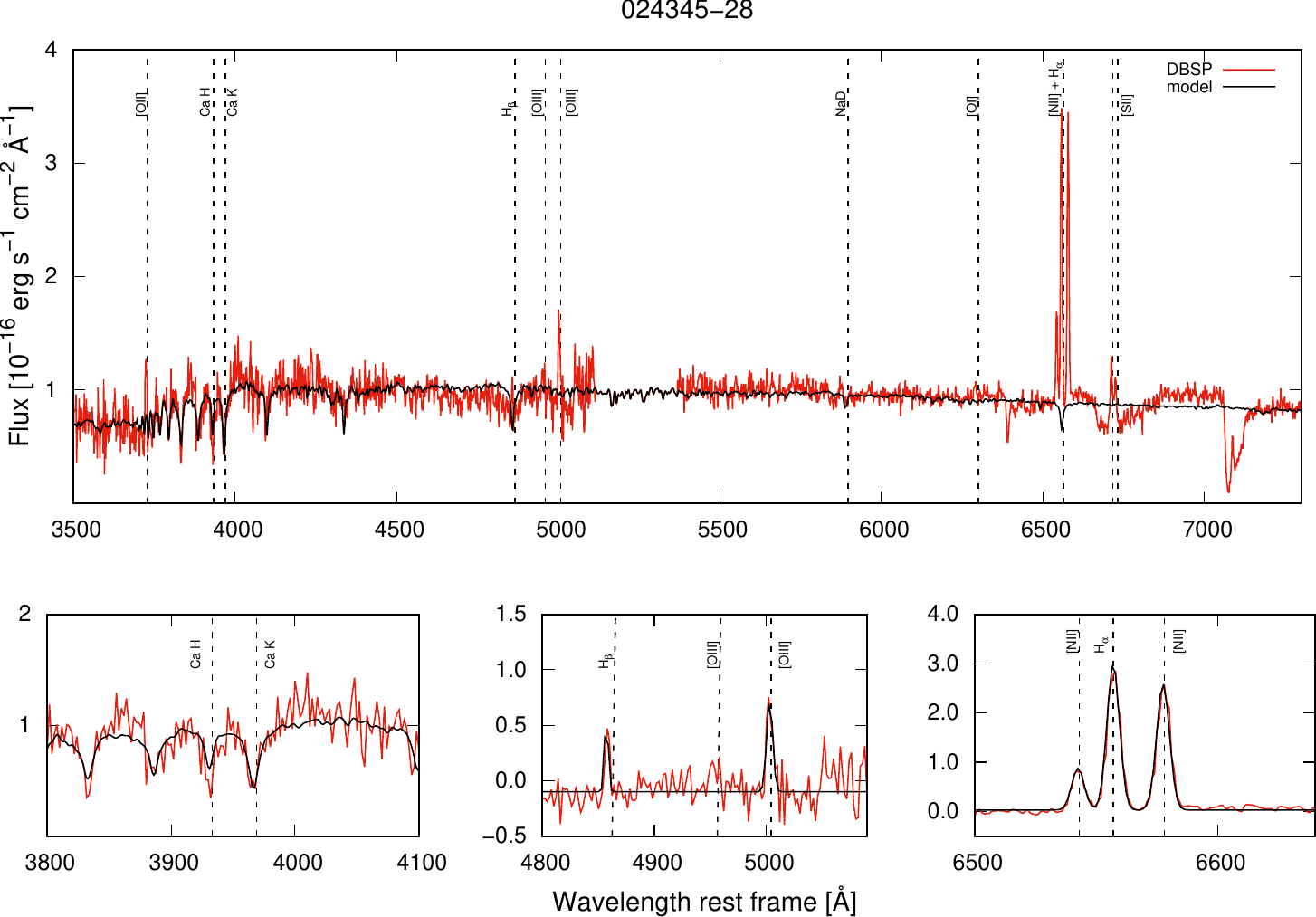}
   \includegraphics[width=0.45\textwidth]{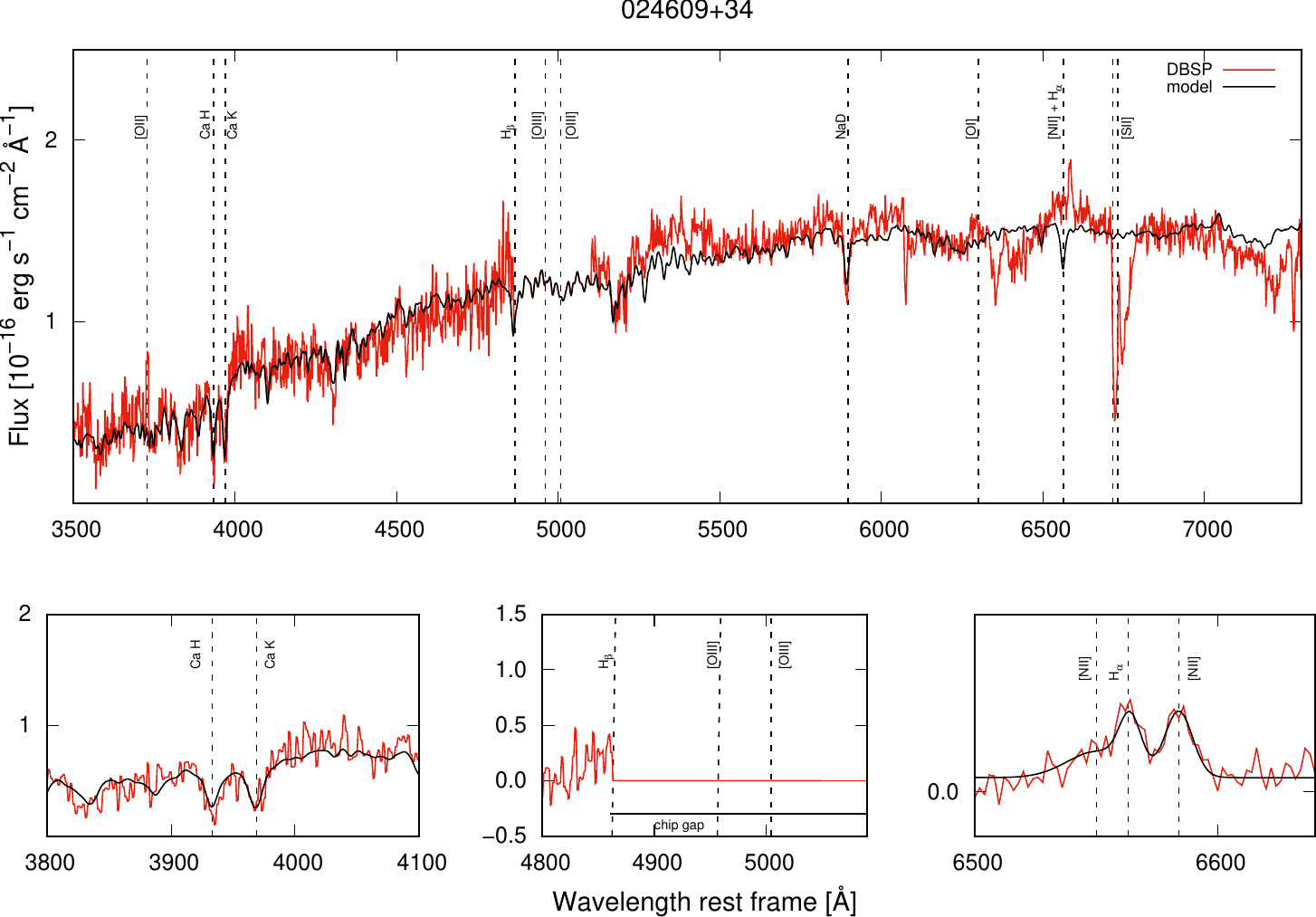}
   \includegraphics[width=0.45\textwidth]{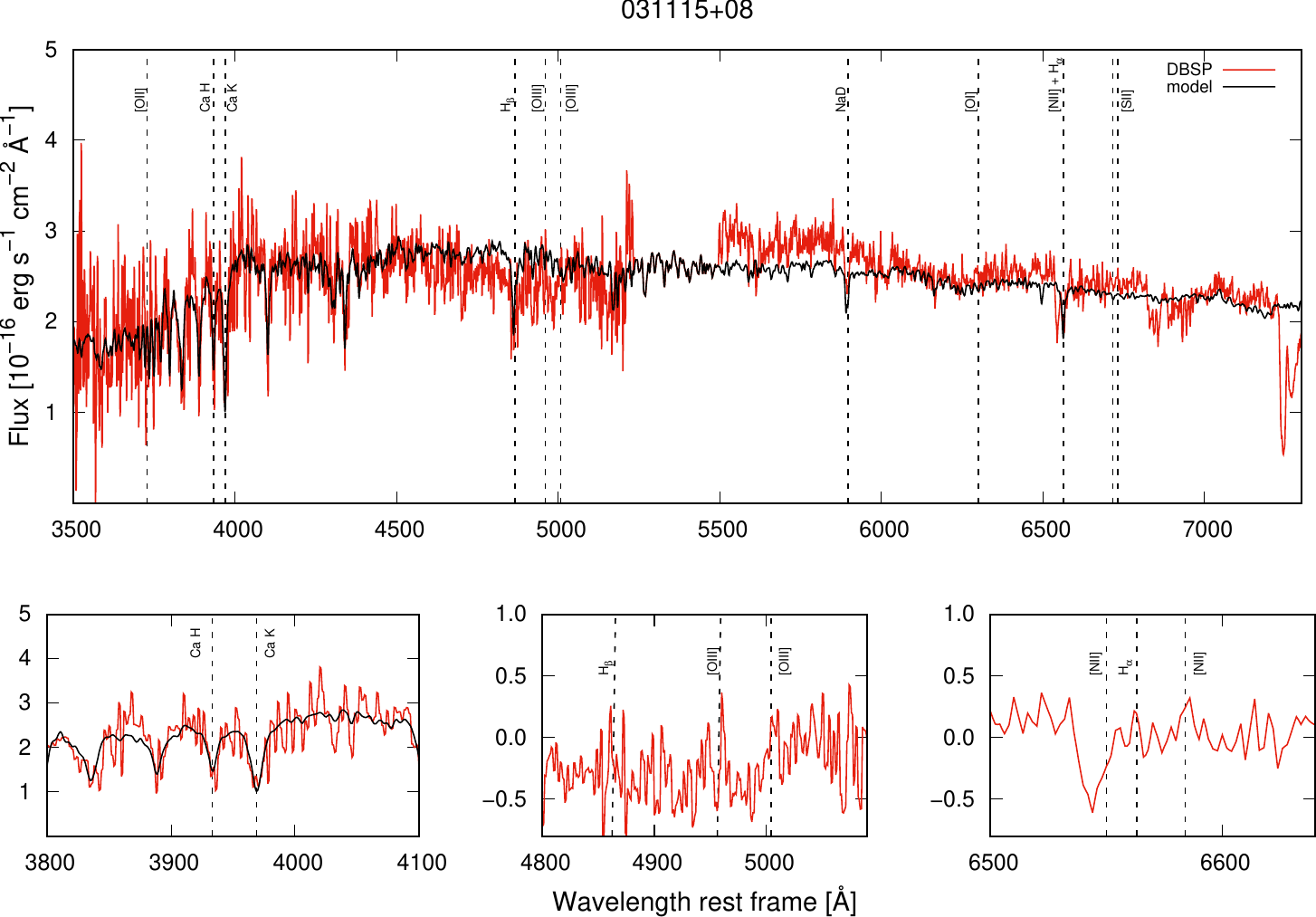}
   \includegraphics[width=0.45\textwidth]{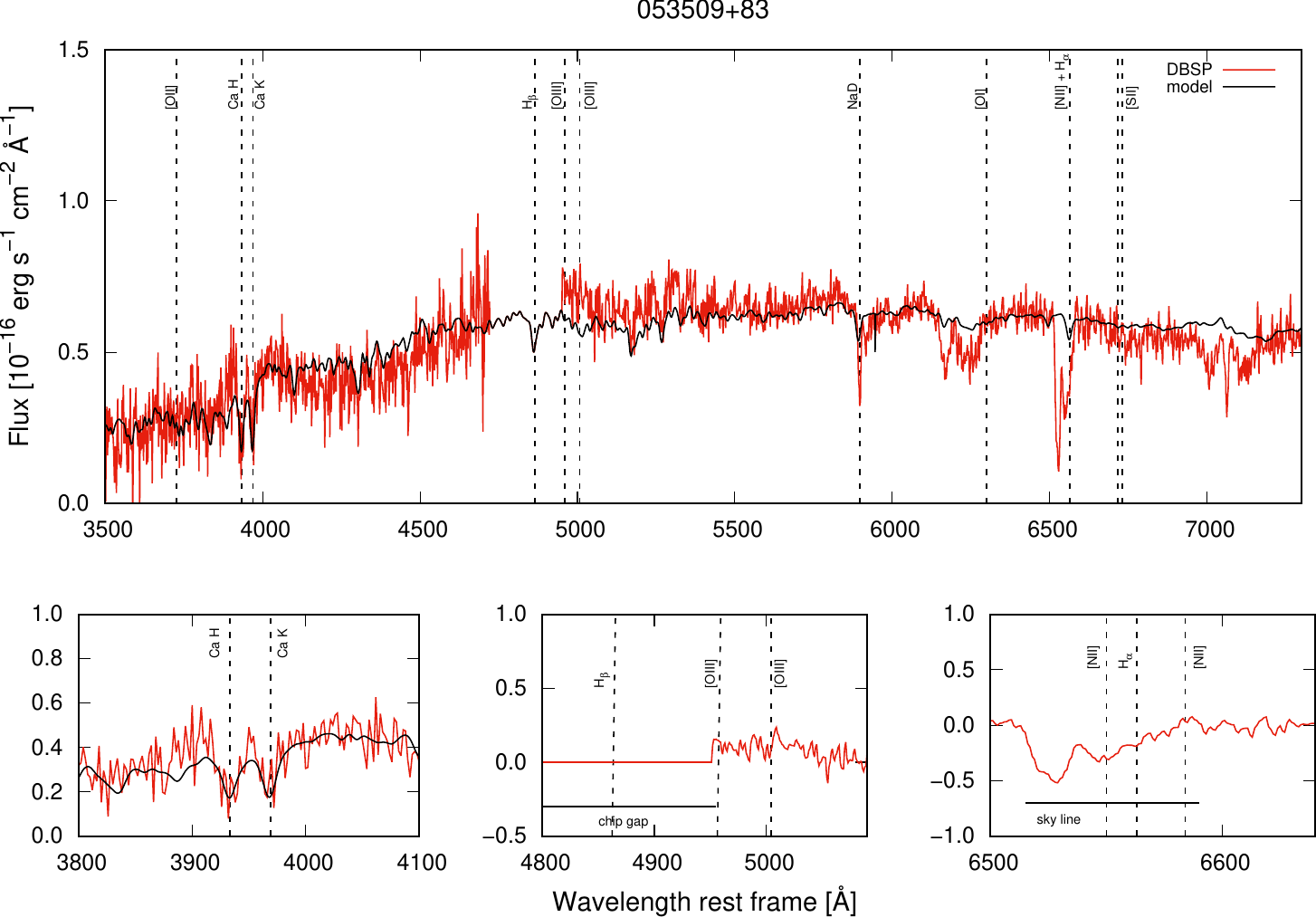}
   \includegraphics[width=0.45\textwidth]{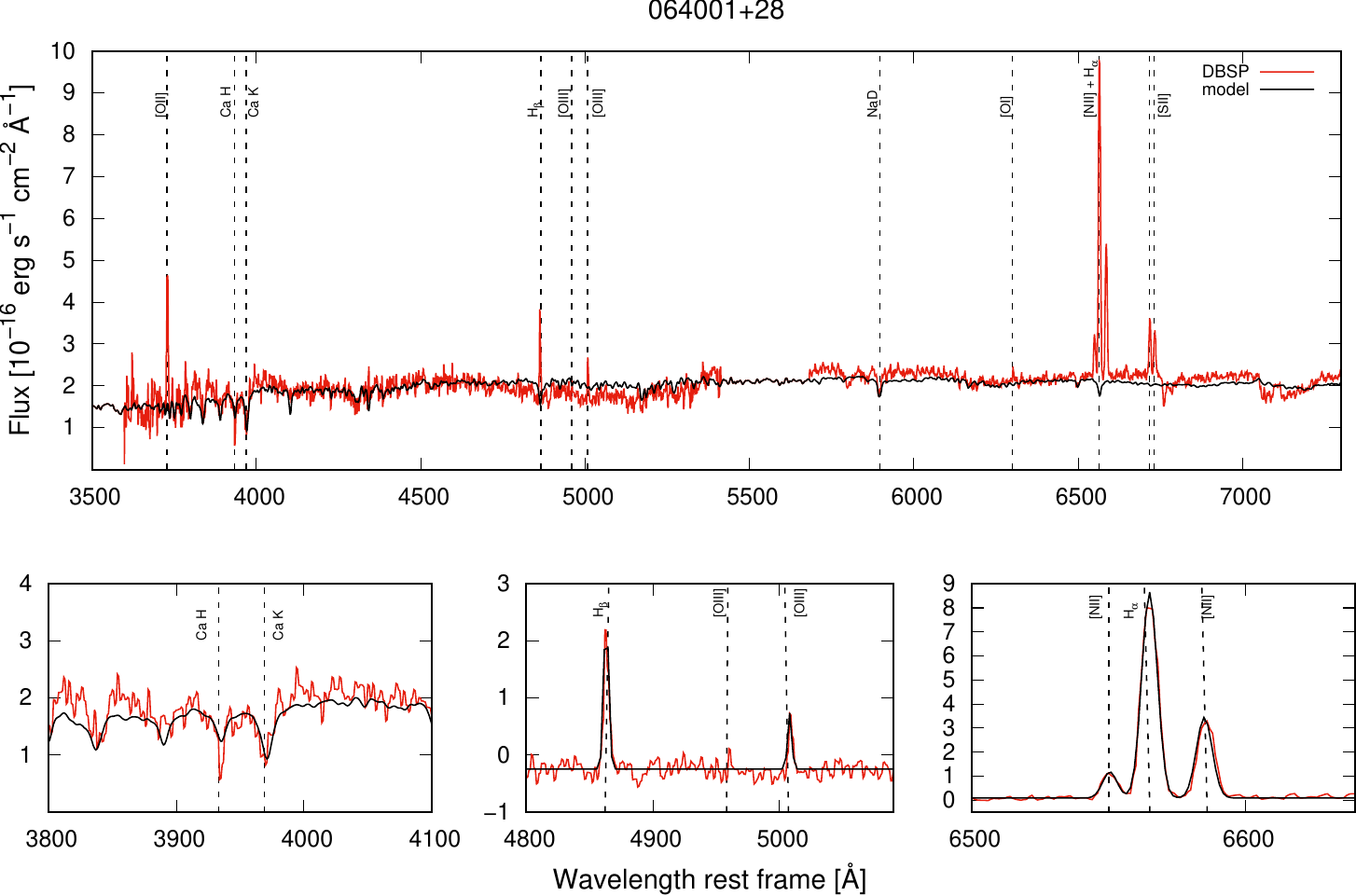}
   \includegraphics[width=0.45\textwidth]{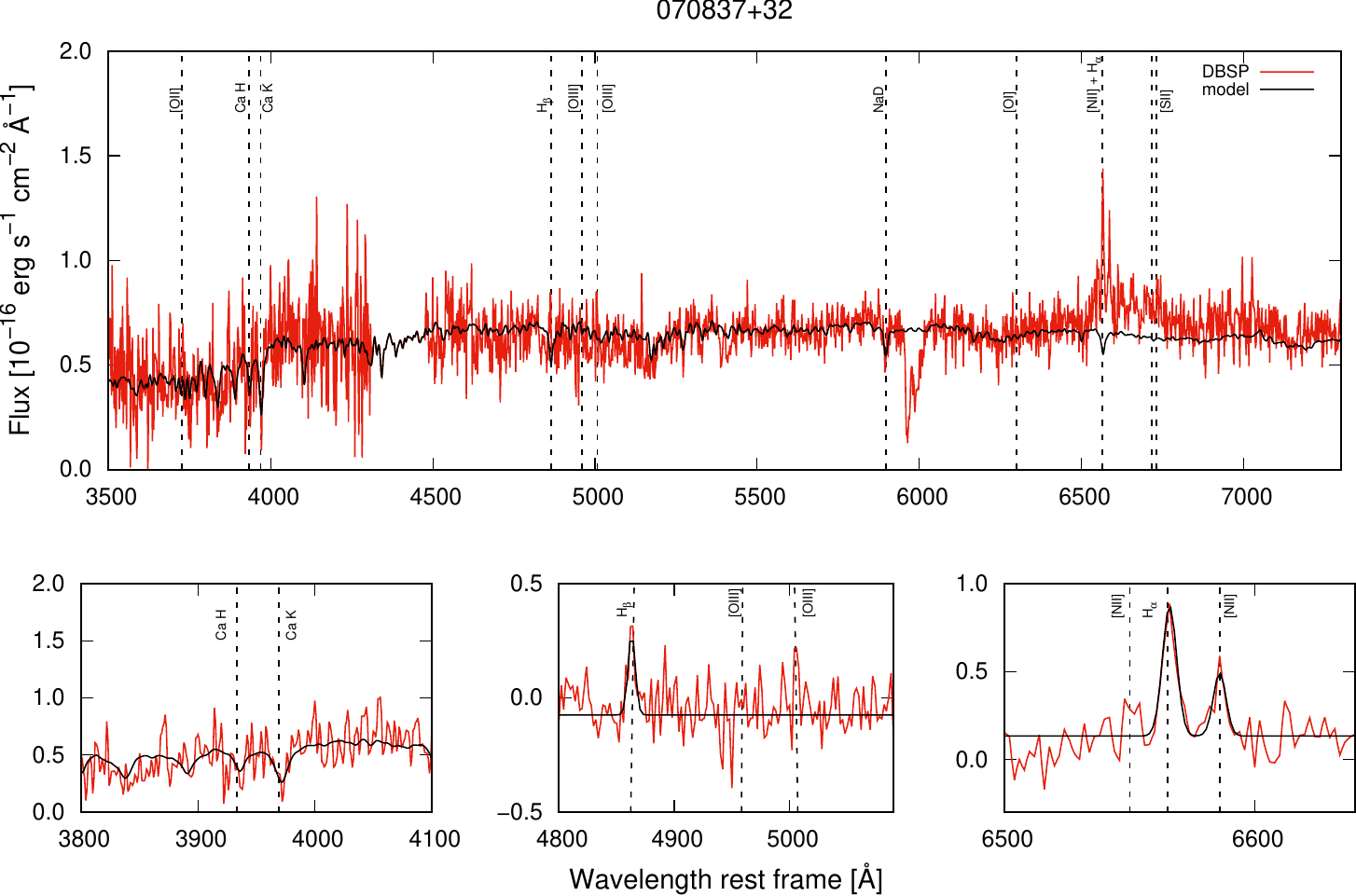}
    \end{figure}

\begin{figure}[h!]
   \centering
   \includegraphics[width=0.45\textwidth]{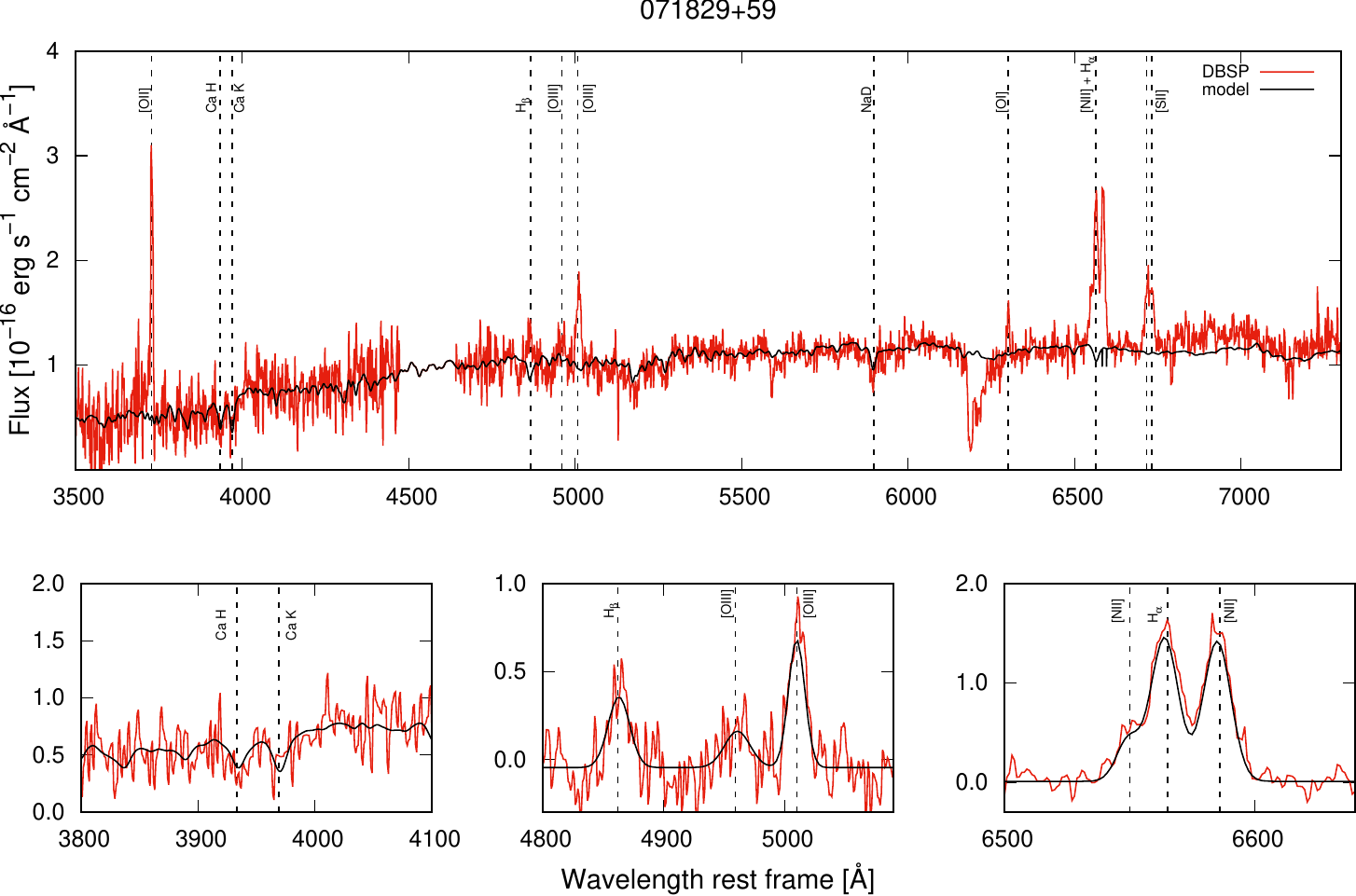}
   \includegraphics[width=0.45\textwidth]{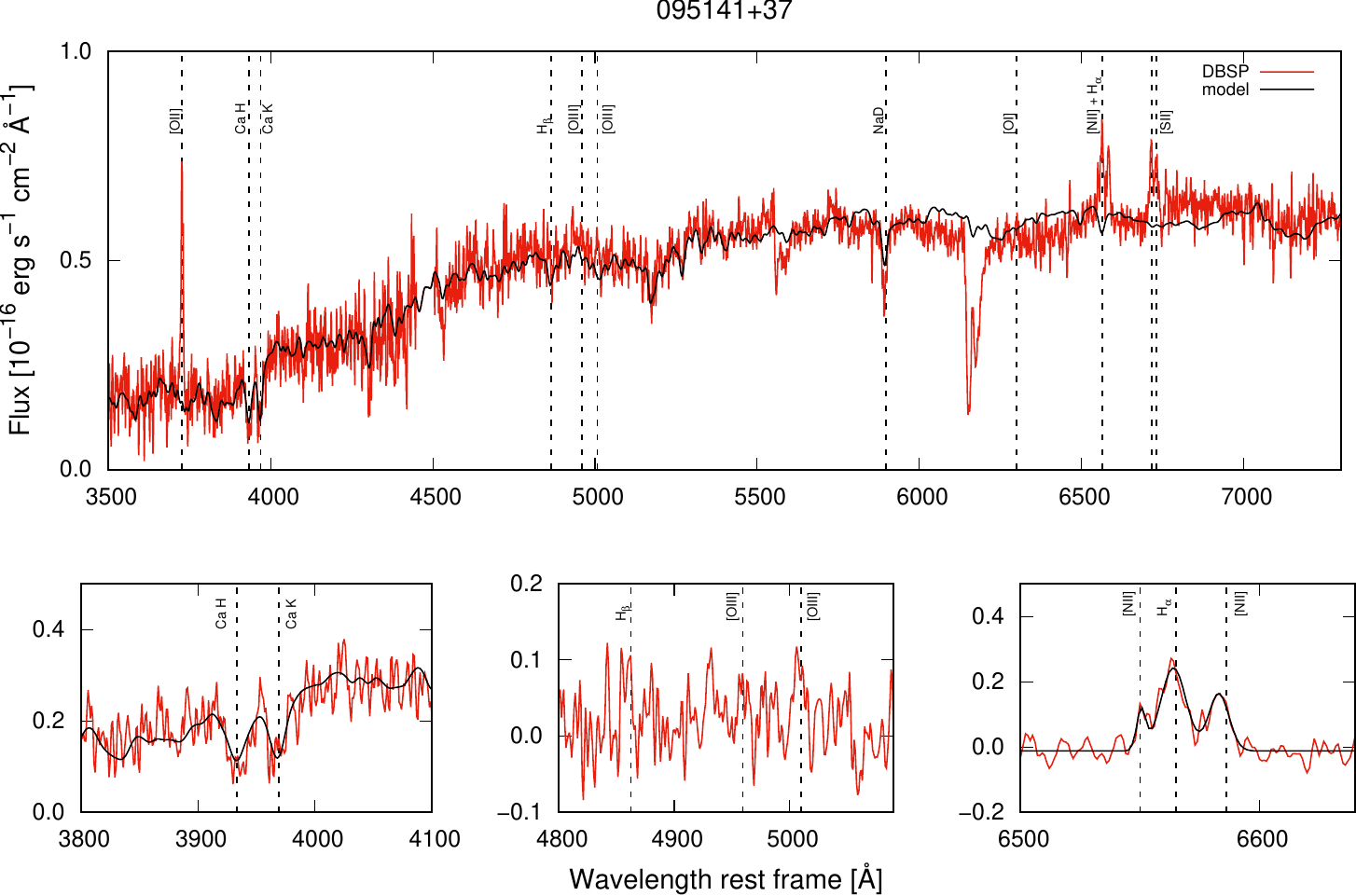}
   \includegraphics[width=0.45\textwidth]{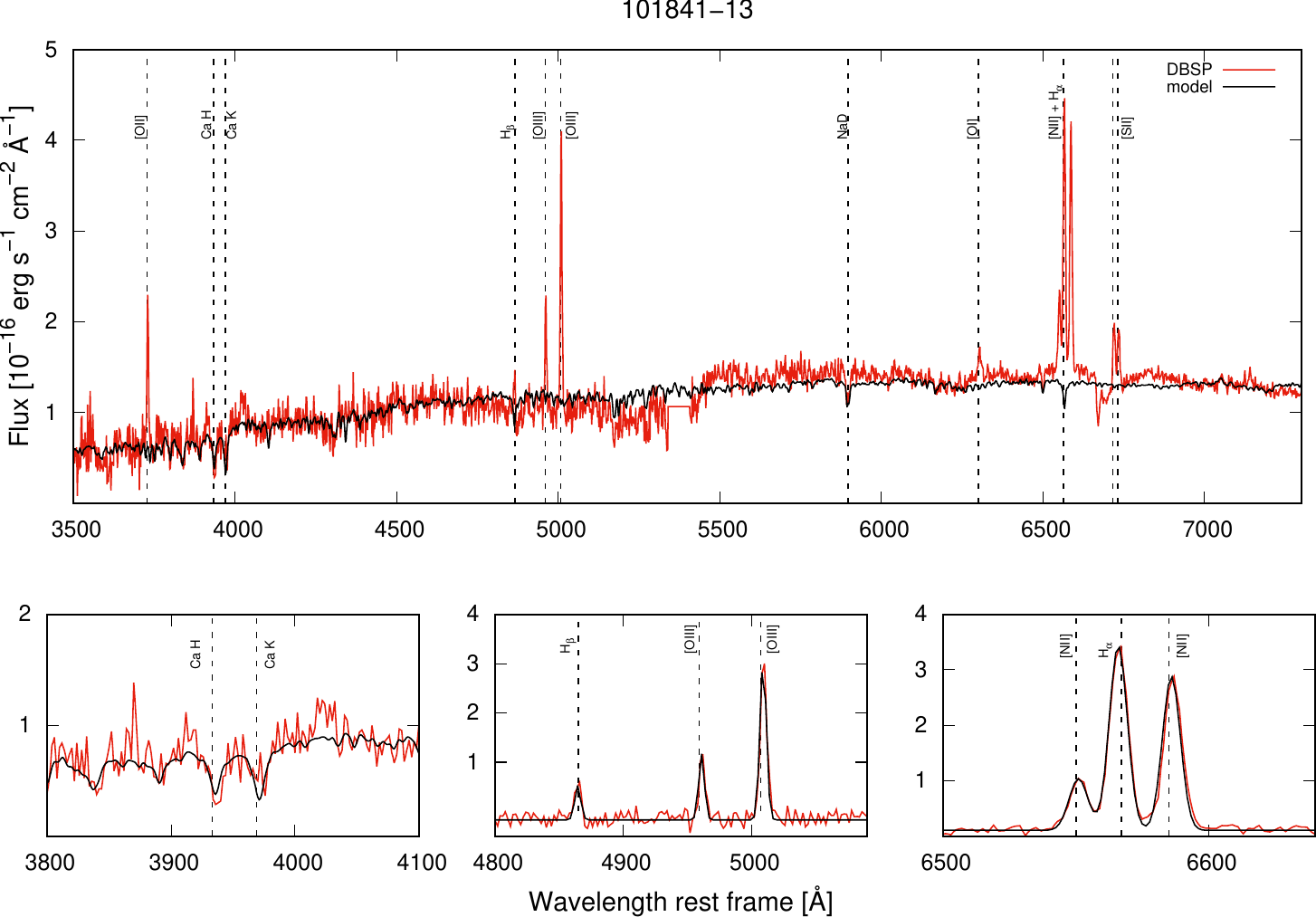}
   \includegraphics[width=0.45\textwidth]{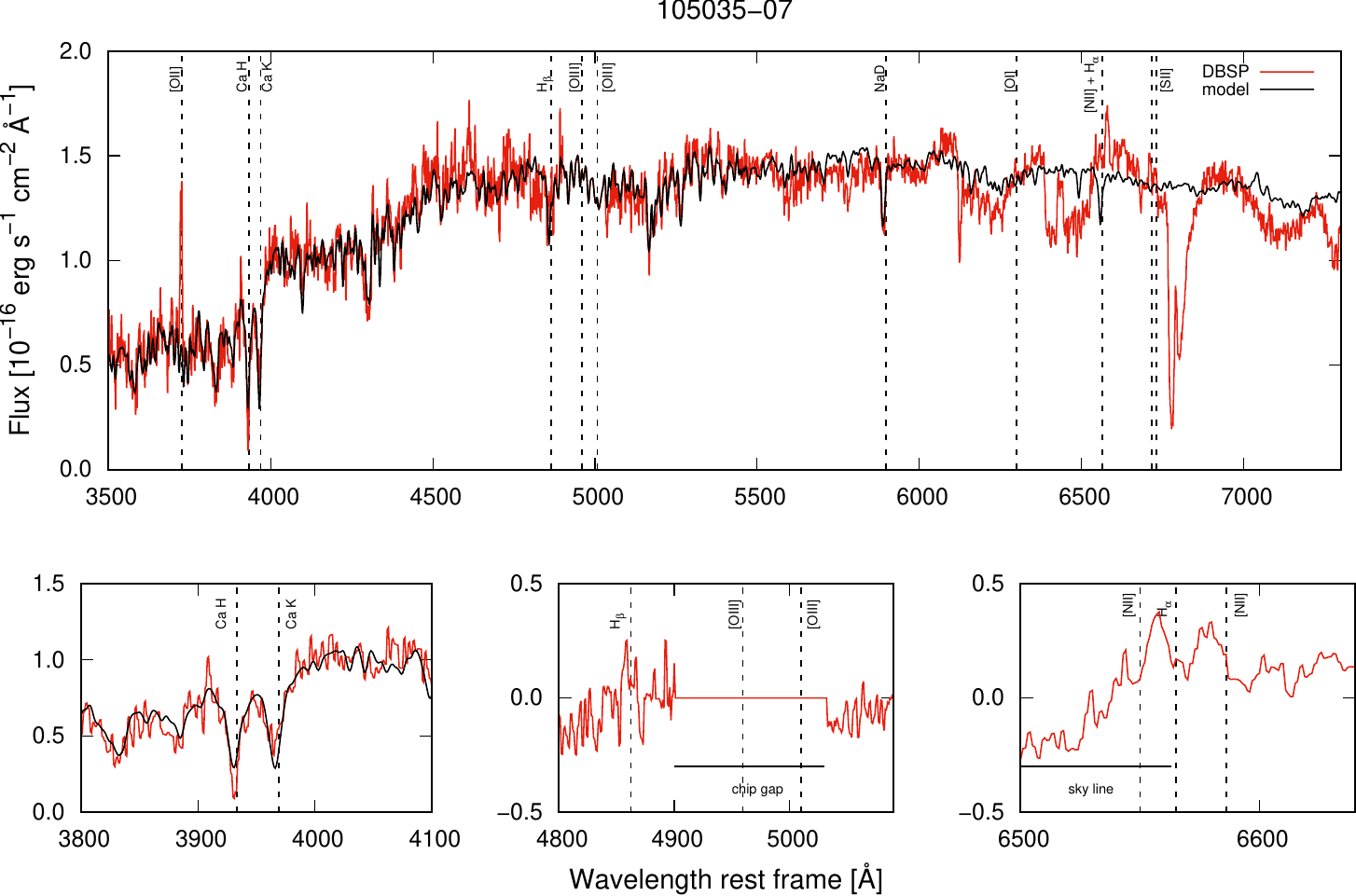}
   \includegraphics[width=0.45\textwidth]{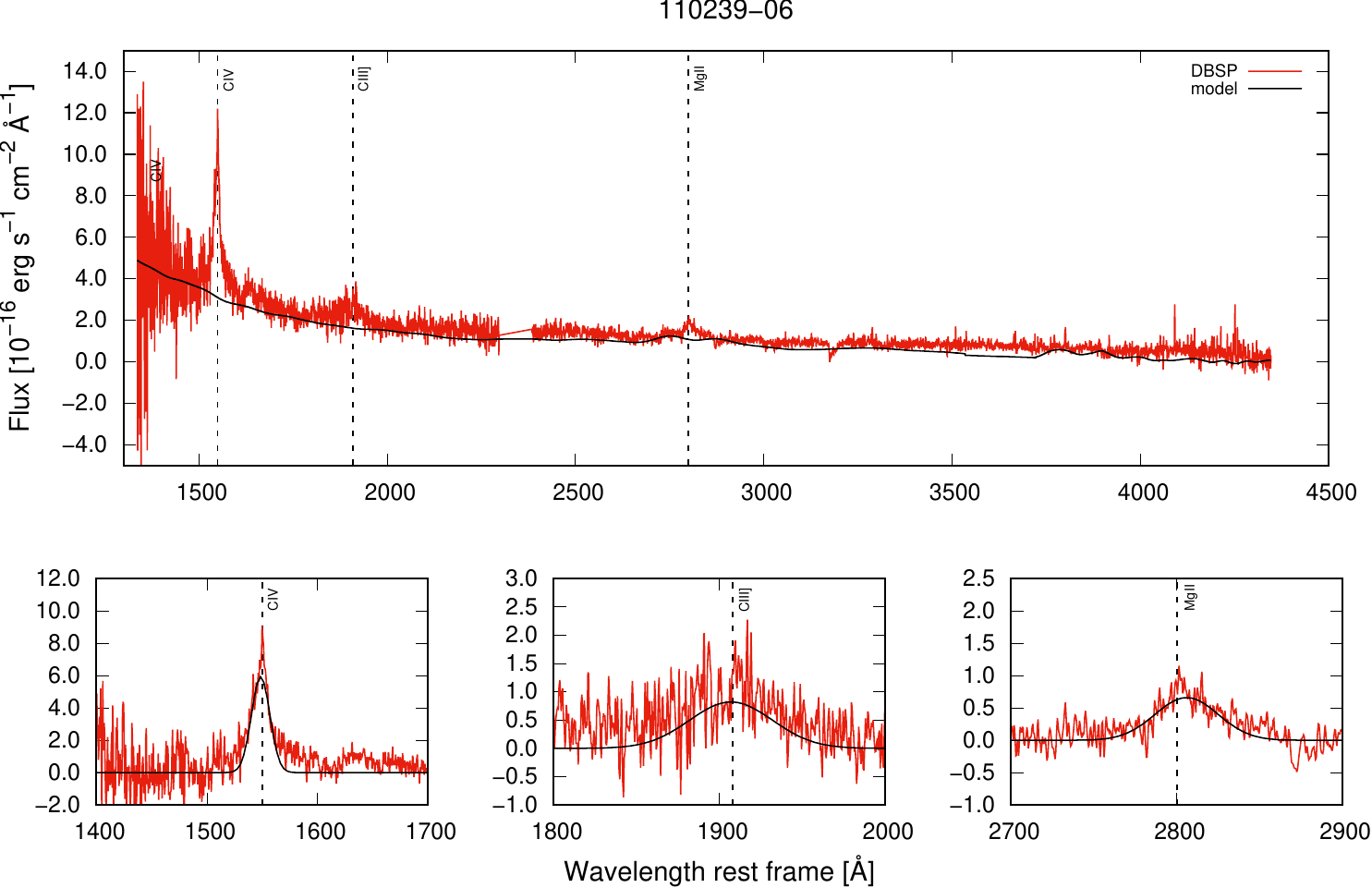}
   \includegraphics[width=0.45\textwidth]{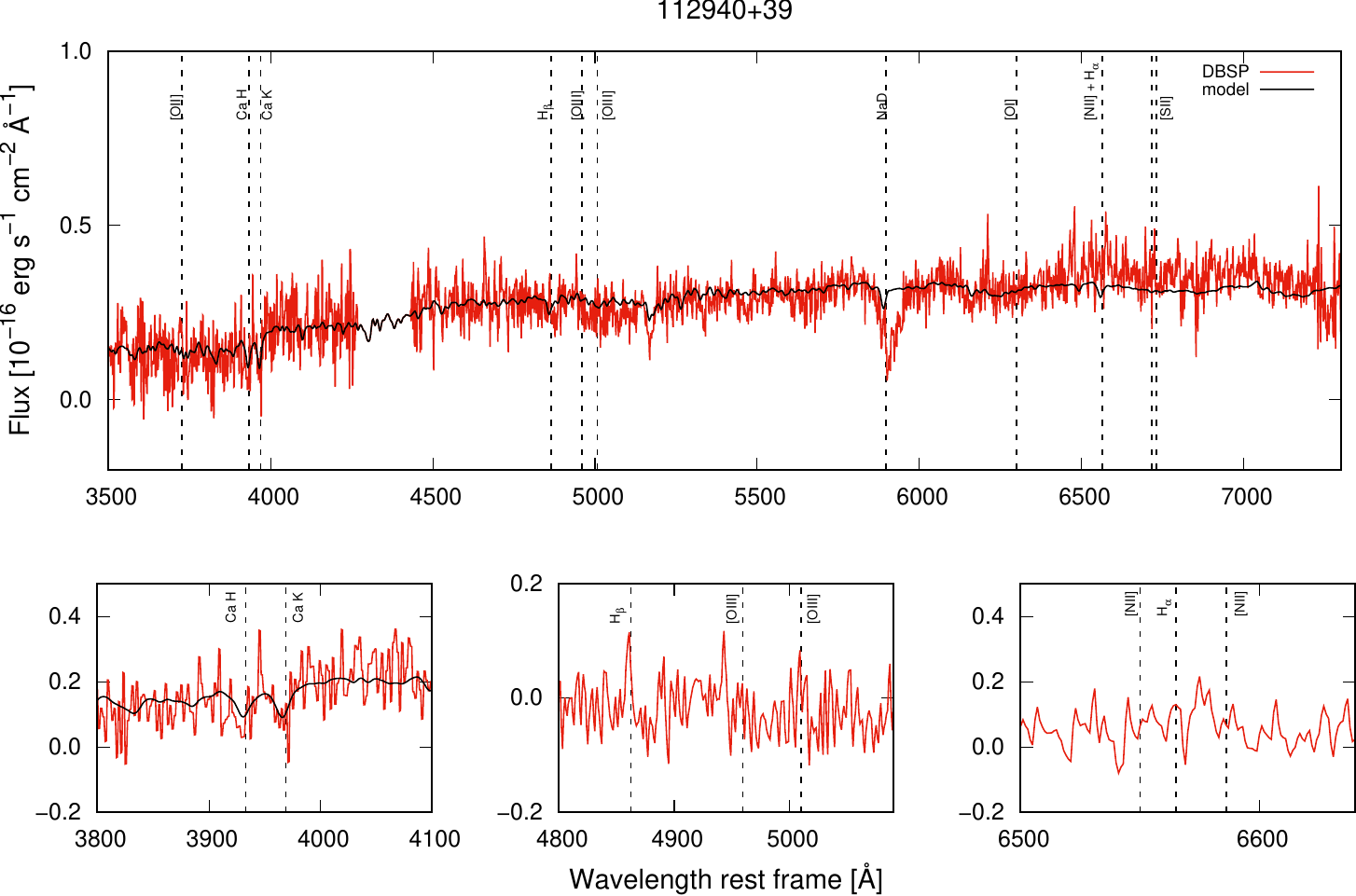}
   \includegraphics[width=0.45\textwidth]{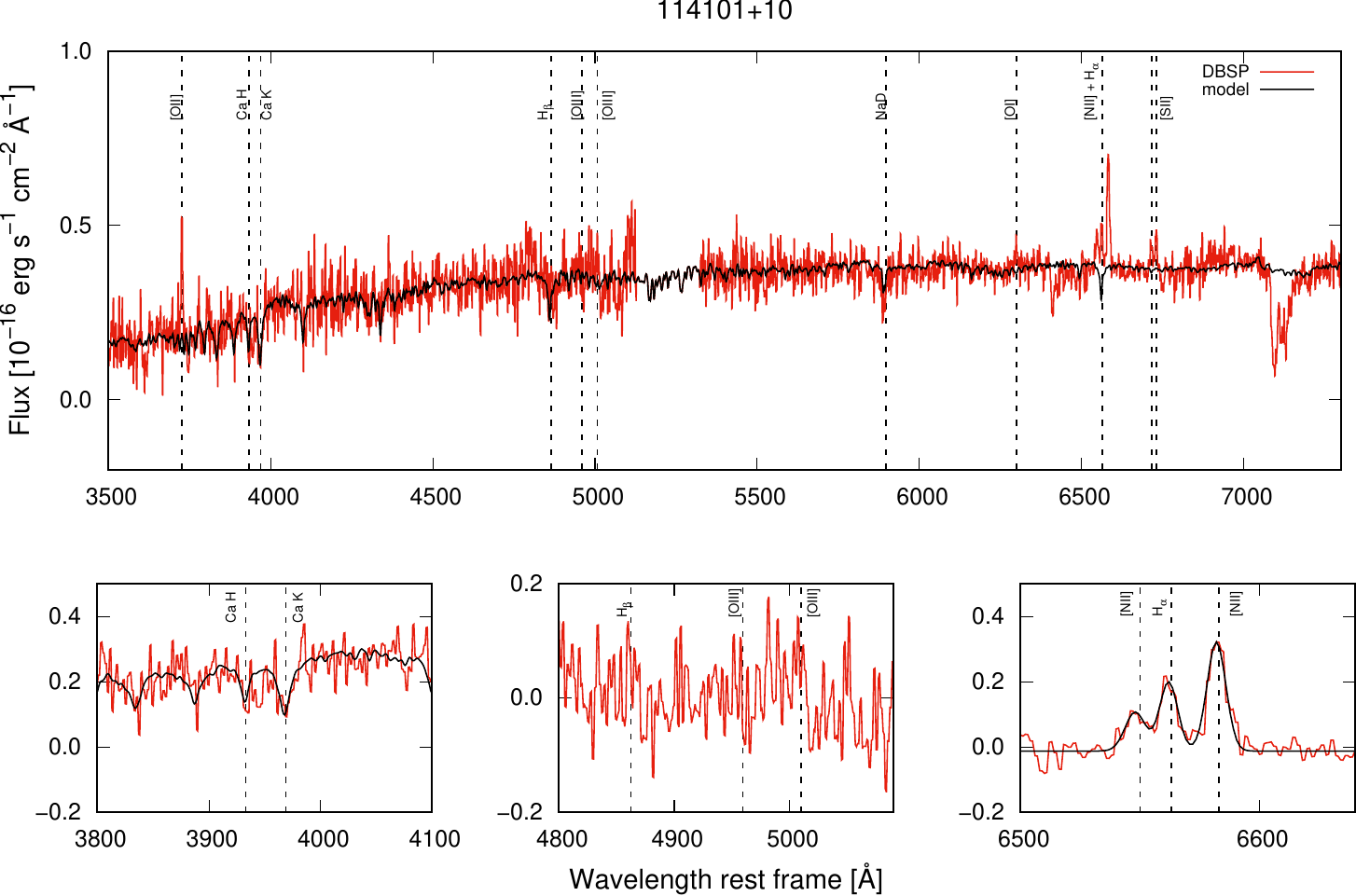}
   \includegraphics[width=0.45\textwidth]{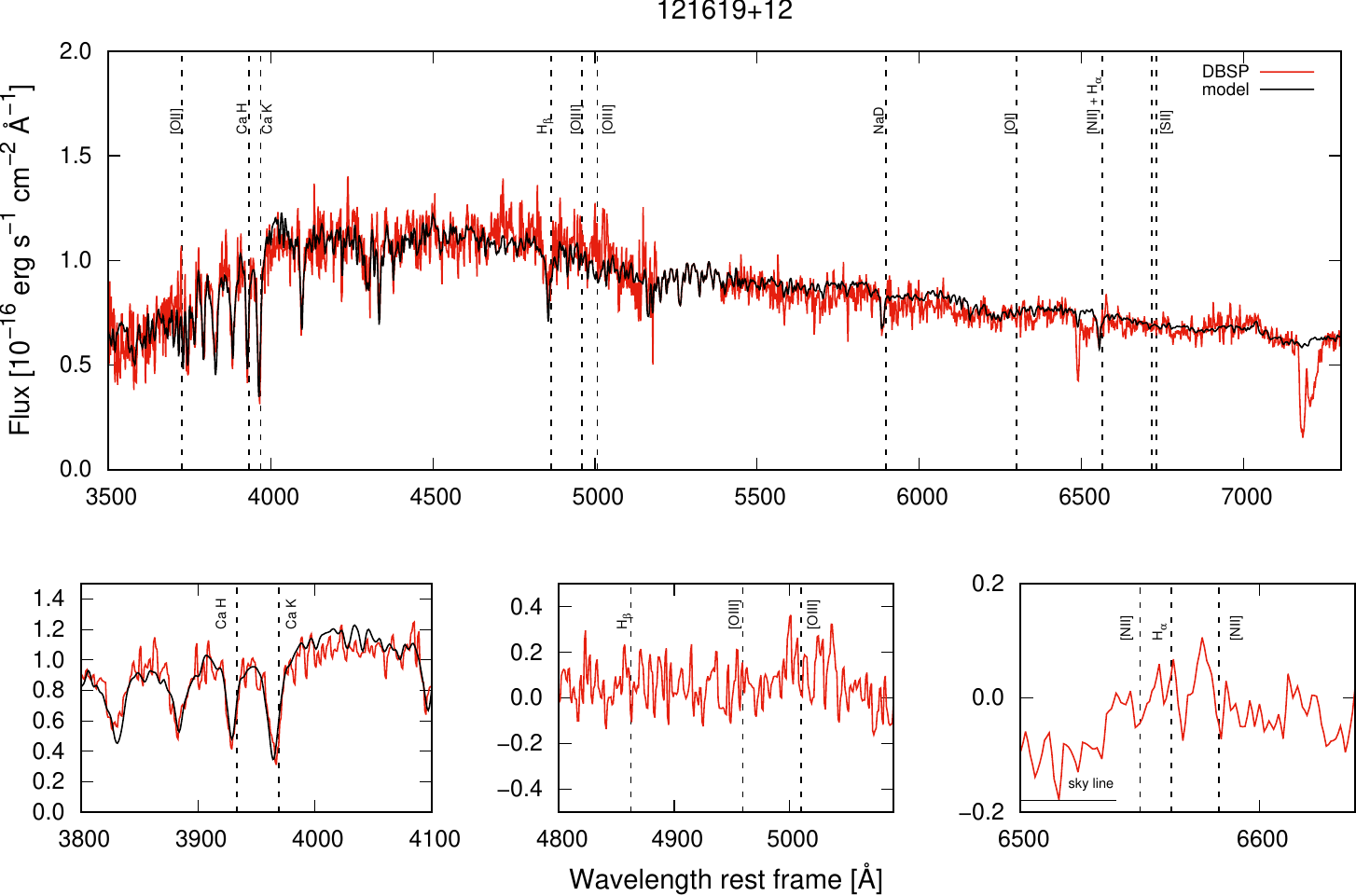}
    \end{figure}

   \begin{figure}[h!]
   \centering
   \includegraphics[width=0.45\textwidth]{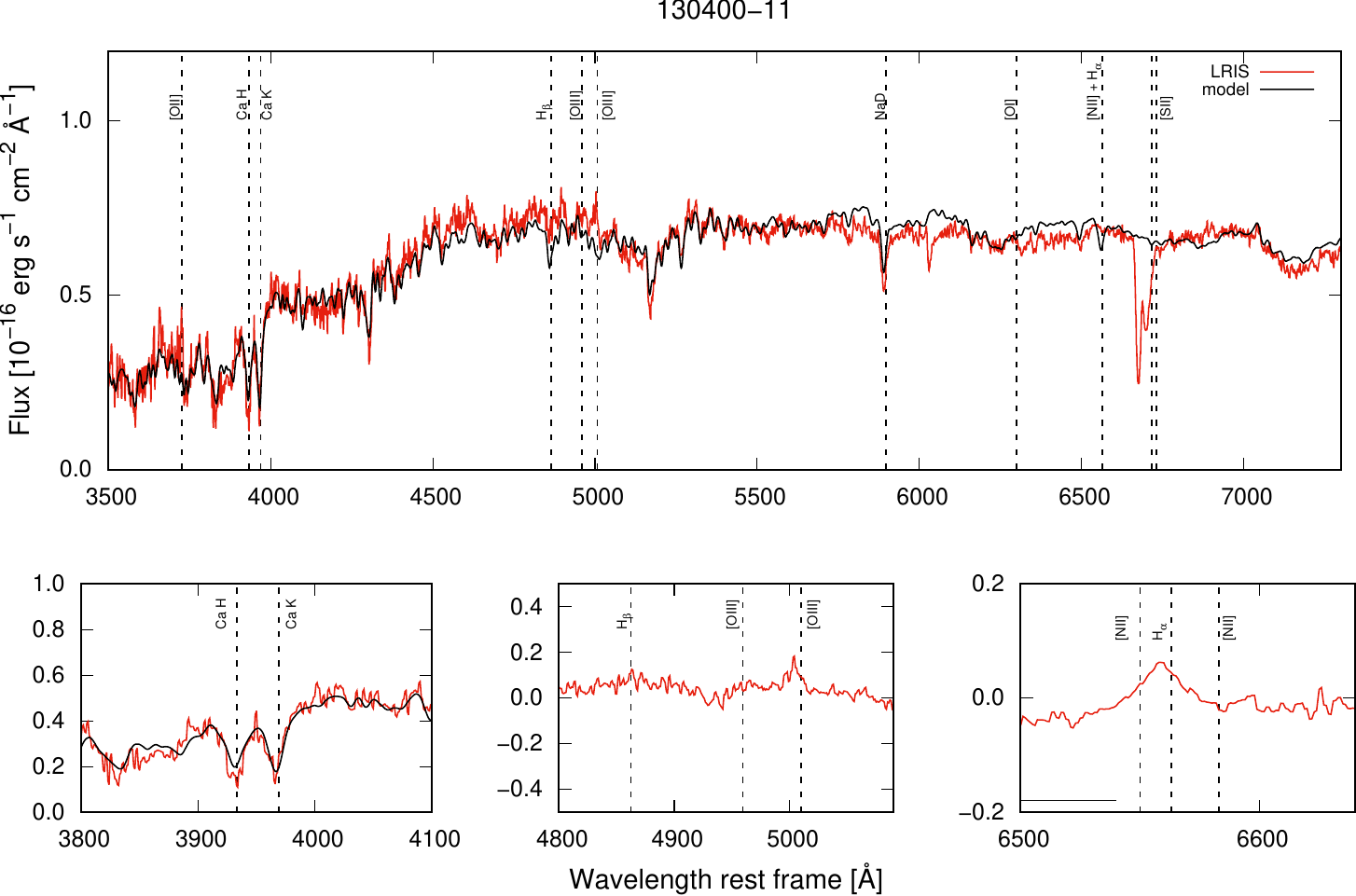}
   \includegraphics[width=0.45\textwidth]{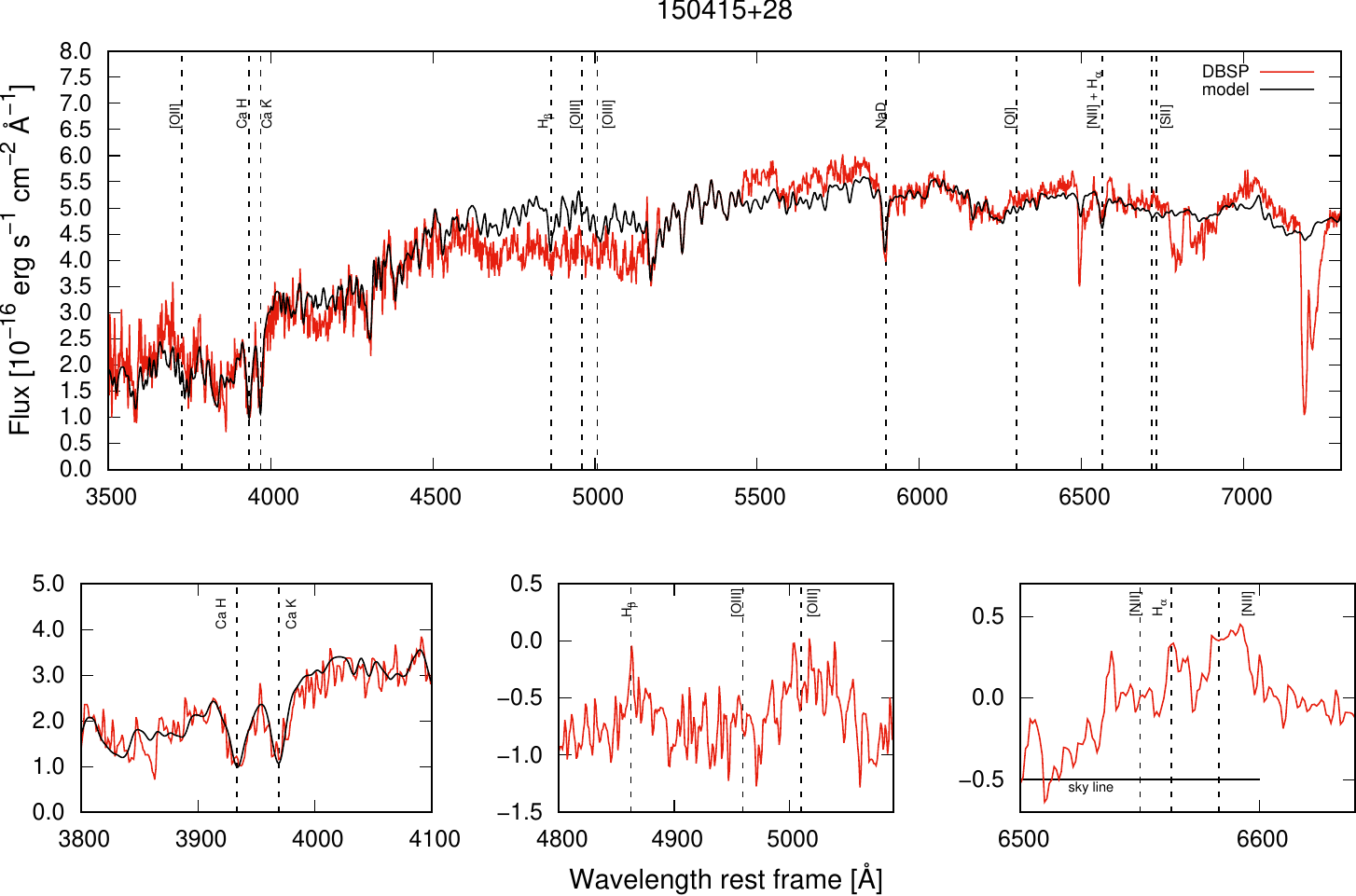}
   \includegraphics[width=0.45\textwidth]{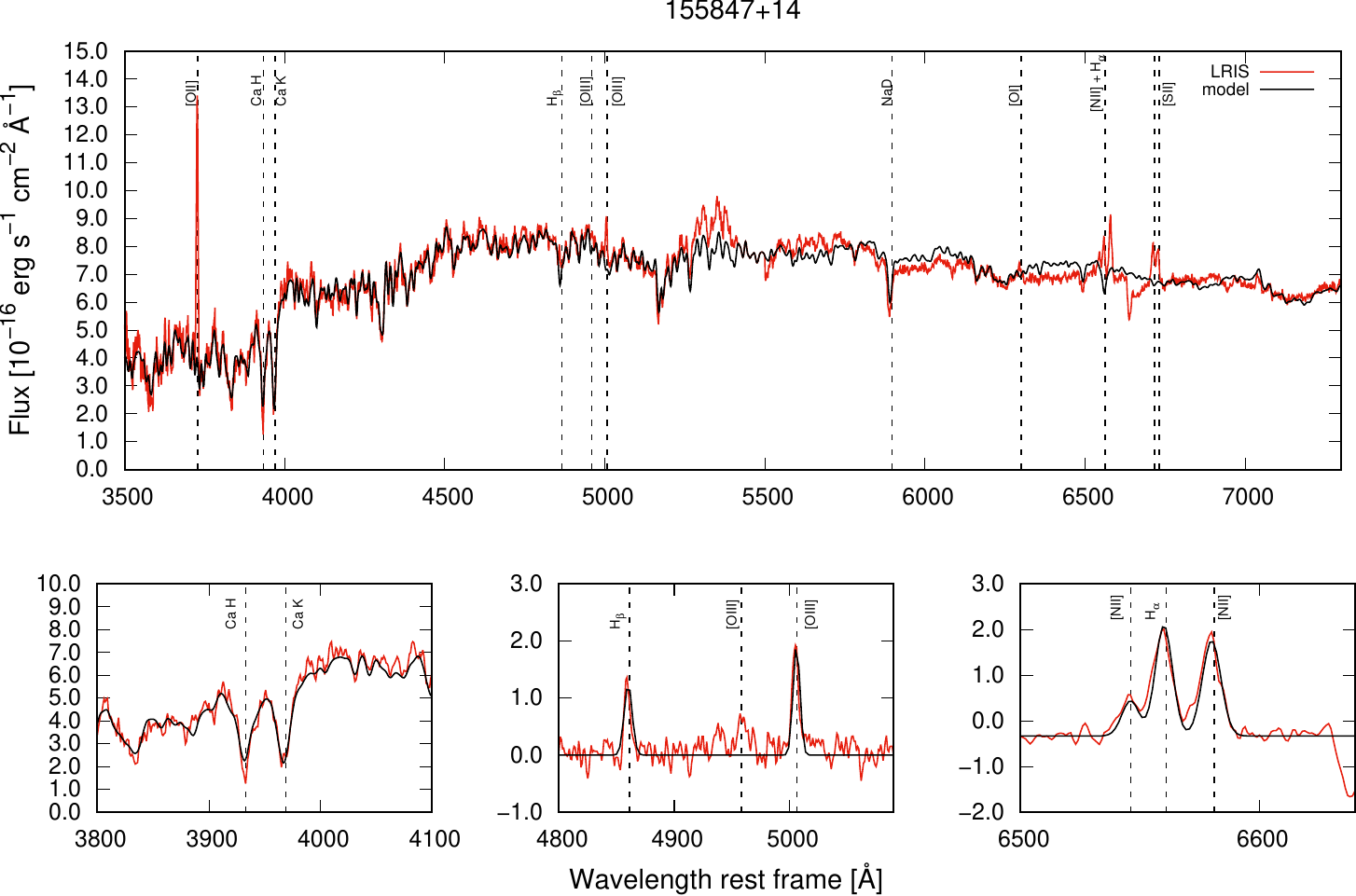}
   \includegraphics[width=0.45\textwidth]{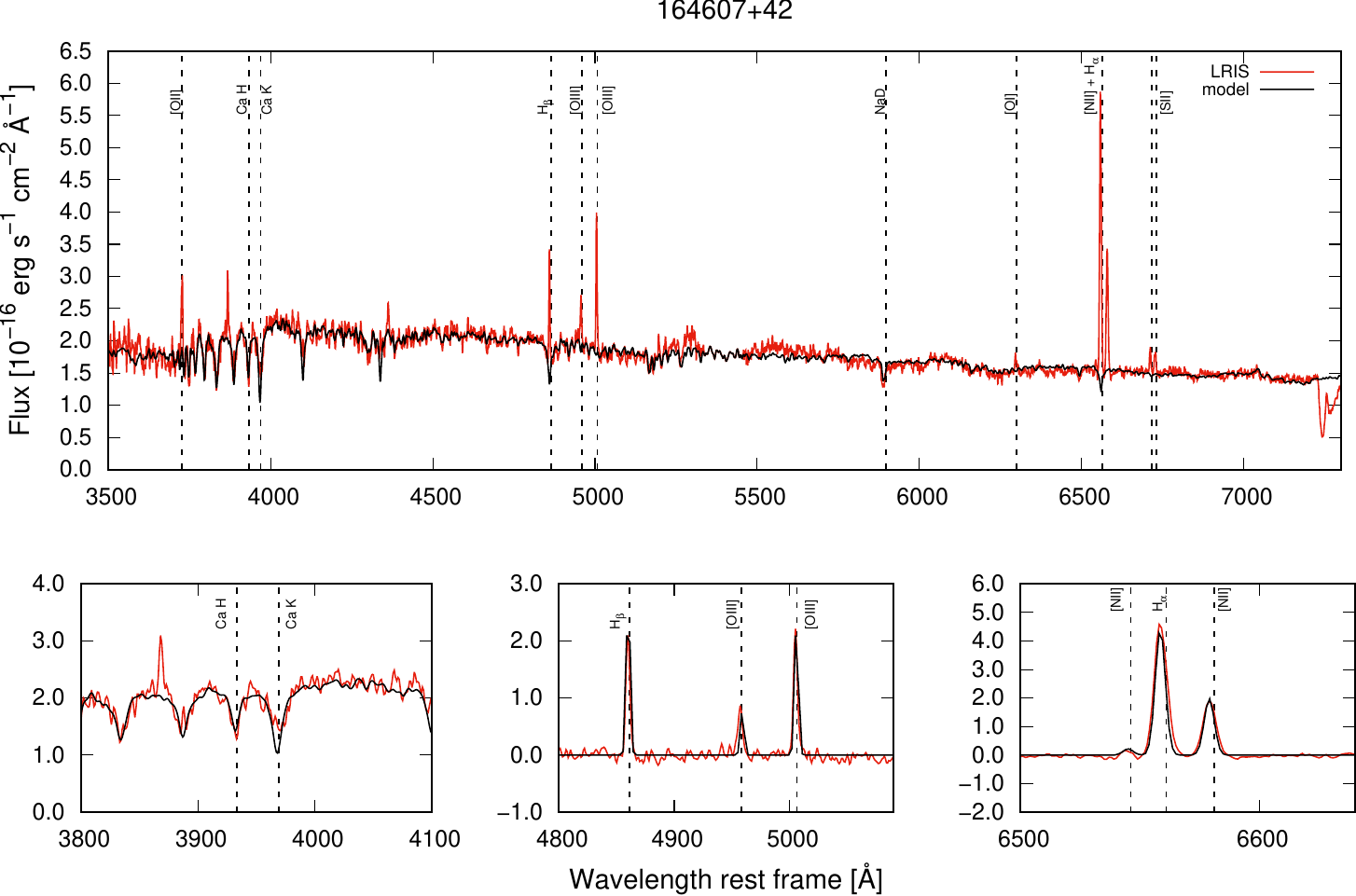}
   \includegraphics[width=0.45\textwidth]{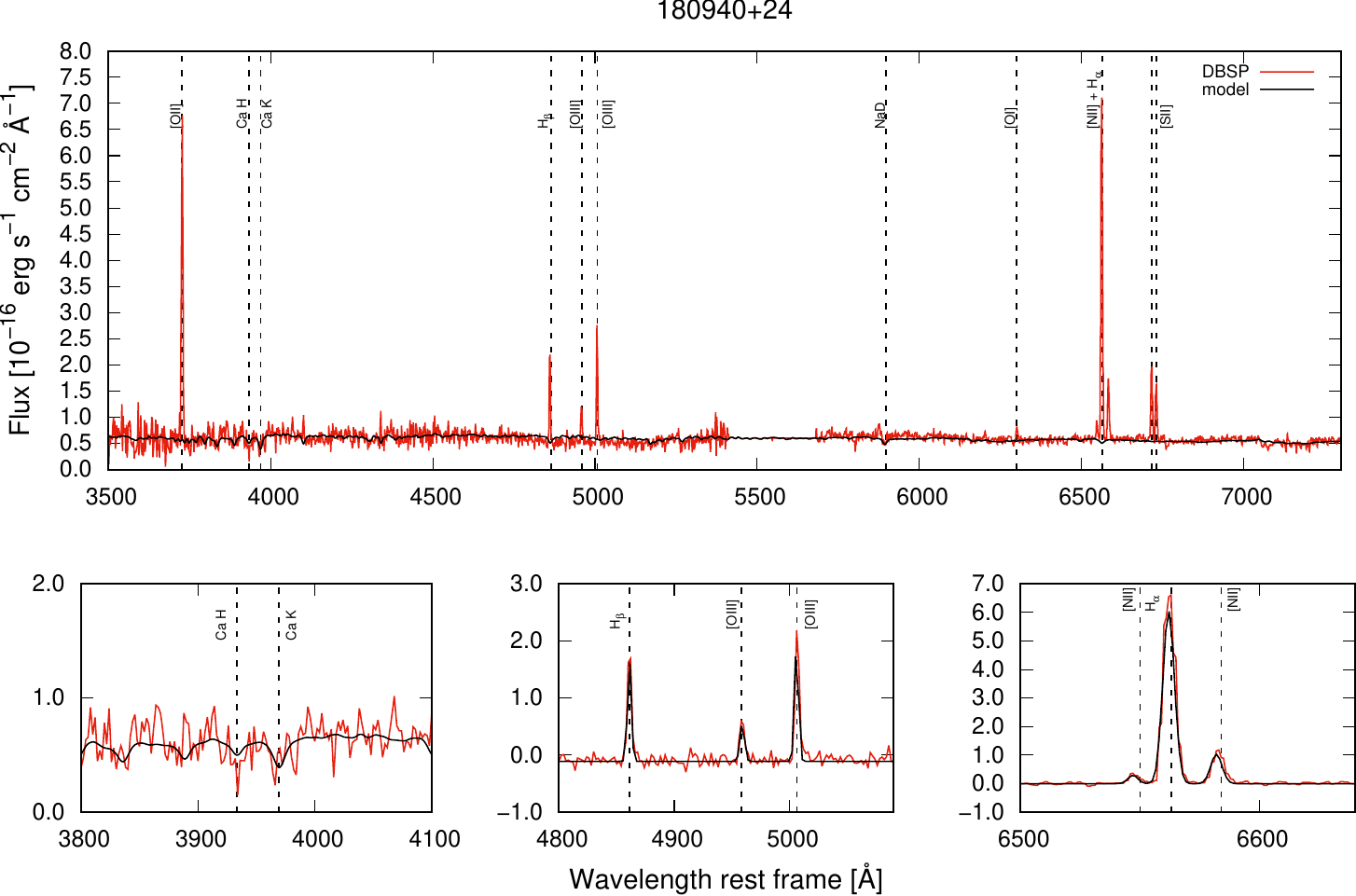}
   \includegraphics[width=0.45\textwidth]{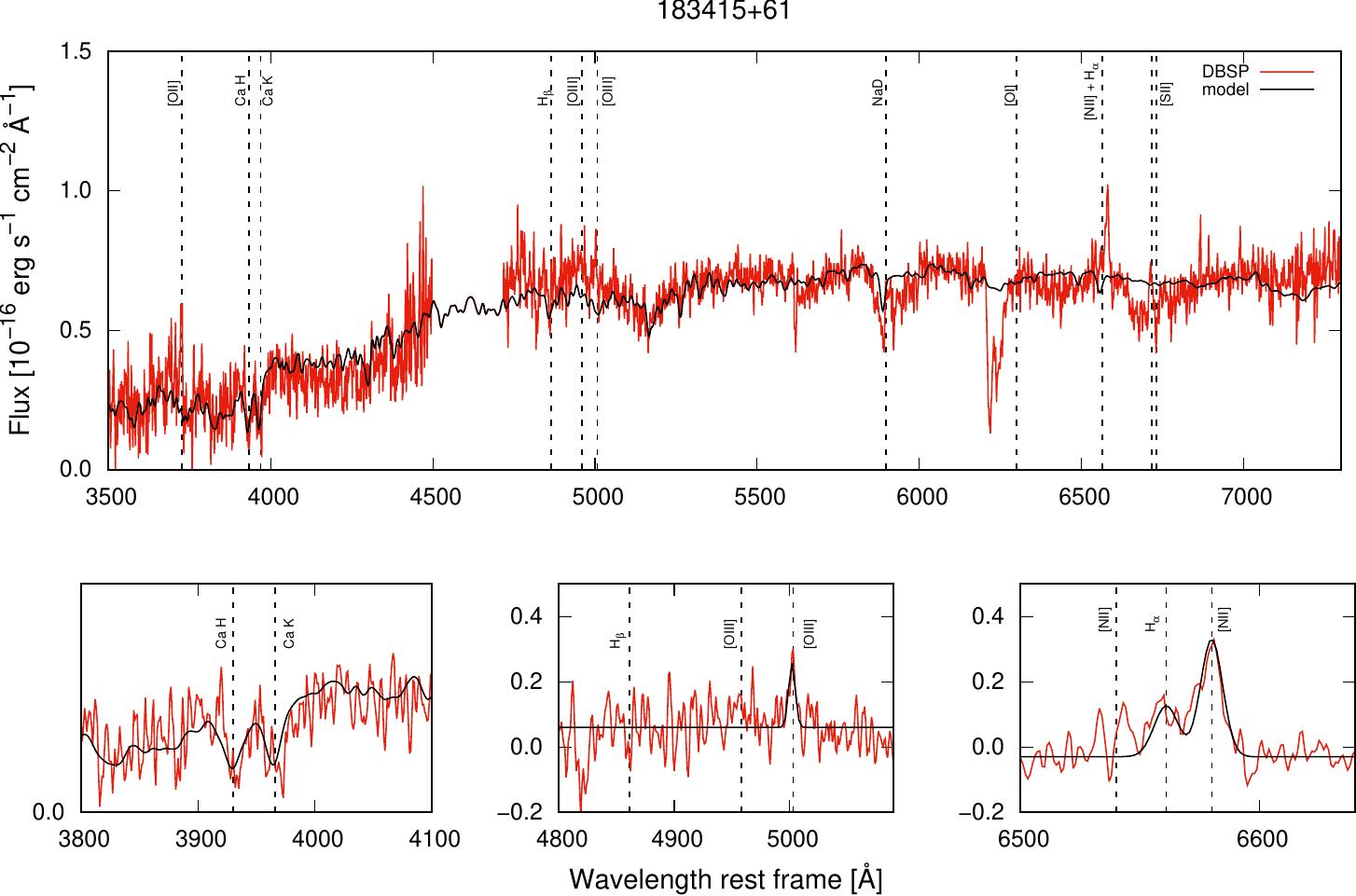}
   \includegraphics[width=0.45\textwidth]{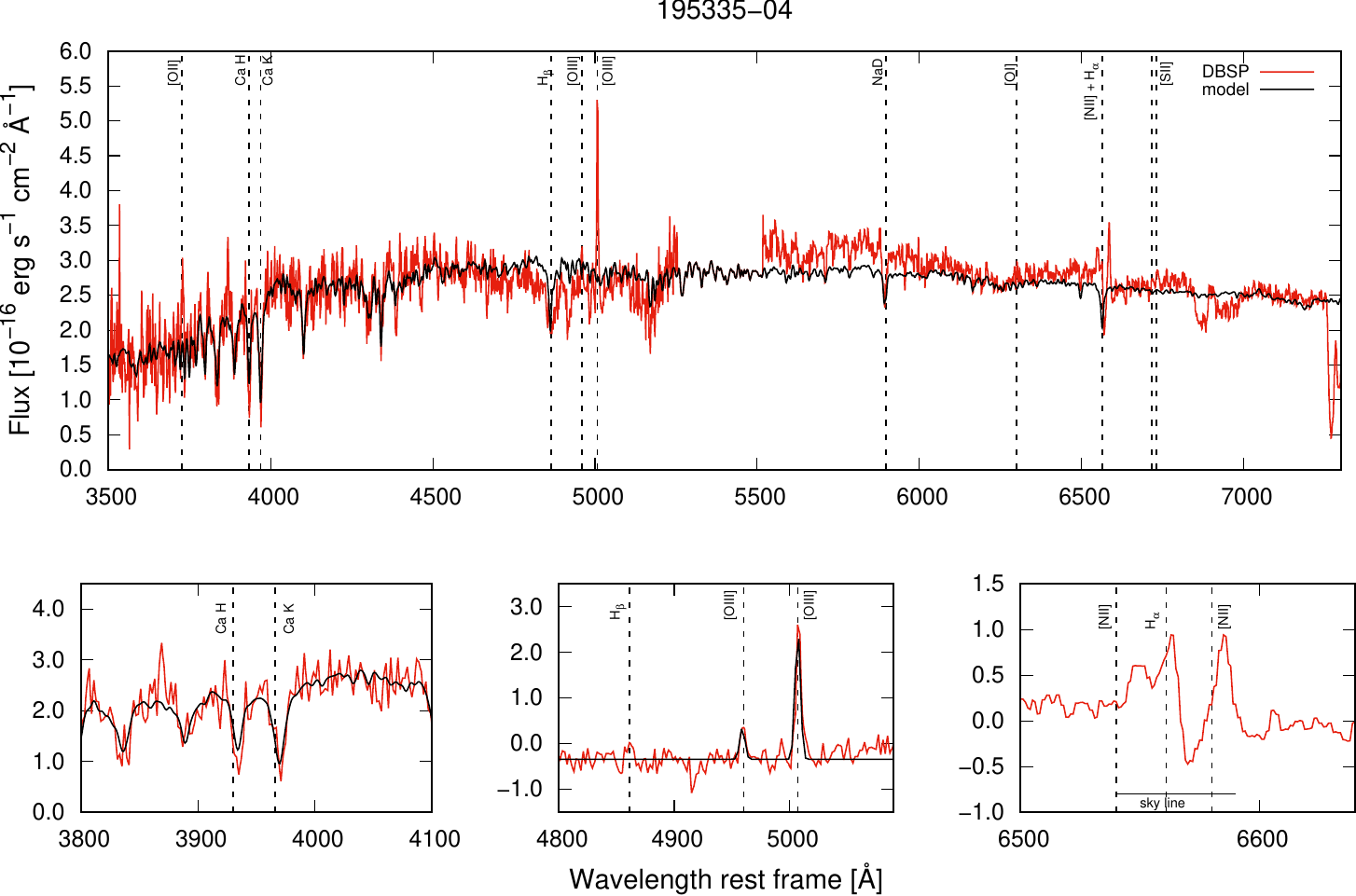}
   \includegraphics[width=0.45\textwidth]{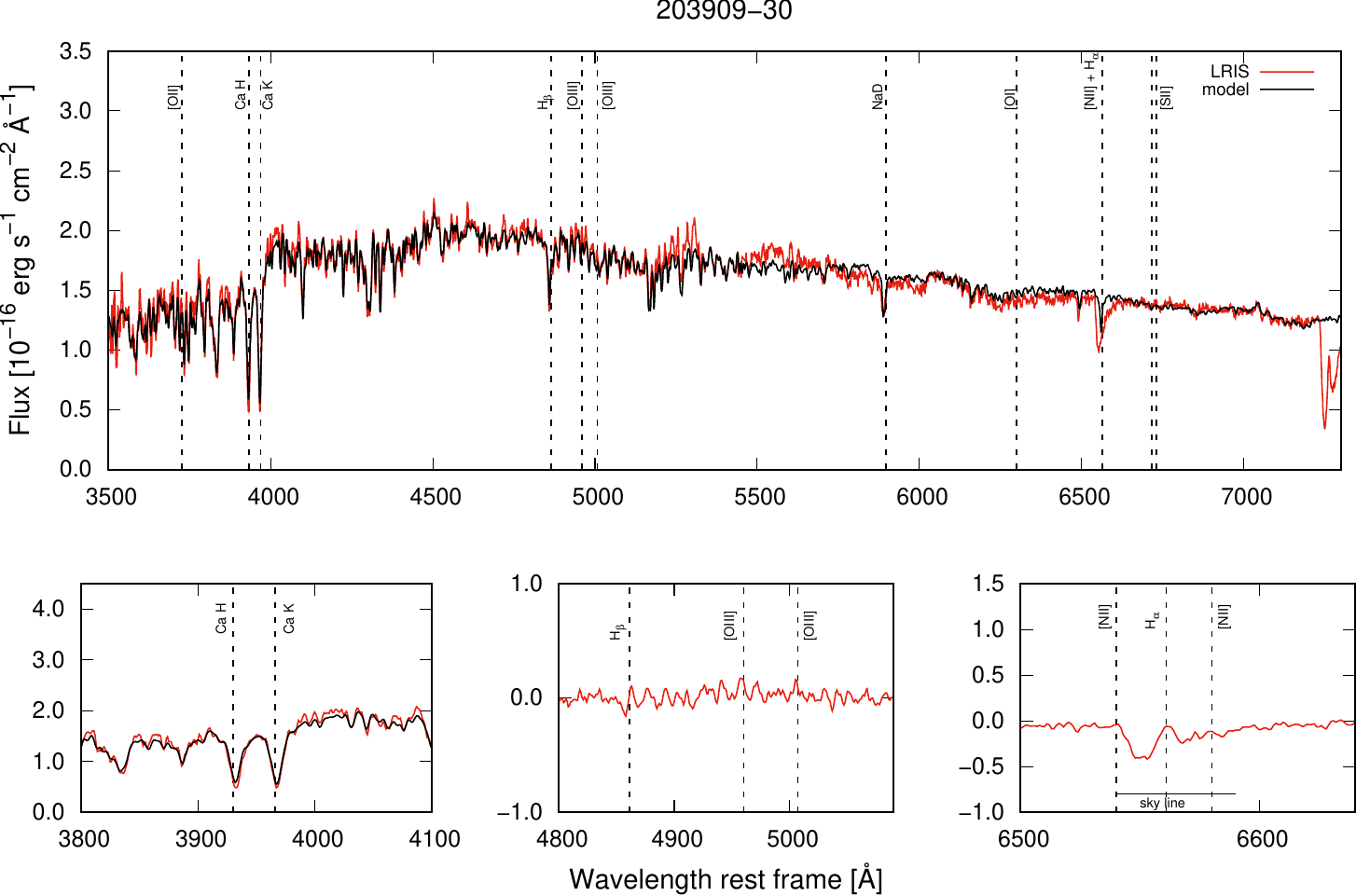}
  \end{figure}

\end{appendix}

\end{document}